# Ordered patches cap the stability of disordered frameworks

Joohan Nam[1]*

[1]Chair of Inorganic Chemistry I, Technische Universität Dresden, Bergstraße 66, 01069 Dresden, Germany.

*e-mail: joohan.nam@mailbox.tu-dresden.de

**Structural disorder can stabilize frameworks against pressure-induced failure, yet the ordered regions it contains limit this gain. We show that, in rigid-jointed frames under hydrostatic pressure, an ensemble containing arbitrarily large patches of a crystal at positive density, whose centres carry a scaled crystal prestress, cannot stay linearly stable once those patches are stressed beyond the crystal's threshold for localizable instabilities. A homogeneous loss of positive definiteness while all Bloch modes, including the acoustic limit, remain stable cannot localize, leaving a window for stabilization. Extending an acoustic criterion to prestress, we show for all point groups that such a window requires a strain that keeps the crystal's symmetry. Unless the patches are shielded from stress, disorder cannot protect a crystal against a failure that breaks its symmetry, and protects a cubic crystal at most against volume collapse. In two-dimensional Truchet frames, planted ordered domains bring the threshold within 1% of the bound, and correlations short of long-range order localize failure on ordered clusters. Across 876 slender-strut nets, 798 fail first by a localizable instability; the other 78 have a median window of 7.6%. Stability gained through disorder is capped by the localizable instabilities of the local order it permits.**

Low-density frameworks fail under pressure long before their struts break: metal–organic frameworks amorphize or transform at a gigapascal or less,[1,2] and architected lattices buckle in modes that are Bloch waves of the periodic cell.[3,4] Disorder can delay this failure: the aperiodic Truchet-tile framework TRUMOF-1 remains crystalline at pressures that collapse its ordered congener, an effect likened to the irregular stones that keep drystone walls from failing collectively.[5] The rule that assembles a disordered framework, however, still admits ordered regions, and their instabilities limit how far disorder can raise its stability.

Locally ordered regions control the extremes of disordered systems. In Lifshitz's model of a binary alloy, the states at the edge of the spectrum live in rare regions made of one kind of atom,[6] and rare regions govern phase transitions in disordered systems.[7] For random Schrödinger operators with independent disorder, the almost-sure spectrum contains that of every periodic configuration in the support of the disorder.[8] The mechanical counterpart of this inclusion is a cap. If an ensemble contains arbitrarily large patches of a crystal at positive density, and the prestress at their centres approaches a multiple of the crystal's, any instability of that crystal that can localize inside a patch ends the linear stability of the infinite ensemble. A finite sample can stay stable above that load, by a margin that vanishes as its largest patch grows. One homogeneous instability cannot localize: loss of positive definiteness of the homogenized tangent while every Bloch and acoustic mode remains stable admits no compactly supported negative-energy field.[9] The cap therefore depends on which homogeneous instabilities can localize.

Here we show that symmetry decides which failures escape: none that breaks the crystal's point-group symmetry. Every non-trivial irreducible strain subspace contains a symmetrized rank-one gradient, so a homogeneous softening that breaks the point-group symmetry closes an acoustic branch at the same load, and a patch of the crystal can localize it. Without prestress, this acoustic statement underlies the classification of elastic phase transitions[10,11] once its table is corrected for repeated representations,[12] and was recently proved for all crystal classes.[13] We extend it to hydrostatic load: it holds exactly for the pressure-controlled Gibbs tangent, whose rank-one restriction is the acoustic tensor,[14] whereas at fixed strain, or under a non-hydrostatic dead load, ellipticity and positive definiteness separate. The result is a selection rule for what disorder can hide: only a totally symmetric softening escapes the patches, and only up to its localizable threshold. At the stress its patches carry, disorder therefore cannot protect a crystal against a failure that breaks its symmetry, and protects a cubic crystal at most against volume collapse; beyond that, it gains only by excluding the weak patches or by shielding them from stress.

We test the bound in decorated Truchet frames. Planted ordered domains lower the threshold of random samples to within 1% of the two-dimensional checkerboard bound, and a rule that forbids the soft motif removes that bound. Independent tiles keep most of their margin at accessible sizes, because large patches are exponentially rare; correlations short of long-range order localize the first mode on ordered clusters, and the excess then falls slowly with sample size N, consistent with 1/ln N. Across 876 crystal nets from the RCSR database,[15] modelled as rigid-jointed frames, 798, including all 95 cubic nets, fail first by an instability that can localize, and in the remaining 78 the intrinsic window, the largest gain disorder can make at equal patch stress, has a median of 7.6%.

## Three thresholds and a localizability criterion

We separate three thresholds of a rigid-jointed strut frame under hydrostatic load $\lambda$. To the microscopic (Bloch) and macroscopic (loss of ellipticity) instabilities of periodic solids[9] we add the loss of positive definiteness of the loaded crystal.[14] The relaxed homogeneous tangent loses positive definiteness at $\lambda_{PD}$. It loses strong ellipticity, so that an acoustic wave speed vanishes, at $\lambda_{LE} \geq \lambda_{PD}$. Bloch modes at non-zero wavevector, and zone-centre modes other than rigid translations, first become non-positive at $\lambda_\mu$; because the acoustic limit $k \to 0$ is included, $\lambda_\mu \leq \lambda_{LE}$ (Methods).

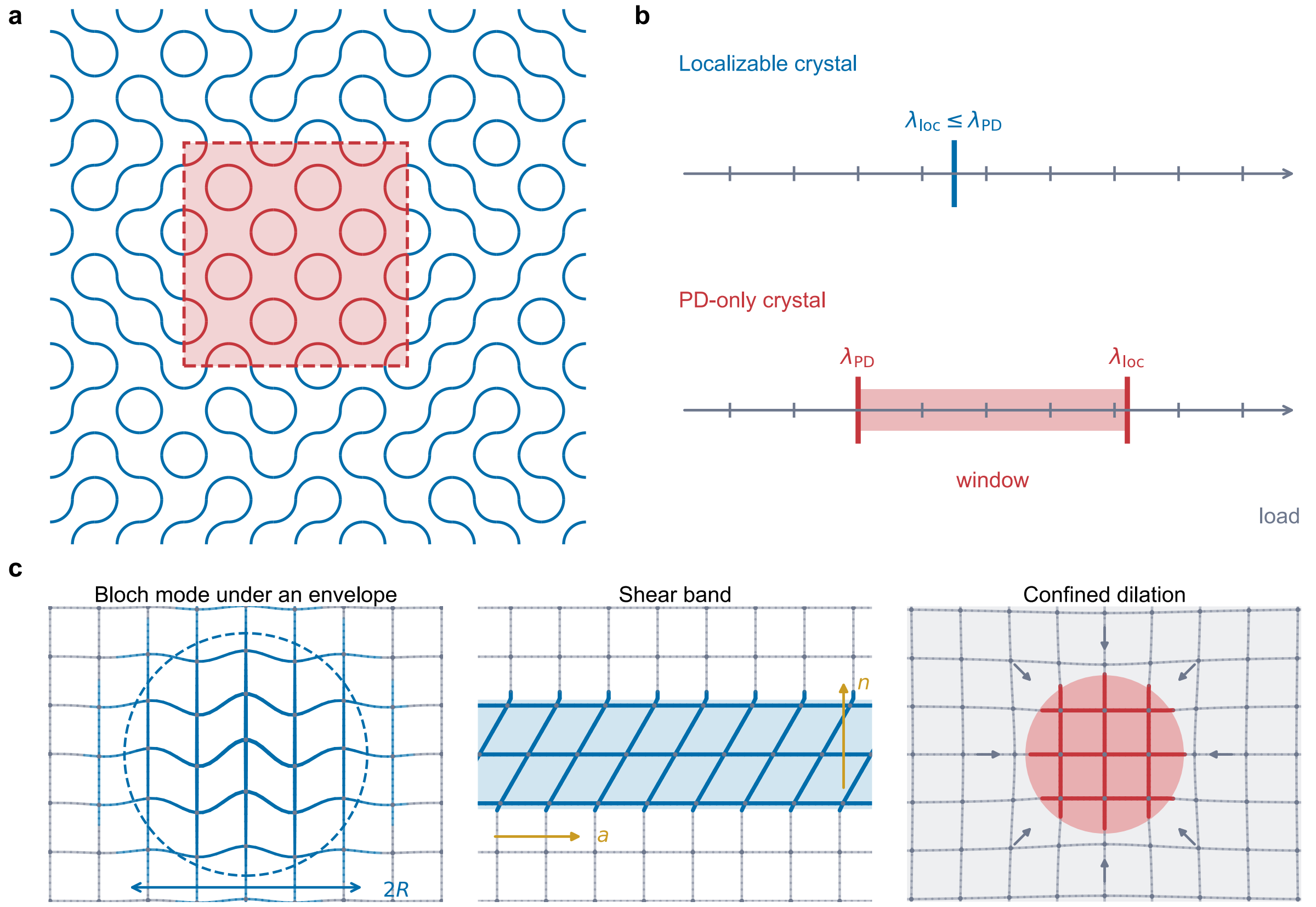

**Fig. 1 | Schematic illustration of ordered patches and the localizability of failure. a**, Decorated Truchet frame, drawn as tile arcs, with independent tile choices (blue) and an ordered checkerboard patch (red); in the computed frame each tile carries four nodes whose environments are identical in both regions (Methods; Fig. 2a). **b**, Thresholds of an allowed crystal under increasing hydrostatic load. A crystal whose first instability localizes has no window ($\lambda_{loc} \leq \lambda_{PD}$); a PD-only crystal fails homogeneously at $\lambda_{PD}$ but localizes only at $\lambda_{loc}$, and disorder can act only in between. **c**, Test fields on a square strut grid (grey, undeformed struts away from the field; dots, nodes). Left: a zone-boundary Bloch mode under an envelope of width 2R (dashed). Middle: a shear band carrying sym($a \otimes n$) (shaded), with n normal to the band. Both patterns illustrate localizable instabilities; with suitable cutoffs and sufficiently large support they yield compact trial fields of negative energy above $\lambda_{loc}$. Right: a homogeneous dilation of a soft region (red) yields no such field; the surrounding matrix (grey) confines it.

By the Bloch-wave theorem of Geymonat, Müller and Triantafyllidis,[9] a periodic frame is stable against compactly supported perturbations exactly as long as its Bloch spectrum is positive, including strong ellipticity as $k \to 0$. Loss of positive definiteness while the whole Bloch spectrum, ellipticity included, stays positive is therefore the only failure of a crystal that no compactly supported field can realize. We call it PD-only failure, and we call $\lambda_{loc} = \lambda_{\mu}$ the localizable threshold (Fig. 1). A crystal first fails at $\lambda_1 = \min(\lambda_{PD}, \lambda_{\mu})$, and it is PD-only when $\lambda_1 = \lambda_{PD} < \lambda_{loc}$. A localization identity for finite-range frames (Methods) makes this quantitative: a Bloch mode of a crystal can be cut to a window of size R at an energy cost that vanishes as R grows, as $R^{-2}$ for finite-wavevector and optical modes.

**The bound**

Let a local rule with identical local environments generate a stationary, ergodic ensemble $\mathscr{E}$. Assume that patches of a periodic configuration P of every size occur with positive probability (hypothesis H); that the rule admits them is necessary but not sufficient. The bound rests on a second hypothesis, H2: near the centre of a large patch the prestress approaches that of P scaled by a stress-concentration factor $s_P$, so that the tangent converges for every trial field of fixed size (Methods). A Bloch mode of P under a smooth envelope then yields a negative field once $s_P\lambda > \lambda_{loc}(P) + \Delta_P(R) + E_P(R)$. Here $\Delta_P(R)$ is the cost of confining the mode to a large patch, at most $\beta_P R^{-2}$ when the critical mode has a finite wavevector or is optical (Methods), and $E_P(R)$ is the error from the patch prestress, which under H2 vanishes for a trial field of fixed size as the patch grows.

The linear stability limit of the ensemble therefore satisfies

$$\lambda_c(\mathscr{E}) \leq \lambda_{cap} = \inf_P[\lambda_{loc}(P)/s_P], \tag{1}$$

where the cap $\lambda_{cap}$ is the infimum over the periodic configurations that satisfy H and H2. For a PD-only crystal, the bound is set by its corrected localizable threshold $\lambda_{loc}/s_P$, not by its lower homogeneous threshold, because loss of positive definiteness alone yields no compact negative-energy field.

**Symmetry decides what disorder can hide**

Which homogeneous failures can lie below the localizable threshold is restricted by symmetry.

Extending the unstressed argument[13] to hydrostatic prestress, we obtain a selection rule for all 32 point groups (Methods): if the failing eigenspace has an invariant part outside the totally symmetric sector, that part contains a traceless strain of zero determinant, a symmetrized rank-one gradient sym(a $\otimes$ n), so the acoustic tensor along n becomes singular at the same load. Symmetry-breaking softening therefore always closes an acoustic branch, and its onset can concentrate in a band normal to n. A totally symmetric failing strain is necessary for PD-only failure, but not sufficient: below cubic symmetry a totally symmetric strain can itself be symmetrized rank-one and close an acoustic branch, so the computed thresholds decide (Methods). In a cubic crystal the totally symmetric sector is pure dilation, which is never symmetrized rank-one, so the only PD-only failure the rule permits is a vanishing bulk modulus. Whether a surrounding matrix confines PD-only failure is a separate stability condition, shown analytically for isotropic volumetric softening[16,17] and predicted for inclusions of negative bulk modulus.[18] We embedded spherical domains of five PD-only nets and of the PD+LE control ncj in a stiffer frame of the same topology, 24 configurations in all (Methods; Supplementary Table 1). Every domain, the control included, survived beyond its own PD threshold, so at these sizes the matrix also confines localizable failure; the test probes confinement by a matrix, not protection by disorder. Only the nets with the widest windows approach their localizable threshold: the largest cco and vne domains failed at 2.41 and 1.95 times the PD threshold and at 1.11 and 1.16 times the localizable one, corrected by their mean interior pressure ratio $s_P$ = 0.77. The cco domain fails inside itself by its localizable strut-buckling mode, with 99.8% of the mode energy in the domain.

**Planted ordered domains approach the bound**

Localizable failure is what the bound caps: one ordered domain should lower the stability of a disordered sample toward the threshold of its crystal, which we tested in a two-dimensional decorated Truchet frame with identical node environments and independent tile choices (Fig.

2a,b). At prescribed macroscopic strain, 20 random samples of 32 × 32 tiles buckle at 1.219 times the load $p_0$ of the checkerboard, the weakest of 35 periodic-cell representatives with periods up to 3 × 3 and of all 65,536 4 × 4 patterns at fixed pressure.

A planted checkerboard domain draws the critical mode into itself: in a 32 × 32 sample, a 16 × 16 domain holds 97% of the mode energy and lowers the threshold from 1.21 to 1.03 times the checkerboard threshold at fixed pressure (Fig. 2a). A 48 × 48 checkerboard domain in a 64 × 64 sample lowers the threshold at prescribed strain to 1.0057 times the checkerboard value, with 99.8% of the critical-mode energy inside the domain. At fixed pressure the same test gives 1.0043, and a mode confined to a 64 × 64 domain in a 96 × 96 sample bounds the threshold by 1.0024 (Supplementary Table 2). The stress-concentration factor of these domains, $s_P = 1.0006$, shifts the cap $p_0/s_P$ by 0.06%.

A three-dimensional frame shows the same trend (Fig. 2c). Its cubes carry seven node types and independent binary states, and the weakest of its 22 periodic orbits up to 2 × 2 × 2, a layer-uniform checkerboard, fails by an in-plane optical mode. Ten random $10^3$ samples buckle at 1.424 times that load. Planting an $8^3$ domain in one background lowers the threshold from 1.411 to 1.138, with 82% of the mode energy inside the domain; $s_P$ of these domains lies within 0.3% of one, and within 0.1% for $R \geq 3$ (Supplementary Table 3). Its compact-bound excess approaches the same $R^{-2}$ scaling: $\alpha_2 R^{-2} + \alpha_3 R^{-3}$ fits the corrected excess for R = 4–10 cells to 1.0% rms (Fig. 2d; Supplementary Table 4).

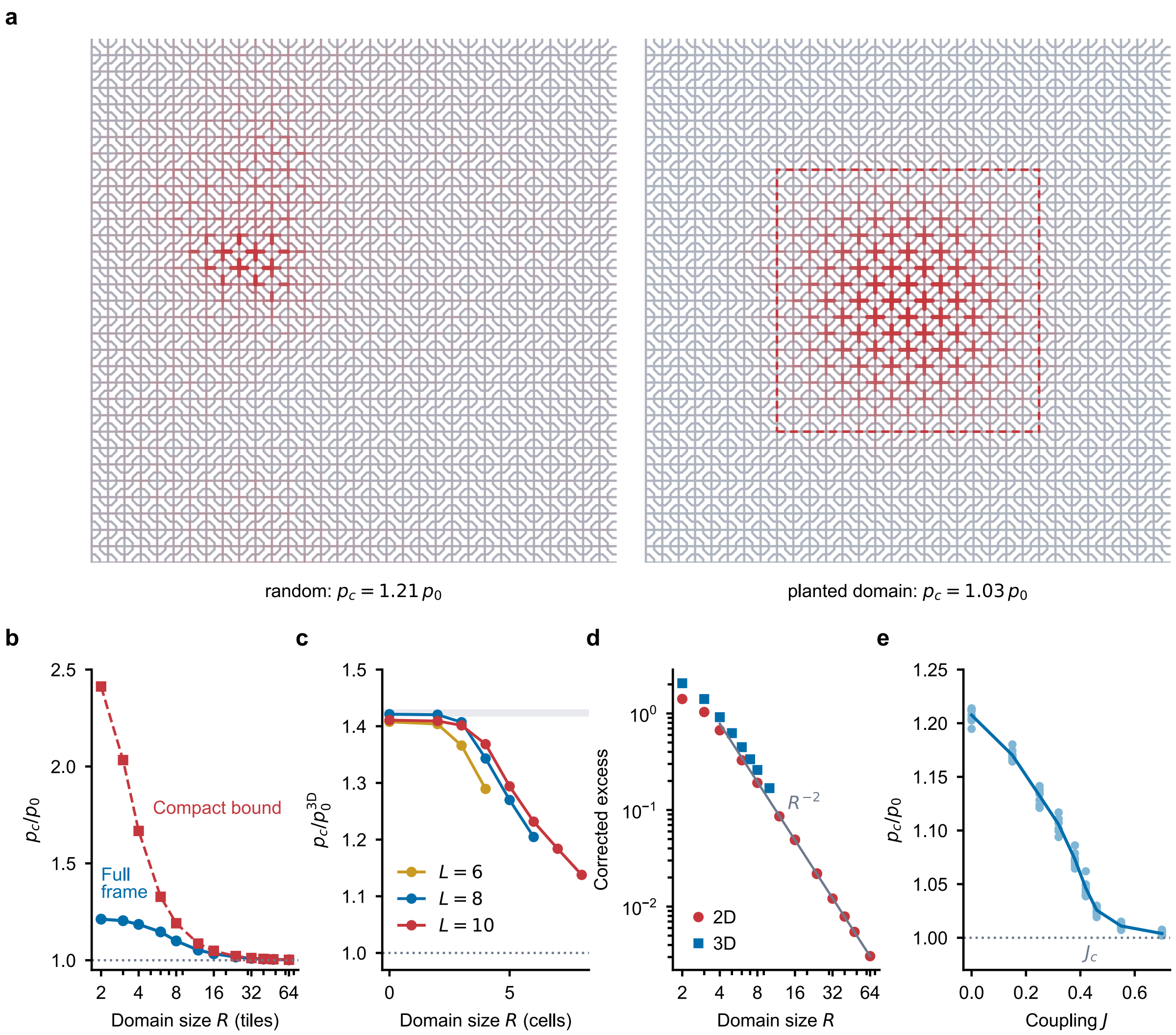

**Fig. 2 | Planted ordered domains approach the bound.** **a**, Critical mode of a random 32 × 32 decorated Truchet frame at fixed hydrostatic pressure (left) and of the same background with a planted 16 × 16 checkerboard domain (right, dashed): the mode fills the domain (97% of its energy) and the threshold drops from 1.21 to 1.03 times the checkerboard value. Colour and line width, mode energy per strut. **b**, Threshold of a sample containing one checkerboard domain of size R, in one background per sample size ($L = 32$ with $R \leq 24$, $L = 64$ with $R = 16$–48, $L = 96$ with $R = 64$; full frame), and the variational upper bound from fields confined to the domain (compact bound). Both are relative to the threshold $p_0$ of the checkerboard, the weakest of all periodic configurations with periods up to 3 × 3 and all 4 × 4 patterns; $p_c$ is a sample's first-instability pressure. **c**, Three-dimensional frame: threshold, relative to the threshold $p_0^{3D}$ of the layer-uniform checkerboard, versus planted domain size ($R = 0$: no domain) for boxes of $L = 6$, 8 and 10 cells; the thin grey band marks the mean ± s.d. of ten random $10^3$ samples. **d**, Excess of the compact bound, corrected by $s_P$, over the threshold of the weakest periodic configuration, in two dimensions (tiles) and three dimensions (cells, one $14^3$ background); the line has slope −2. **e**, Threshold of 64 × 64 samples whose tile states carry Ising correlations favouring the checkerboard, versus the coupling J (dots, eight samples each; line, mean); the label $J_c$ marks the Ising critical coupling on the axis. No sample falls below the checkerboard threshold.

**Why random samples look stronger**

Random samples of accessible size do not fail inside ordered patches. In three dimensions their first modes extend over the disordered network, occupying 88–98% of the cells, and their mean threshold stays at 1.42–1.43 times that of the layer-uniform checkerboard from $4^3$ to $10^3$ cells. In two dimensions the first mode of independent-tile samples localizes, on about 170 tiles at L = 192, but not in the largest checkerboard square, which holds 8% of its energy; the mean threshold falls only from 1.214 to 1.197 between L = 32 and 192 (Supplementary Table 5). Natural patches large enough to approach the checkerboard threshold are exponentially rare. For independent binary tiles a patch costs h = ln 2 of entropy per tile, so the largest patch in N tiles grows only as $R_{max}^d \simeq \ln N/h$. With the corrected compact-bound excess of planted domains, $\Delta_{pl}(R) \simeq 12.3\ R^{-1.99}$ in two dimensions (Fig. 2d), the rare-patch estimate undercuts the margin of 32 × 32 samples only for $N \gtrsim 10^{17}$ tiles.

Spatial correlation removes most of this margin without long-range order. We drew tile states from an Ising model of the staggered tile variable, with a coupling J that favours the checkerboard; every configuration remains an allowed sample of the same rule. Below the Ising critical coupling $J_c \approx 0.441$, where no long-range order forms, the first mode localizes: at J = 0.38 and 0.42 it involves 1–1.5% of the tiles of 128 × 128 samples. The mean threshold falls from 1.087 and 1.059 times the checkerboard threshold at L = 32 to 1.063 and 1.042 at L = 128, and at L = 64 and 128 ten times longer chains from random or ordered starts reproduce these values within their scatter (Supplementary Tables 6 and 7). Multiplied by ln N, the excess stays at 0.59–0.63 and 0.38–0.41 from L = 32 to 128, as expected if the weak regions grow in area as ln N with an $R^{-2}$ confinement cost; a weighted fit $\alpha_0 + \alpha_1/\ln N$ gives $\alpha_0 = 0.006 \pm 0.013$ and $0.001 \pm 0.014$, although a weak power law fits equally well. In 64 × 64 samples the threshold falls from 1.21 times the checkerboard threshold at J = 0 to 1.07 at J = 0.38, about two thirds of the margin gone, and to 1.004 at J = 0.7, above $J_c$ (Fig. 2e; Supplementary Table

8).

**Removing a soft crystal's bound**

The cap is the infimum of the corrected thresholds of the allowed crystals, so a rule changes this upper bound by removing soft crystals from its support or by changing the stress concentration of those it keeps. A matching lower bound would follow from a certificate that every allowed configuration is stable (Supplementary Theorem 4), which the rules below lack. In two dimensions, forbidding every 2 × 2 checkerboard block keeps extensive entropy and raises the weakest of 180 enumerated allowed candidates to 1.224 times the checkerboard threshold (Figs. 3 and 4; Supplementary Table 9); the weakest of seven candidates of a 3 × 3 sparsity rule lies at 1.345 (Fig. 3). Under H and H2 with $s_P = 1$, both are upper estimates of the cap of their rule. Twelve random 64 × 64 samples under the first rule buckle at 1.268 times the checkerboard threshold, against 1.207 without the rule, a finite-size indication.

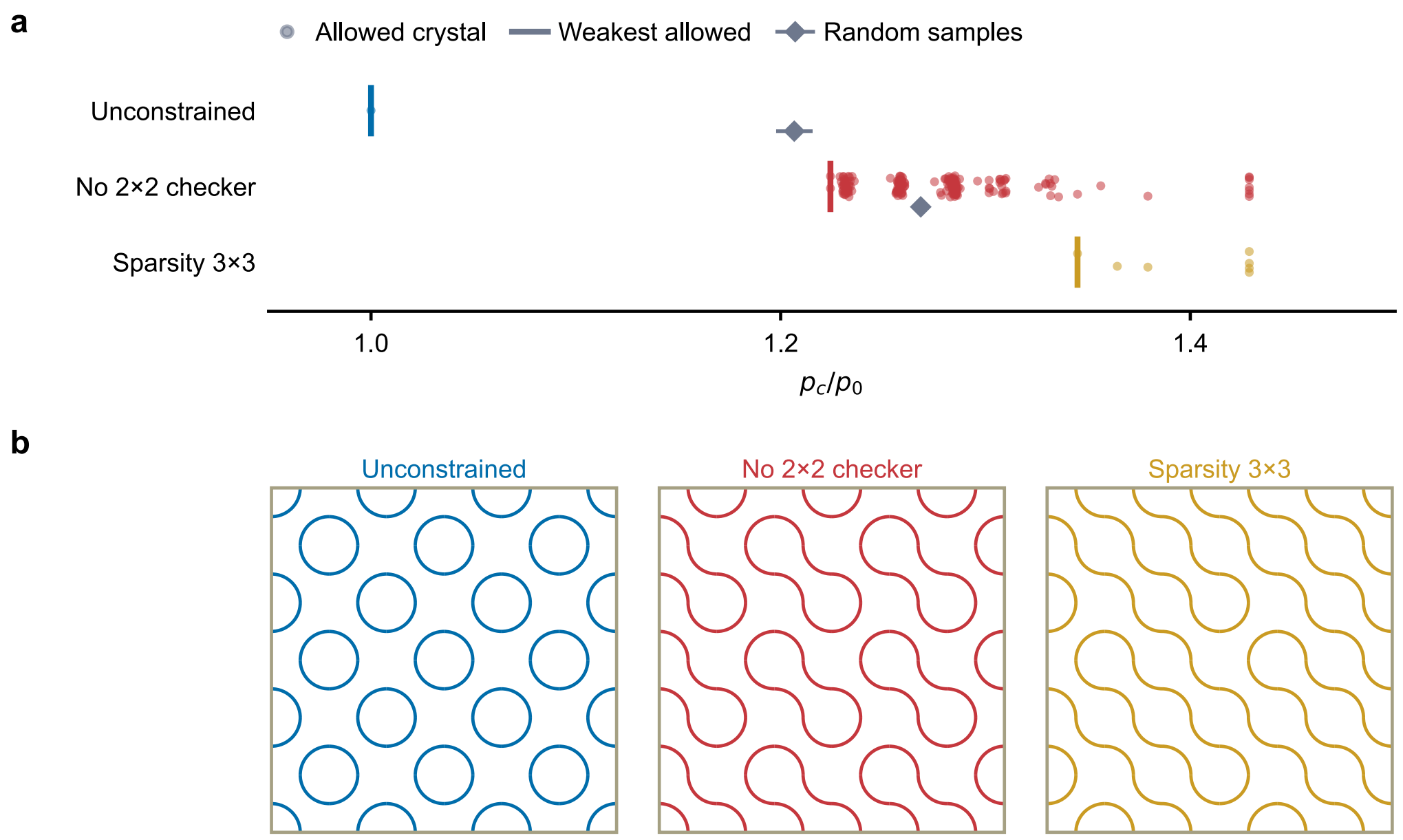

**Fig. 3 | A rule removes the checkerboard bound.** At $s_P = 1$ the weakest enumerated allowed crystal gives an upper estimate of the cap of each rule; twelve random samples of the no-checker rule fail 5% higher than without it. **a**, First-instability pressure of the allowed periodic members (dots, coloured by ensemble as in **b**) of the unconstrained ensemble and of two local rules in the two-dimensional frame, relative to the checkerboard threshold $p_0$, the weakest enumerated member without a rule; for the unconstrained ensemble only its weakest member, the checkerboard, is shown. Vertical bars mark the weakest allowed member of each ensemble; diamonds and lines give the mean and range of twelve random 64 × 64 samples, for the two ensembles that were sampled (unconstrained and no-checker). **b**, Weakest allowed member of each ensemble, drawn as Truchet tiles.

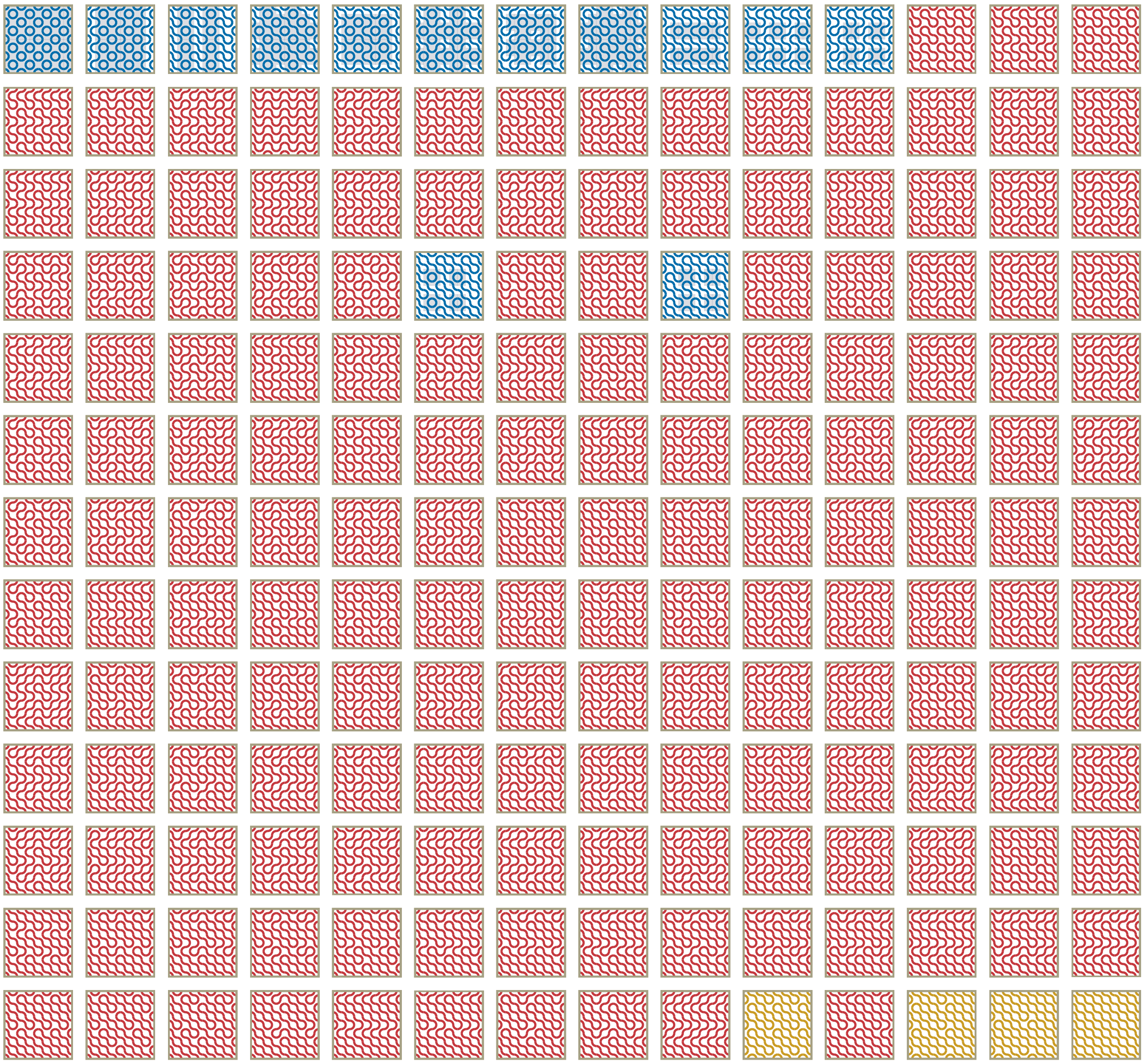

**Fig. 4 | Tilings a rule blocks and keeps.** The 182 periodic tilings of the two-dimensional decorated Truchet frame in the three rule catalogues (periods up to 3 × 3, the allowed 4 × 4 patterns and, for the sparsity rule, 3 × 4 cells). They are counted once up to translation and the eight symmetries of the square, including the exchange of tile states that those symmetries induce, and each is drawn as 8 × 8 tiles. The catalogue is finite and does not contain every periodic configuration. The tilings are ordered by their threshold at fixed hydrostatic pressure, from the weakest at the top left, along each row from left to right and then row by row downwards. Blue: the tiling contains a 2 × 2 checkerboard block (shaded) and is blocked by the no-checker rule (13 tilings). Red: allowed by the no-checker rule but not by the sparsity rule (165). Ochre: allowed by the sparsity rule as well (4). The eleven weakest tilings are all blocked. The 180 and 7 allowed candidates of Fig. 3 count periodic-cell representatives, including repeated supercell descriptions of smaller cells. The same tilings with their thresholds are in Supplementary Fig. 1.

**Crystal nets**

A decoration or rewiring rule built on a parent net whose ensemble contains, at positive density,

arbitrarily large patches of the uniformly decorated parent inherits, under H2, the parent's corrected localizable threshold $\lambda_{loc}/s_P$ as an upper bound on its cap. Of the 938 RCSR nets[15] with at most twelve vertices per cell, 876 give valid, full-symmetry graphs that are stable at zero load; we classified their first pressure instability in maximum-symmetry embeddings with uniform slender rigid-jointed struts (Fig. 5, Supplementary Table 10, Methods). At r/l = 0.02, 508 fail first at a finite-wavevector or optical mode, 290 lose positive definiteness and ellipticity together, and 78 lose positive definiteness alone. The class is unchanged at r/l = 0.05 for 867 nets. The PD-only count falls from 104 to 54 as the tie tolerance between $\lambda_{PD}$ and $\lambda_{loc}$ rises from $10^{-8}$ to 3% (Supplementary Table 11).

Because every allowed crystal bounds the cap at $\lambda_{loc}/s_P$, disorder can raise the stability limit of a parent crystal above its homogeneous threshold in two ways: by the intrinsic window $\lambda_{loc}/\lambda_{PD} - 1$, the largest gain at equal patch stress, or by shielding its patches from stress ($s_P < 1$). Shielding, $1/s_P - 1$, reaches 38% for the soft inclusions above but stays below 0.3% in the Truchet frames. The intrinsic window of the PD-only nets has a median of 7.6%, interquartile range 2.4% to 22%, and maximum 118%, for cco.

As the selection rule requires, every PD-only net, 78 of 78 at r/l = 0.02 and 74 of 74 at r/l = 0.05, and every PD-only net at every tie tolerance, fails with a strain in the totally symmetric sector of its point group, a check of the implementation (Fig. 6c; rule in Fig. 6a). In all 496 nets whose homogeneous failing strain breaks the point-group symmetry, ellipticity is lost at the same load as positive definiteness, to within $2 \times 10^{-11}$ relative, and none of the 95 cubic nets reaches a vanishing bulk modulus before a localizable mode at the 0.3% tie tolerance (Supplementary Table 11). In pcu the computed acoustic null vector is a pure shear with n and a along the cube axes (Fig. 6b, top, schematic), although a small soft pcu domain in a stiffer frame fails first by an extended shear mode (Supplementary Section S5.3); the soft cco domain fails inside itself (Fig. 6b, bottom).

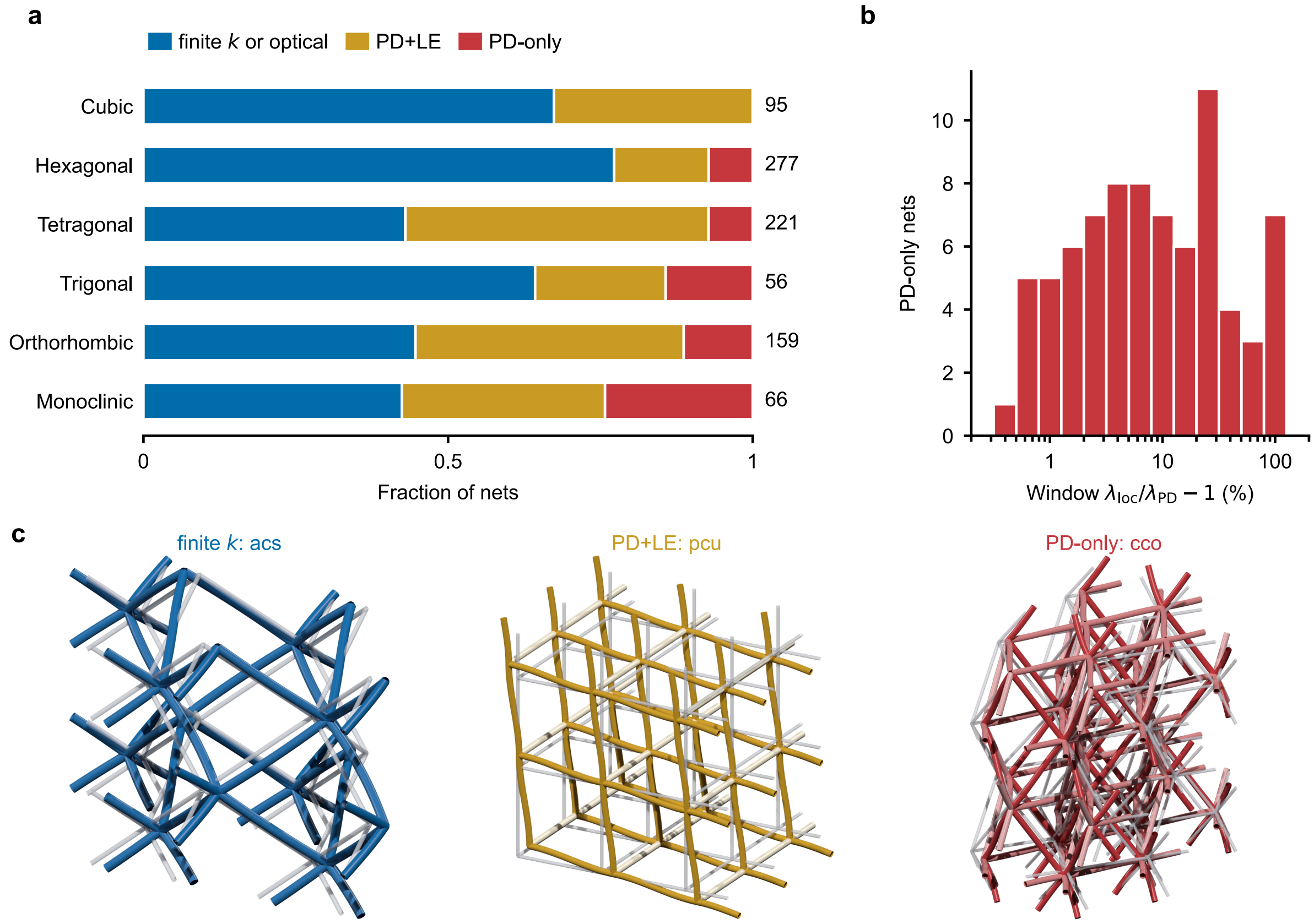

**Fig. 5 | First pressure instability of 876 crystal nets.** Class fractions and windows are statistics of the idealized frame model (r/l = 0.02, maximum-symmetry embeddings). **a**, Fraction of nets in each crystal system that fail first by a finite-wavevector or optical mode, by simultaneous loss of positive definiteness and ellipticity (PD+LE), or by loss of positive definiteness alone (PD-only); numbers give the nets per system (r/l = 0.02). Two triclinic nets are omitted from panel a but included in the overall total of 876. **b**, Window $\lambda_{loc}/\lambda_{PD} - 1$ of the PD-only nets. **c**, Computed first-instability modes (grey, undeformed frame): acs fails by a Bloch mode with wavevector (0, 1/2, 0) (real part shown); pcu by a shear that closes an acoustic branch; cco by a totally symmetric strain that keeps the point group. Colour, mode energy per strut, normalized in each panel; displacement amplitudes are arbitrary.

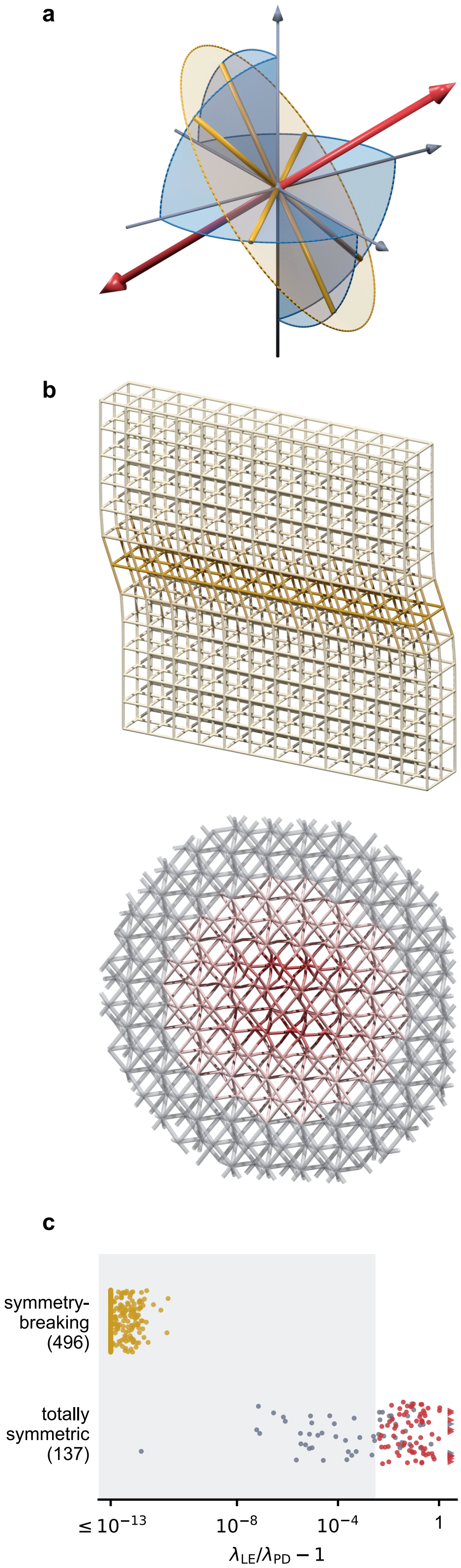
a
b
c
symmetry-
breaking
(496)
totally
symmetric
(137)
≤ $10^{-13}$
$10^{-8}$
$10^{-4}$
1
$\lambda_{LE}/\lambda_{PD} - 1$

**Fig. 6 | Symmetry decides which failures disorder can hide.** Panel a is a geometric construction; panel c checks the implementation of the proven rule across the atlas. **a**, Principal-strain space: grey arrows, the three principal-strain axes; red arrow, the dilation axis; ochre disc, the traceless plane. Blue sheets, the strains sym(a ⊗ n) that a localized band can carry (one zero principal strain and two of opposite sign, or a uniaxial strain); ochre rods, the six traceless symmetrized rank-one strain directions given by the permutations of (1, −1, 0), where the sheets meet the plane. Every non-trivial irreducible strain subspace contains such a strain; the dilation axis, the totally symmetric sector of a cubic crystal, meets no sheet. **b**, Top (schematic): pcu, whose failing strain breaks symmetry, carrying a symmetrized rank-one field sym(a ⊗ n) built from its computed acoustic null vector at $\lambda_{LE} = \lambda_{PD}$, localized in a band (envelope illustrative; colour, local shear). Bottom (computed): onset mode of a soft spherical cco domain (r/l = 0.02) in a cco frame with r/l = 0.03, central slab; the domain fails by strut buckling inside itself at 2.4 times its $s_P$-corrected PD threshold (colour, mode energy; displacements and strut radii exaggerated). **c**, $\lambda_{LE}/\lambda_{PD} - 1$ for every net with a homogeneous threshold, split by the symmetry of the failing strain (r/l = 0.02); ochre, symmetry-breaking (496 nets); red, PD-only; grey, other totally symmetric nets (137 totally symmetric nets in all); shaded, ties within 0.3%. Symmetry-breaking strains close an acoustic branch at the same load in all 496 nets, all within $2 \times 10^{-11}$; only totally symmetric strains open a window. Triangles at the right edge mark totally symmetric nets whose $\lambda_{LE}$ lies beyond an earlier zone-centre optical instability.

### Designing with disorder

Stability gained through disorder is capped by the localizable instabilities of the local order that disorder still permits. Three levels must be kept apart: the margin of accessible finite samples, carried by extended modes in three dimensions and by modes localized outside the largest checkerboard patches in two; the bound on the infinite ensemble, set by the localizable instabilities of the admitted crystals; and changes of that bound by rules that forbid weak motifs. The bound turns the design of disorder-stabilized materials into a question about crystals and their patches: which crystals a rule admits, how large their patches grow and how much stress the patches carry. At the stress its patches carry, a rule can beat its weakest admitted crystal only if that crystal fails by a totally symmetric softening that cannot localize, and then only up to its localizable threshold. In our idealized frame model, in which failure is strut buckling rather than node rupture, the nets of MOF-5, ZIF-8 and UiO-66 (pcu, sod and fcu) fail first by localizable modes, so an ensemble that contains large patches of them is capped by them, and stabilization requires excluding their weak motifs or shielding them from stress. For TRUMOF-1, whose rule admits ordered patches of a weaker approximant in a reconstructed coarse-grained model, that approximant would set an upper bound under the corresponding tangent-transfer assumptions (Supplementary Section S7). The bound is testable: under prestress transfer, the threshold of a sample cannot stay a finite amount above the crystal's corrected localizable threshold as its largest patch grows; for finite-wavevector and optical instabilities the confinement part of the bound is then of order $\beta_P R^{-2}$. A family of samples that keeps a finite margin would show that prestress transfer fails for it.

We establish the bound for linear thresholds of rigid-jointed beam frames, under the hypotheses that the ensemble contains arbitrarily large patches of the periodic configuration and that the prestress near their centres approaches a scaled crystal prestress (Methods). The catalogue of periodic configurations searched is finite. The bound holds along any proportional load path (Supplementary Theorem 1′), whereas the selection rule needs hydrostatic load. We treat bulk ensembles; free surfaces, which can soften before the bulk, lie outside the bound. Post-buckling and collapse, finite temperature, adsorption-coupled loading and the node chemistry of real frameworks lie outside the model.

**Methods**

**Frame model.** Each strut is a circular Euler–Bernoulli beam with Young's modulus $E_s = 1$ and Poisson ratio $\nu_s = 0.3$, and all struts meeting at a node share its degrees of freedom (rigid joints). All thresholds are linear bifurcation onsets of the loaded frame in its initial geometry under a preload linear in the load; post-buckling and collapse lie outside the model. Rigid-jointed beam frames are the standard model for slender cellular and architected lattices.[3,4] Whether such a frame is stretching- or bending-dominated is set by its connectivity through Maxwell's count.[19] In the RCSR atlas every strut has radius r proportional to its own length l (r/l = 0.02 or 0.05), nodes have six degrees of freedom, and each parent strut is divided into two Hermite elements. For 138 nets, all 78 PD-only nets and 60 others, refining to eight elements changed the thresholds by 0.15% in the median and at most 1.7%, and changed no class (Supplementary Table 12). The Truchet frames use tiles and cubes of unit side, a uniform strut radius of 0.01 and one Hermite element per member. Planar nodes have two translations and one rotation, and spatial nodes have three translations and three rotations. The port-stub length is 0.2 tile units, and the three-dimensional frame adds unit-length vertical members. The inclusion tests use r/l = 0.02 in the domain and 0.03 in the surrounding frame (Supplementary Table 1).

**Prestress and tangent.** The macroscopic load is a non-penetrating hydrostatic pressure $p = \lambda p_{ref}$ applied through the Gibbs potential $\Pi = U + pV$, with $p > 0$ compressive, stresses positive in tension and the cell shape free. For each configuration we first solve the periodic, relaxed preload for a unit isotropic stress, with all internal translations and rotations relaxed. The second variation of $\Pi$ about this state is $K + \lambda G + \lambda p_{ref}VD$, where K is the linear stiffness, D the volume Hessian and G the initial-stress tangent of all elements. G is the Hessian of the work of the axial force, both shear forces, the torque and the end moments on the exact strains of a Kirchhoff beam with semi-tangential end rotations.[20] This tangent is conservative and objective: a rigid rotation of an equilibrated frame costs no energy to machine precision ($< 4 \times 10^{-17}$ relative in every validation and atlas net). The long-wavelength limit of the Bloch problem reproduces the acoustic tensor of the homogenized tangent to within $2.5 \times 10^{-14}$. The preload equilibrium residual is below $1.4 \times 10^{-12}$ of the member end forces in every net. Convergence and invariance tests are listed in Supplementary Tables 13–15.

**Thresholds.** $\lambda_{PD}$ is the first load at which the relaxed homogenized Gibbs tangent, obtained by Schur complementation of the internal degrees of freedom, loses positive definiteness on symmetric strains. This is the Born criterion[21] generalized to a loaded crystal through the stress–strain coefficients of the stressed state.[14] $\lambda_{LE}$ is the first load at which the acoustic tensor $A_{ik}(n) = \bar{C}_{ijkl}n_j n_l$ of that tangent becomes singular for some direction n, searched over 3,000 quasi-uniform directions on the unit sphere and refined from the ten lowest. The production search, over 64 directions (26 symmetry axes and 38 quasi-uniform) refined from the best two, missed the minimum or overestimated $\lambda_{LE}$ by more than $10^{-6}$ relative, by up to 34%, for 8 nets at $r/l = 0.02$ and 5 at 0.05, and misclassified two of them at $r/l = 0.02$. $\lambda_{\mu}$ is the infimum of the loads at which a Bloch mode with non-zero wavevector, or a zone-centre mode on the complement of the rigid translations, becomes non-positive (Supplementary Section S1.2).[3,9,22] Because the infimum includes the limit $k \to 0$, mathematically $\lambda_{\mu} \leq \lambda_{LE}$ and $\lambda_{loc} = \lambda_{\mu}$. Numerically we sample wavevectors on a $4^3$ grid of the irreducible zone outside a small zone-centre exclusion and refine locally; this estimate is written $\lambda_{\mu}^{num}$ (Supplementary Table 10). We evaluate $\lambda_{loc}$ as $\min(\lambda_{LE}, \lambda_{\mu}^{num})$, with $\lambda_{LE}$ computed from the acoustic tensor. Every net that is PD-only at either radius and 40 PD+LE controls were rechecked on an $8^3$ grid with a four times smaller zone-centre exclusion. The finer grid, which also seeds the acoustic search from near-zone-centre minima, reclassified one net, ncj, and lowered $\lambda_{LE}$ for 32 net–radius pairs; nets that would be PD-only solely at tolerances below 0.3% were not rechecked. A net is PD-only when $\lambda_{PD}$ precedes both other thresholds by more than 0.3% (Supplementary Table 11). The failing strain is projected onto the totally symmetric subspace of the point group $G_{pt}$ with the projector $P_{inv} = |G_{pt}|^{-1}\sum_g g \otimes g$ over its elements g. Strains with more than 99% of their norm in that subspace are counted as totally symmetric. Among the 633 nets with a defined $\lambda_{PD}$, no failing strain has a fraction between $10^{-9}$ and 99%.

**Selection rule.** Let $e_0$ be a failing strain in a non-trivial irreducible representation $\Upsilon$ of the point group $G_{pt}$. Schur's lemma gives tr e = 0 on $\Upsilon$. If $\Upsilon$ is one-dimensional, some operation maps $e_0$ to $-e_0$, so det $e_0$ = −det $e_0$ = 0; take $e_1 = e_0$. If $\Upsilon$ is two- or three-dimensional, det is an odd cubic form on the irreducible subspace and vanishes at some non-zero $e_1$ in it; the invariant critical eigenspace contains this subspace, so $e_1$ fails too. A traceless symmetric strain with zero determinant has eigenvalues $(\zeta, -\zeta, 0)$ and equals sym($a \otimes n$) for orthogonal a, n along the bisectors of its non-zero principal axes. Then $e_1{:}\bar{C}{:}e_1 = a\cdot A(n)a = 0$, so the acoustic tensor is singular at the same load. Under hydrostatic pressure the Gibbs second variation is rotation invariant and the volume Hessian vanishes on rank-one gradients, so the argument holds exactly with prestress. The condition is necessary, not sufficient: the totally symmetric sector contains the rank-one uniaxial strain in hexagonal, trigonal and tetragonal crystals and a codimension-one set of symmetrized rank-one strains in orthorhombic, monoclinic and triclinic crystals, and if the critical eigenspace mixes representations its non-trivial part closes an acoustic branch. For a repeated representation the critical strain is a constant-dependent combination of basis strains whose subspace still contains a zero-determinant strain, so the zero-velocity wave travels along a direction that rotates with the elastic constants. Full proofs, including repeated representations and mixed eigenspaces, are in the Supplementary Information, together with elastic matrices (Supplementary Tables 16 and 17) on the stability boundaries of the repeated representations that Cowley listed as type 0 in his Table I.[10,12]

**Localization bound.** For a finite-range quadratic form Q on a lattice and a partition of unity $\{\chi_W\}$ of windows of radius R, the Ismagilov–Morgan–Sigal identity (Lemma 3.1)[23] gives $Q = \sum_W \chi_W Q \chi_W + 1/2 \sum_W [\chi_W, [\chi_W, Q]]$. Adapting it to finite-range lattice forms, we show that the remainder is bounded by $c^2 a_0^2 \kappa\, R^{-2}\|u\|^2$, with c a window constant, $a_0$ the lattice spacing and κ the largest site stiffness (Supplementary Lemma 1). A Bloch mode of an allowed crystal under a smooth envelope therefore yields a compactly supported negative field once the load exceeds its localizable threshold by the confinement cost $\Delta_P(R)$. Because the second variation is affine in the load and the unloaded frame is stable, this holds beyond every localizable threshold without assumptions on the dispersion (Supplementary Lemma 2). For a finite-wavevector or optical critical mode $\Delta_P(R)$ is at most the envelope cost $\beta_P R^{-2}$ for large R; for an acoustic one no general $R^{-2}$ rate holds, and an anisotropic example gives $R^{-1}$ (Supplementary Lemma 3). Inside a patch the elastic stiffness equals that of the crystal; only the prestress differs. Hypothesis H2 states that, near the centre of large patches, the tangent approaches that of the crystal at $s_P$ times the load; the infinite-volume bound needs no convergence on the growing patch interior. Averaged over eight independent backgrounds, the largest deviation of the member end wrenches in a fixed central 4 × 4 window falls from 1.6% to 0.27% (L = 64) and from 1.5% to 0.55% (L = 96) as the planted domain grows from 16 to 48 tiles. The mean quadratic-form error on fields in that window falls to 0.07–0.16% (Supplementary Tables 18 and 19). For independent tiles a planted domain samples exactly the ensemble conditioned on that window being a patch. Any finite-R estimate carries both $\Delta_P(R)$ and the prestress error $E_P(R)$, for which H2 fixes no rate. If the member forces near patch centres converge to those of P under a non-hydrostatic cell stress $\Sigma_P$ (hypothesis H2′), the bound holds with the localizable threshold of P loaded along $\Sigma_P$ in place of $\lambda_{loc}(P)/s_P$ (Supplementary Section S3). The rare-patch argument bounds the threshold from above; whether the first instability of a large sample is carried by its largest patch is not decided by it. The rare-region argument follows Lifshitz[6] and Vojta;[7] d denotes the spatial dimension. Statements and proofs are in the Supplementary Information.

**Decorated Truchet ensembles.** In two dimensions, each square tile of the in-plane frame carries four internal nodes at fixed positions connected to the edge midpoints, and the tile state chooses between the two non-crossing pairings of its ports (Figs. 1a and 3b). Out-of-plane buckling of the planar sheet is excluded. Every node star is congruent across configurations, which we verified by canonical Gram matrices of the incident vectors. The three-dimensional ensemble stacks these layers with vertical struts and independent cube states. Random samples are drawn by independent tile choices or, for constrained rules, by exact conditional row Gibbs sampling. Periodic catalogues comprise 35 periodic-cell representatives with periods up to 3 × 3 and all 65,536 4 × 4 patterns (669 symmetry orbits) in two dimensions, and all 22 orbits of 2 × 2 × 2 patterns (256 assignments) in three dimensions. The 4 × 4 catalogue is evaluated at fixed hydrostatic pressure, as for the rule catalogues. The no-checker rule forbids every 2 × 2 checkerboard block. The sparsity rule requires every translated 3 × 3 window to contain at most one tile in state 1. The next-weakest 4 × 4 pattern fails at 1.054 times the checkerboard; local wavevector refinement of the 20 weakest changed no threshold (Supplementary Table 20). Compact-support bounds minimize the Rayleigh quotient over fields that vanish outside the planted domain. Mode energy is the positive elastic energy $u^{*}k_{e}u$ of each strut in the critical mode u; energy fractions are normalized by its sum over the sample, and map colours by its maximum.

**RCSR atlas.** Nets are taken from the RCSR database[15] and expanded with gemmi.[24] The origin setting is chosen so that every RCSR edge centre is the exact midpoint of its edge. One representative edge per edge centre, the shortest vertex pair with that midpoint, is completed to its space-group orbit, so every graph carries the full symmetry of its net; vertex coordinates are averaged over their site stabilizers. Of 938 nets with at most twelve vertices per conventional cell, 878 give valid graphs and 876 are stable at zero load.

**Data availability**

All computed data underlying the figures and tables are deposited at Zenodo with the code (https://doi.org/10.5281/zenodo.23682287). Crystal nets were taken from the RCSR database (https://rcsr.net), and the TRUMOF-1 cluster connectivities from the supplementary data of Meekel et al.[25]

**Code availability**

Computation scripts, numerical results (JSON) and numerical starting states (NPZ) are deposited at Zenodo (https://doi.org/10.5281/zenodo.23682287) and GitHub (https://github.com/joohan-git/ordered-patch-stability). The RCSR net file and the structure files of Meekel et al. are not redistributed; the cluster connectivities extracted from the latter are included. The computational scripts used in this study were developed with assistance from Claude Code (Anthropic). All AI-assisted code was reviewed and tested by the author.

**Author contributions**

J.N. conceived the study, performed the calculations and wrote the paper.

**Competing interests**

The author declares no competing interests.

**Supplementary Information**

# Ordered patches cap the stability of disordered frameworks

Joohan Nam[1]*

[1]Chair of Inorganic Chemistry I, Technische Universität Dresden, Bergstraße 66, 01069 Dresden, Germany.

*e-mail: joohan.nam@mailbox.tu-dresden.de

## S1. Setting and thresholds

**S1.1 Frames and loading.** A frame is a periodic or disordered network of beams with rigid joints. A configuration ω is drawn from an ensemble $\mathscr{E}$ generated by a local rule; we assume $\mathscr{E}$ is stationary and ergodic under lattice translations and that all local environments are congruent. The macroscopic load is a hydrostatic pressure $p = \lambda\, p_{ref}$ with the Gibbs potential $\Pi = U + pV$, with $p > 0$ compressive and stresses positive in tension, so that the preload stress is $-p_{ref}\mathbf{1}$. The second variation about the relaxed, equilibrated preload is

$$Q^{\omega}{}_{\lambda}[u] = u^{T}(K^{\omega} + \lambda\, G^{\omega} + \lambda\, p_{ref}VD)u,$$

with $K^{\omega}$ the linear stiffness, $G^{\omega}$ the initial-stress tangent of all members, and D the volume Hessian on macroscopic gradients. The pre-buckling solution is linear in λ, so $Q^{\omega}{}_{\lambda}$ is affine in λ. We assume (A1) that $K^{\omega}$ and $G^{\omega}$ have finite range $r_0a_0$, with $a_0$ the lattice spacing, and bounded site stiffness $\kappa_{site} = \max_i\sum_{e\ni i}\|k_e\|$, with $k_e$ the stiffness block of member e, and (A2), a modelling assumption that no proof below uses, that elastic moduli fluctuate on short ranges only. (A0) The unloaded configuration is Bloch-stable: $Q_0(\mathbf{k}) \succ 0$ for every $\mathbf{k} \neq 0$, its zone-centre form is positive on the complement of the rigid translations, and the unloaded relaxed tangent is positive definite on symmetric strains. For a connected rigid-jointed beam frame (A0) follows from connectivity, because every beam has stretching, bending and torsional stiffness, so a zero-energy motion moves the whole frame rigidly. A frame with disconnected components, such as an interpenetrated net, violates (A0). For spring networks (A0) is an additional hypothesis. For an infinite configuration ω we assume that a stationary equilibrated preload, with member end wrenches $\lambda\, n^{\omega}{}_{e}$, exists, for example as the limit of periodic approximants; $G^{\omega}$ and hypothesis H2 refer to it.

**S1.2 Three thresholds of a periodic member.** For a periodic configuration P, write $Q_{\lambda}(\mathbf{k}) = K(\mathbf{k}) + \lambda\, G(\mathbf{k})$ for the Bloch matrix of $Q^{P}{}_{\lambda}$ and $f(\mathbf{k};\lambda)$ for its smallest eigenvalue. At $\mathbf{k} = 0$ the rigid translations are null vectors at every load; $f_0^{\perp}(\lambda)$ denotes the smallest eigenvalue of $Q_{\lambda}(0)$ on their orthogonal complement. Following Geymonat, Müller and Triantafyllidis[1] we distinguish the macroscopic threshold $\lambda_{LE}$ from the microscopic threshold $\lambda_\mu$, and add the Born threshold $\lambda_{PD}$[2] of the loaded homogenized tangent:[3]

• $\lambda_{PD}(P)$, the first load at which the relaxed homogenized Gibbs tangent $\bar{C}_P(\lambda)$ loses positive definiteness on symmetric strains;

• $\lambda_{LE}(P)$, the first load at which strong ellipticity fails, $a \cdot A_P(n;\lambda)\, a = 0$ for some a, n ≠ 0, with $A_{ik}(n) = \bar{C}_{ijkl} n_j n_l$;

• $\lambda_\mu(P)$, the infimum of the loads at which $f(\mathbf{k};\lambda) \leq 0$ for some $\mathbf{k} \neq 0$ or $f_0^{\perp}(\lambda) \leq 0$.

Because the acoustic branches behave as $\xi^2$ times the acoustic tensor (Lemma 2), $\lambda_\mu \leq \lambda_{LE}$, and the localizable threshold is $\lambda_{loc} = \min(\lambda_{LE}, \lambda_\mu) = \lambda_\mu$. The computed threshold of the Methods samples wavevectors outside a small neighbourhood of the zone centre and takes the minimum with $\lambda_{LE}$ from the acoustic tensor; it is a numerical estimate of $\lambda_{loc}$, whose dependence on the exclusion we checked by refinement (Methods). Positive definiteness implies ellipticity, so $\lambda_{PD} \leq \lambda_{LE}$. We write $\lambda_\Gamma$ for the first zero of $f_0^{\perp}$.

**S1.3 Absorbed and unabsorbed optical modes.** Partition the zone-centre Hessian into macroscopic strain e and internal degrees of freedom w, with blocks $C_{aff}$, B and $D_w$, where Bv is the macroscopic stress exerted by an internal mode v. Then $\bar{C}_P = C_{aff} - BD_w^{-1}B^T$. A softening internal mode v with Bv ≠ 0 drives an eigenvalue of $\bar{C}_P$ to $-\infty$ as $D_w$ becomes singular, so it crosses zero earlier and loss of positive definiteness comes first; a softening mode with Bv = 0 is a pure internal mode and enters $\lambda_\mu$. Beyond $\lambda_\Gamma$, $\bar{C}_P$ denotes the Schur complement continued through uncoupled zone-centre modes; $\lambda_{PD}$ and $\lambda_{LE}$ then refer to it, and Theorem 5 is a statement about it.

## S2. Localization lemmas

**Lemma 1 (localization identity).** Let $\{\chi_W\}$ be a partition of unity by windows of radius R with $\sum_w \chi_{w,i}^2 = 1$ and $\sum_w(\chi_{w,i} - \chi_{w,j})^2 \le c^2 a_0^2/R^2$ whenever $|x_i - x_j| \le r_0 a_0$. For any finite-range quadratic form Q of the type above, restricted to fields without macroscopic gradient, on which the pressure term vanishes,

$$Q[u] = \sum_w Q[\chi_w u] - \mathscr{R}[u],\ |\mathscr{R}[u]| \le c^2 a_0^2 R^{-2} \kappa \|u\|^2,$$

with $\kappa = \kappa_{site}$ for the elastic part and $\kappa = \lambda\kappa_G$, $\kappa_G = \max_e|n_e|\max_i\sum_{e \ni i}\|g_e\|$, with $n_e$ the reference force of member e and $g_e$ its initial-stress block per unit force, for the initial-stress part. In operator form $Q = \sum_w(\chi_w Q \chi_w + 1/2\ [\chi_w, [\chi_w, Q]])$; the individual terms need not be positive.

*Proof.* A member e = (i, j) contributes $u_i^T M^{ii}_e u_i + 2u_i^T M^{ij}_e u_j + u_j^T M^{jj}_e u_j$, where the $M_e$ are the blocks of its stiffness or initial-stress matrix. Because $\sum_w \chi_{w,i}^2 = 1$, the windows leave the diagonal blocks unchanged, and $\sum_w \chi_{w,i}\chi_{w,j} = 1 - 1/2\ \Theta_e$ with $\Theta_e = \sum_w(\chi_{w,i} - \chi_{w,j})^2 \le c^2 a_0^2/R^2$. Hence $\mathscr{R}[u] = -\sum_e \Theta_e\, u_i^T M^{ij}_e u_j$. Bounding $|u_i^T M^{ij}_e u_j| \le 1/2\ \|M^{ij}_e\|(\|u_i\|^2 + \|u_j\|^2)$ and summing over sites gives the estimate, with a spare factor 1/2. This is the Ismagilov–Morgan–Sigal identity[4] for lattice forms. ■

**Lemma 2 (compact negative fields beyond $\lambda_{loc}$).** Assume (A0), (A1) and hydrostatic load, and let P be periodic.

(i) Each $f(\mathbf{k};\cdot)$ and $f_0^{\perp}$ is concave in λ. For $\lambda < \lambda_\Gamma$ the acoustic function $g_{a,n}(\lambda) = a\cdot A_P(n;\lambda)a$ is concave in λ, and $f(\xi\, n;\lambda) = c_P \xi^2 \min_{|a|=1} g_{a,n}(\lambda) + O(\xi^3)$ with $c_P > 0$, uniformly in n and in λ on compact subsets of $[0, \lambda_\Gamma)$.

(ii) At least one of the following holds: (a) $\lambda_{loc} = \lambda_{LE} < \lambda_\Gamma$; (b) $f(\mathbf{q};\lambda_{loc}) = 0$ for some $\mathbf{q} \neq 0$; (b′) $\lambda_{loc} = \lambda_\Gamma$ and every null vector of $Q_\lambda(0)$ at $\lambda = \lambda_\Gamma$ off the translations satisfies $Bv = 0$.

(iii) For every $\lambda > \lambda_{loc}$ there is a compactly supported v with $Q^P_\lambda[v] < 0$. In case (b) the real or imaginary part of $\psi(x/R)\exp(i\mathbf{q}\cdot x)\,\varphi$, with φ a unit eigenvector of $Q_\lambda(\mathbf{q})$ for $f(\mathbf{q};\lambda)$, has Rayleigh quotient at most $f(\mathbf{q};\lambda) + c_1\kappa_\lambda\, a_0^2 R^{-2}$ for large R, where $\kappa_\lambda = \kappa_{site} + \lambda\kappa_G$ and $c_1$ depends only on ψ and $r_0$, and $f(\mathbf{q};\lambda) \le -(\lambda - \lambda_{loc})f(\mathbf{q};0)/\lambda_{loc}$. In case (b′) the same holds with $\mathbf{q} = 0$, $\varphi_0$ a unit eigenvector for $f_0^{\perp}(\lambda)$, and $f_0^{\perp}$ in place of $f(0;\cdot)$, which the translations pin at zero for $\lambda \le \lambda_\Gamma$.

(iv) For a fixed compactly supported v, $Q^P{}_\lambda[v]$ is affine in λ with $Q^P{}_0[v] > 0$, so once negative it stays negative at every larger load.

*Proof.* (i) $f(\mathbf{k};\lambda) = \min_{|\varphi|=1}\varphi^* Q_\lambda(\mathbf{k})\varphi$ is a minimum of affine functions of λ, hence concave; the same holds for $f_0^{\perp}$. For $\lambda < \lambda_\Gamma$ the internal block $D_w(\lambda)$ is positive definite, so $e{:}\bar{C}_P(\lambda){:}e = \inf_w \delta^2\Pi_\lambda[(e, w)]$ is an infimum of affine functions and concave. By S4.3 and Lemma A, $g_{a,n} = e{:}\bar{C}_P{:}e$ with $e = \mathrm{sym}(a \otimes n)$, because the volume term vanishes on rank-one gradients and $\partial\Pi/\partial E = 0$ at the equilibrated preload. The pressure term acts on macroscopic gradients only and is absent from $Q_\lambda(\mathbf{k})$ for $\mathbf{k} \neq 0$. The eigenvalues of $Q_\lambda(\xi\, n)$ below $f_0^{\perp}(\lambda)/2$ agree to $O(\xi^3)$ with those of its static Schur complement, which is analytic in ξ by finite range and equals $c_P\xi^2 A_P(n;\lambda) + O(\xi^3)$;[1] the remainder is uniform because the internal block is uniformly positive, and Weyl's inequality makes degeneracies harmless.

(ii) If a null vector v of $D_w(\lambda_\Gamma)$ has $Bv \neq 0$, choose a unit n with $(Bv){:}(n \otimes n) \neq 0$; the tensors $n \otimes n$ span all symmetric tensors. Then the relaxed energy of $e = n \otimes n = \mathrm{sym}(n \otimes n)$ tends to $-\infty$ as $\lambda \uparrow \lambda_\Gamma$, so $g_{\min}(\lambda) = \min_{|a|=|n|=1} g_{a,n}(\lambda)$ vanishes before $\lambda_\Gamma$ and $\lambda_{LE} < \lambda_\Gamma$. If $\lambda_\Gamma < \infty$ then $f_0^{\perp}(\lambda_\Gamma) = 0$, so $\lambda_\Gamma$ lies in the set defining $\lambda_\mu$ and $\lambda_\mu \le \lambda_\Gamma$. If $\lambda_{LE} < \lambda_\Gamma$, concavity of $g_{\min}$ on $[0, \lambda_\Gamma)$ with $g_{\min}(0) > 0$ gives $g_{\min} < 0$ on $(\lambda_{LE}, \lambda_\Gamma)$, attained at a direction $n^*$, and (i) gives $f(\xi\, n^*;\lambda) < 0$ there for small $\xi \neq 0$; if $\lambda_{LE} \ge \lambda_\Gamma$ then $\lambda_\mu \le \lambda_\Gamma \le \lambda_{LE}$. In both cases $\lambda_\mu \le \lambda_{LE}$. Take $\lambda_j \downarrow \lambda_\mu$ with $f(\mathbf{k}_j;\lambda_j) \le 0$ or $f_0^{\perp}(\lambda_j) \le 0$ and a subsequence $\mathbf{k}_j \to \mathbf{k}^*$. If $\mathbf{k}^* \neq 0$, continuity gives case (b). If $\mathbf{k}^* = 0$ and $f_0^{\perp}(\lambda_\mu) = 0$, a coupling null vector would give $\lambda_{LE} < \lambda_\Gamma = \lambda_\mu \le \lambda_{LE}$, so case (b′) holds. Otherwise the uniform expansion of (i) gives $\lambda_{LE} \le \lambda_\mu$: case (a).

(iii) A concave F with $F(0) > 0$ and $F(\lambda_*) = 0$ satisfies $F(\lambda) \le -(\lambda - \lambda_*)F(0)/\lambda_* < 0$ for $\lambda > \lambda_*$. Apply this to $f(\mathbf{q};\cdot)$ in case (b), to $f_0^{\perp}$ in case (b′), and to $g_{\min}$ in case (a); in case (a) fix λ in $(\lambda_{LE}, \lambda_\Gamma)$ and then choose $\xi_0 \neq 0$ with $f(\xi_0 n^*;\lambda) < 0$; larger loads follow from (iv). Let $\varphi_B$ be the Bloch eigenfunction with eigenvalue $\vartheta < 0$ and $v_R = \psi_R\varphi_B$. The identity $\psi_R Q \psi_R = 1/2\ (\psi_R^2 Q + Q\psi_R^2) - 1/2\ [\psi_R, [\psi_R, Q]]$ gives $Q[v_R] = \vartheta\|v_R\|^2 - 1/2\ \langle\varphi_B, [\psi_R, [\psi_R, Q]]\varphi_B\rangle$, and as in Lemma 1 the double commutator is bounded by $c^2 a_0^2 R^{-2}\kappa_\lambda$ times the weight of $\varphi_B$ on the $r_0a_0$-neighbourhood of the support, so any $c_1 > 1/2\ c^2|\mathrm{supp}\ \psi|/\int\psi^2$ works for large R. Since Q is real symmetric, $Q[\mathrm{Re}\ v_R] + Q[\mathrm{Im}\ v_R] = Q[v_R]$ and the norms add, so one real part has Rayleigh quotient at most that of $v_R$, which is negative for large R. In case (a) the order is: load, then $\xi_0$, then R.

(iv) $Q^P{}_\lambda[v] = Q^P{}_0[v] + \lambda\ v^T G^P v$, and $Q^P{}_0[v] > 0$ for $v \neq 0$ compactly supported by (A0). ■

**Lemma 3 (rates of the patch excess).** Let $\Delta_P(R)$ be the infimum of the loads $\delta \geq 0$ for which some v supported in $B_R$ has $Q^P{}_\lambda[v] < 0$ at $\lambda = \lambda_{loc} + \delta$. By Lemma 2, $\Delta_P(R)$ is finite for $R \geq R_0(P)$, nonincreasing in R, and tends to zero; this uses (A0), (A1) and the hydrostatic load only. For a finite-wavevector or zone-centre optical instability (cases b, b′), the envelope field of Lemma 2(iii) gives, for large R,

$$\Delta_P(R) \leq \Delta^{env}{}_P(R) = \beta_P/R^2,\ \beta_P = c_1\kappa_{\bar\lambda}\ a_0^2\lambda_{loc}/f(\mathbf{q};0),$$

with $f_0^\perp(0)$ in place of $f(\mathbf{q};0)$ in case (b′) and $\kappa_{\bar\lambda} = \kappa_{site} + \bar\lambda\kappa_G$ for a fixed load $\bar\lambda > \lambda_{loc}$. The slope $\tau$, minus the right derivative of $f(\mathbf{q};\cdot)$ (of $f_0^\perp$ in case b′) at $\lambda_{loc}$, satisfies $\tau \geq f(\mathbf{q};0)/\lambda_{loc}$ by concavity and needs no transversality assumption; with this $\tau$ the Rayleigh quotient is at most $-\tau(\delta - \Delta^{env}{}_P(R))$ for large R and $\lambda_{loc} + \delta \leq \bar\lambda$. A sharper constant in terms of the band curvature requires a separate remainder estimate, and a vanishing band Hessian does not make the excess zero. For an acoustic instability (case a), Lemma 2 gives existence and $\Delta_P(R) \to 0$ but no rate: a one-directional expansion $-\delta\xi^2 + \xi^4$, up to positive factors, does not imply an $R^{-2}$ excess for a multidimensional patch, and a finite-range example with a square patch has an excess of order $R^{-1}$ (Supplementary Table 19 note). The weakest periodic members of the decorated Truchet frames fail at zone-centre modes: the checkerboard at $\mathbf{k} = 0$, where $\lambda_{PD}$, $\lambda_{LE}$ and $\lambda_\Gamma$ coincide to $10^{-13}$ relative (Supplementary Table 20), and the layer-uniform checkerboard by a Γ-optical mode (Supplementary Table 21). Where such a mode couples weakly to macroscopic strain, $\lambda_{LE}$ lies slightly below $\lambda_\Gamma$, by $1.8 \times 10^{-7}$ relative for the weakest no-checker candidate and by at most 0.11% among the candidates of Supplementary Table 9 classed as zone-centre optical. The bound $\beta_P R^{-2}$ then applies above $\lambda_\Gamma$, that is, up to this offset from $\lambda_{loc}$.

**Lemma 4 (PD-only modes do not localize).** By Lemma 2(iii) and the Bloch decomposition $Q^P_\lambda[v] = \int\langle \mathscr{F}v(\mathbf{k}),\ Q_\lambda(\mathbf{k})\mathscr{F}v(\mathbf{k})\rangle\ d\mathbf{k}$, with $\mathscr{F}$ the Bloch transform,[1] the quadratic form of a periodic configuration is nonnegative on all compactly supported perturbations for $\lambda \le \lambda_{loc}$ and has compactly supported negative fields for every $\lambda > \lambda_{loc}$. For $\lambda_{PD} < \lambda < \lambda_{loc}$ the crystal is therefore unstable only against macroscopic homogeneous strain: no compactly supported field of negative energy exists. PD-only failure is the only failure with this property. Lemma 1 makes the converse quantitative: a homogeneous strain field in a sample of linear size L has Rayleigh quotient of order $(a_0/L)^2$ times the lowest eigenvalue of $\bar{C}_P$ on symmetric strains, which is negative but lies below any fixed window bound $-c^2a_0{}^2\kappa/R^2$ only when $L \lesssim R$.

## S3. Theorems

**Definitions.** For an infinite configuration $\omega$, $\lambda_c(\omega)$ is the infimum of the loads at which $Q^\omega{}_\lambda[v] < 0$ for some compactly supported v. It is invariant under lattice translations, so by ergodicity it equals a constant $\lambda_c(\mathscr{E})$ almost surely. Requiring the homogenized tangent to stay positive definite can only lower the threshold, so every upper bound on $\lambda_c(\mathscr{E})$ also bounds that threshold. For a finite periodic sample of N cells, $\lambda_c(N)$ is defined in the same way with fields periodic on the sample. A configuration $\omega$ has a P-patch of radius R if $\omega = P$ on a window $\Omega_R$; its interior $I_R$ excludes a boundary layer of width w(R), with $r_0a_0 \le w(R) \le R/2$ and $w(R) \to \infty$; a fixed width would not do, see the status of H2 below. Member forces are $\lambda\ n^\omega{}_e$ and $\lambda\ n^P{}_e$. For a dilute isotropic spherical patch, Eshelby's solution of the inhomogeneity problem[5] gives the stress-concentration factor $s_P$ as $K_P(3\bar{K} + 4\bar{G})/[\bar{K}(3K_P + 4\bar{G})]$, with $K_P$, $G_P$ the bulk and shear moduli of the patch and $\bar{K}$, $\bar{G}$ those of the surrounding medium. The prestress deviation $\varepsilon_P(R)$ is the largest $|n^\omega{}_e - s_P n^P{}_e|$ over the members with at least one end in $I_R$ divided by $\max_e|n^P{}_e|$, with $s_P$ the prestress factor of hypothesis H2 below. Numerically we estimate $s_P$ by $s_R$, the mean pressure of a planted domain of size R divided by the applied pressure; the main text quotes $s_R$ of the largest domain as $s_P$. In the numerical sections R denotes the side of a square or cubic domain, twice its inscribed radius; rates are unaffected.

**Hypothesis H (patch support).** For each admitted periodic member P and every radius R, the event that a given site is the centre of a P-patch of radius R, including the collar of every interaction touching it, has positive probability under the stationary ergodic measure of $\mathscr{E}$. Ergodicity then places such patches, for every integer radius at once, in almost every sample. Local admissibility of the patch by the rule alone is not this statement; it is a property of the measure.

**Hypothesis H2 (local prestress transfer).** For each admitted periodic member P there is a constant $s_{P,\mathscr{E}} > 0$ with the following property. Almost surely, P-patches of radii tending to infinity have centres $t_j$ such that $Q^{\omega}_{\lambda}[T_j v] - Q^{P,s}_{\lambda}[v] \to 0$ for every fixed compactly supported field v and every load λ, where $Q^{P,s}_{\lambda}$ is the form of P at load $s_{P,\mathscr{E}}\lambda$. Here $T_j$ is the translation to $t_j$, and the patch contains the support of v and every interaction touching it. Convergence of all tangent coefficients on each fixed finite neighbourhood of the centre is sufficient. Uniform convergence on the growing interior $I_R$ is not required; it is the stronger hypothesis $H2_{\infty}$, $\varepsilon_P(R) \to 0$ with the layer w(R) of the Definitions, used only for the finite-patch estimates below. We write $s_P$ for $s_{P,\mathscr{E}}$; a rule changes it together with the ensemble.

**Lemma 5 (prestress perturbation).** For supp $v \subset I_R$ the elastic stiffness and the member geometry inside the patch are those of P, and the pressure term vanishes on compactly supported fields, so $|Q^{\omega}_{\lambda}[v] - Q^{P,s}_{\lambda}[v]| \le \lambda\varepsilon_P(R)\kappa_G\|v\|^2$, where $Q^{P,s}_{\lambda}$ is the form of P at load $s_P\lambda$.

*Proof.* Bound each member term by $\|g_e\|(\|v_i\|^2 + \|v_j\|^2)$ and sum over sites, as in Lemma 1. ■

**Theorem 1 (bound).** Assume (A0) for every periodic member, (A1), and that $K^{\omega}$ is positive on non-zero compactly supported fields, as for a connected rigid-jointed frame. (i) Let P fail by a finite-wavevector or zone-centre optical mode, let $\omega$ have a P-patch of large radius R, and let $s_P\lambda = \lambda_{loc}(P) + \delta \leq \bar{\lambda}$, with $\bar{\lambda}$ the load of Lemma 3. The envelope field $v_R$ of Lemma 2 supported in $I_R$ then satisfies

$$(Q^{\omega}_{\lambda}[v_R])/(\|v_R\|^2) \leq -\tau(\delta - \Delta^{env}_P(R - w)) + \lambda\varepsilon_P(R)\kappa_G,$$

where $\kappa_G$ is built with the member forces of P. Whenever $s_P > \varepsilon_P(R)\kappa_G/\tau$ and the following value is at most $\bar{\lambda}/s_P$, therefore, $\lambda_c(\omega) \leq [\lambda_{loc}(P) + \Delta^{env}_P(R - w)]/[s_P - \varepsilon_P(R)\kappa_G/\tau]$. We write this bound as $[\lambda_{loc}(P) + \Delta^{env}_P(R - w) + E_P(R)]/s_P$, which defines $E_P(R) \geq 0$; under $H2_\infty$, $E_P(R) \to 0$. The upper bound $\beta_P R^{-2}$ applies to $\Delta^{env}_P$ only, with $\Delta^{env}_P(R - w(R)) \leq 4\beta_P R^{-2}$ because $w(R) \leq R/2$; $H2_\infty$ fixes no rate for $E_P$. (ii) For almost every $\omega$, $\lambda_c(\omega) \leq \lambda_{loc}(P)/s_P$ for every P that satisfies H and H2. Hence $\lambda_c(\mathscr{E}) \leq \lambda_{cap} = \inf_P[\lambda_{loc}(P)/s_P]$, with the infimum over these P.

*Proof.* (i) is Lemma 2(iii) for $Q^{P,s}_{\lambda}$ plus Lemma 5. (ii) Fix P and $\delta > 0$. By Lemma 2(iii) there is a compactly supported v with $Q^P_{\lambda}[v] < 0$ at $\lambda = \lambda_{loc}(P) + \delta$. Ergodicity and H place P-patches of every radius in almost every $\omega$; by H2, at a patch centre $t_j$ far enough along the sequence, $Q^{\omega}_{\lambda}[T_j v] < 0$ at $\lambda = (\lambda_{loc}(P) + \delta)/s_P$. $Q^{\omega}_{\lambda}[T_j v]$ is affine in $\lambda$ and positive at $\lambda = 0$ for a connected rigid-jointed frame, so as in Lemma 2(iv) it stays negative at larger loads and $\lambda_c(\omega) \leq (\lambda_{loc}(P) + \delta)/s_P$. Let $\delta \to 0$; the admitted periodic patterns form a countable set, so the almost-sure events can be intersected. ■

*Status of H2.* We measured member forces in planted checkerboard domains at fixed hydrostatic pressure, with a common normalization by the largest reference member force (Supplementary Table 19). For two-dimensional domains of side 48, the maximum deviation in the core with two boundary tiles excluded is 1.9% and 2.5% in samples of side 64 and 96 and does not decrease with R; boundary effects decay inward, but a core with a fixed boundary layer keeps acquiring members near the edge. For a random background this is expected: if the deviation at a fixed distance from the edge has positive spread, its maximum over a perimeter of growing length tends to the supremum of that distribution, not to zero. $H2_\infty$ is therefore stated with a layer w(R) that grows with R, and the fixed-centre column is the diagnostic of H2. The maximum deviation in a fixed central 4 × 4 region falls from 1.7–1.9% at R = 8 to 0.35–0.41% at R = 48. The signed correction of the actual compact critical mode v, the difference between the forms of the sample at its threshold $p_c$ and of the crystal at $s_R$ times $p_c$, divided by $v^TKv$, is −0.003% and −0.013% of its elastic energy at R = 48. The member-wise and operator-norm bounds on it are 0.2–0.3% and 6–7%. Deviations thus largely cancel under the mode weighting. These finite-sample data are consistent with H2 but prove neither H2 nor $H2_\infty$; both remain assumptions, including the existence and scalar form of the limiting prestress. Over eight independent backgrounds (Supplementary Table 18), the maximum fixed-window wrench deviation, as mean ± s.d., falls from 1.62 ± 0.58% at R = 16 to 0.27 ± 0.08% at R = 48 for L = 64 and from 1.51 ± 0.37% to 0.55 ± 0.16% for L = 96, in every background. At R = 48, L = 128 gives 0.52 ± 0.07%, so the smaller value at L = 64 is a finite-box effect. The mean error of the quadratic form on fields supported in that window falls to 0.07–0.16%, and the mean domain stress is hydrostatic to within 0.07%. For independent tiles, planting a domain in a fresh random background samples exactly the ensemble conditioned on that window being a P-patch, so the planted tests probe natural patches; this does not hold for correlated or biased ensembles and does not prove H2. For an anisotropic P in a frame of different stiffness the interior stress need not be hydrostatic; see Theorem 1′.

**Theorem 1′ (tensor form).** Let P be preloaded by a cell-averaged stress $\lambda\Sigma$ with free cell shape and relaxed internal degrees of freedom, and let $\lambda_{loc}(P;\Sigma)$ be the first load at which its Bloch spectrum, including the acoustic limit, becomes non-positive on that path. If H2 holds with $Q^P$ preloaded along $\Sigma_P$ (hypothesis H2′), then $\lambda_c(\mathscr{E}) \leq \inf_P \lambda_{loc}(P;\Sigma_P)$. The preload is linear, so $\lambda_{loc}(P;s\Sigma) = \lambda_{loc}(P;\Sigma)/s$, and $\lambda_{loc}(P;\Sigma_P) = \lambda_{loc}(P)/s_P$ when $\Sigma_P = -s_P p_{ref}\mathbf{1}$. Otherwise the scalar bound is an approximation whose sign is not fixed, and $\lambda_{loc}(P;\Sigma_P)$ must be computed under $\Sigma_P$, with the acoustic tensor of the homogenized second variation including the prestress work; it coincides with the Gibbs acoustic tensor only for hydrostatic $\Sigma$ (S4.3). The bound holds along any proportional load path: for $\lambda > \lambda_{loc}(P;\Sigma)$ some Bloch branch $f(\mathbf{k};\cdot)$, $\mathbf{k} \neq 0$, or $f_0^{\perp}$ is non-positive at a lower load, so by concavity it is negative at $\lambda$, and the envelope field of Lemma 2(iii) is negative.

**Theorem 2 (planted domains).** (i) If the first instability of P is a finite-wavevector or zone-centre optical mode, a planted P-domain of large radius R lowers the threshold of any sample to at most $[\lambda_{loc}(P) + \Delta^{env}{}_P(R - w) + E_P(R)]/s_P$ under the assumptions of Theorem 1(i). For any localizable first instability, compact fields of the fixed size given by Lemma 2 transfer under the planted-domain analogue of H2, convergence of the tangent near the domain centre to that of P at $s_P$ times the load, so the planted threshold tends to at most $\lambda_{loc}(P)/s_P$ as $R \to \infty$. (ii) If it is PD-only with isotropic volumetric softening and the matrix tangent is positive definite and strongly elliptic, a dilute domain does not lower the threshold while $K_P > -4/3 \min(G_P, \bar{G})$;[6] the anisotropic case is tested numerically (main text and Methods; Supplementary Table 1).

**Theorem 3 (rare-patch upper estimate).** Let a weakest member $P_*$ fail by a finite-wavevector or optical mode, and write $s_*$, $\beta_*$, $E_*$ and $\Delta^{env}{}_*$ for its quantities, so that $\lambda_{cap} = \lambda_{loc}(P_*)/s_*$. Suppose the events that disjoint boxes of side 2R contain a $P_*$-patch of radius R are independent, as for independent tiles, each with probability at least $\exp(-hR^d)$ in d dimensions. For every $\eta > 0$ a sample of N cells then contains, with probability $1 - o(1)$, such a patch with $R_{max}{}^d \geq (1 - \eta)\ln N/h$, because the probability that no box contains one is at most $\exp[-N^{\eta}/(2R_{max})^d]$. By Theorem 1(i), under its assumptions at $R = R_{max}$, and the rare-region argument,[7,8] with $w = w(R_{max})$, the sample then satisfies

$$\lambda_c(N) \leq \lambda_{cap} + (\Delta^{env}{}_*(R_{max} - w) + E_*(R_{max}))/s_*.$$

If the prestress term were subleading and $w(R) = o(R)$, the right-hand side would approach $\lambda_{cap} + (\beta_*/s_*)(h/\ln N)^{2/d}$; we do not know its rate, and the maximum deviations of Supplementary

Table 19 do not show it to vanish. The measured $s_P$-corrected compact-bound excess of a planted domain nevertheless follows an $R^{-2}$ scaling in two and three dimensions, consistent with the proven upper bound $\beta_P R^{-2}$ (Supplementary Tables 2 and 4). That excess contains the actual, signed prestress contribution of the critical field, which the mode-weighted diagnostic of Supplementary Table 19 shows to be far below the bound of Lemma 5. This bounds the rare-patch contribution from above. It becomes the threshold only under rare-patch dominance: the first threshold of modes not confined to the largest patch must exceed the right-hand side. We have no lower bound of this kind. In independent-tile samples of accessible size the first mode is not carried by the largest patch: it extends over the network in three dimensions and, in two, localizes on about 170 tiles at L = 192 outside the largest checkerboard square (Supplementary Table 5). It lies below the rare-patch estimate, though above $\lambda_{cap}$, so these modes fail first. In the two-dimensional frame the measured $s_P$-corrected compact-bound excess $\Delta_{pl}(R) \simeq 12.3R^{-1.99}$ falls below the margin of random 32 × 32 samples (0.219 at fixed strain; 0.214 at fixed pressure) only for $R \gtrsim 7.6$, that is $N \gtrsim 10^{17}$ tiles (Supplementary Tables 2 and 5). In the 32 × 32 scan of Supplementary Table 22, the corrected compact excess of 0.3275 at R = 6 and 0.1911 at R = 8 brackets this crossing. In three dimensions the mean threshold of random samples is constant within 0.5% from $4^3$ to $10^3$ cells, with 88–98% of the cells participating (Supplementary Table 3). Spatially correlated configurations below the Ising transition already localize their first mode at accessible sizes (Supplementary Table 6). The planted-domain experiments, not sample size, test the bound.

**Theorem 4 (conditional stability certificate).** Let $Q^{\omega}_{\lambda} = Q^{\omega}_{0} + \lambda\, G^{\omega}$ with $Q^{\omega}_{0} \geq 0$ for every allowed configuration. Suppose finite-window forms $B^{\omega}_{W}$, each including the interaction collar of its window, satisfy $Q^{\omega}_{\lambda}[u] = \sum_{W} B^{\omega}_{W}[u]$ at $\lambda = \lambda^*$ with $B^{\omega}_{W} \geq 0$ for every compactly supported u, every allowed ω and some $\lambda^* > 0$. Then $Q^{\omega}_{\lambda} \geq 0$ for $0 \leq \lambda \leq \lambda^*$ by affine interpolation; this certifies the absence of negative fluctuation modes, not a spectral gap. Pressure-controlled stability additionally requires the relaxed homogeneous Gibbs tangent of the ensemble to stay positive definite on this interval. Every periodic member that satisfies H and H2 then obeys $\lambda_{loc}(P)/s_P \geq \lambda^*$, since otherwise, by Lemma 2(iii), an enveloped Bloch mode of P would make the form negative. The windows of Lemma 1, $\chi_W Q \chi_W + 1/2\ [\chi_W, [\chi_W, Q]]$, are not positive even at zero load for the no-checker rule, so they give no certificate (Supplementary Table 19 note); we have no positive certificate for any rule. The values 1.224 and 1.345 of the main text are minima over finite catalogues with $s_P = 1$, not certified bounds, and changing a rule changes both its periodic support and the prestress factors $s_{P,\mathscr{E}}$ of the members it keeps.

## S4. Selection rule

**S4.1 Lemmas.** Let Sym be the space of symmetric strains with the action $e \mapsto geg^T$ of the elements g of the point group $G_{pt}$, let $Sym_{inv}$ be its totally symmetric subspace, and let $\bar{C}$ be symmetric and $G_{pt}$-invariant.

- *Lemma A (rank-one identity).* For $e = sym(a \otimes n)$, $e{:}\bar{C}{:}e = a\cdot A(n)\, a$.

- *Lemma B (symmetrized rank-one strains).* We call a non-zero e symmetrized rank-one if $e = sym(a \otimes n)$ for some a, n; this holds if and only if $\det e = 0$ and e is indefinite or of rank one. In particular, $\operatorname{tr} e = \det e = 0$ implies $e = \zeta(m \otimes m - m' \otimes m') = sym(\zeta(m + m') \otimes (m - m'))$ with m, m′ orthonormal.

- *Lemma C (tracelessness).* An invariant subspace without trivial component lies in $\operatorname{tr}^{-1}(0)$ (Schur's lemma).

- *Lemma D (zeros of the determinant).* For linearly independent $e_1, e_2 \in$ Sym, $\det(\cos t\, e_1 + \sin t\, e_2)$ is odd under $t \mapsto t + \pi$, so it vanishes at some $t_0$, where $\cos t_0\, e_1 + \sin t_0\, e_2 \neq 0$.

- *Lemma E (one-dimensional irreps).* If some g maps e to −e, then $\operatorname{tr} e = \det e = 0$.

**S4.2 Theorem 5 (selection rule).** Let $E_0$ be the eigenspace of $\bar{C}$ that loses positive definiteness at $\lambda_{PD}$, and let $E_0^{nt}$ be its component without trivial representation. If $E_0^{nt} \neq 0$, then $\lambda_{LE} = \lambda_{PD}$. Consequently PD-only failure requires $E_0 \subset Sym_{inv}$: every failing strain is totally symmetric. The condition is necessary, not sufficient (S4.4).

*Proof.* $E_0$ is $G_{pt}$-invariant, so $E_0^{nt}$ is a $G_{pt}$-invariant subspace. If dim $E_0^{nt} = 1$, some g acts on it as −1 and Lemma E gives tr $e_0$ = det $e_0$ = 0 for $e_0$ spanning it; otherwise Lemmas C and D give such an $e_0 \in E_0^{nt}$; Lemma B writes $e_0 = \mathrm{sym}(a \otimes n)$ and Lemma A gives $a \cdot A(n)a = e_0{:}\bar{C}{:}e_0 = 0$.

■

Complex-conjugate pairs (point groups 3, 3̄, 4, 4̄, 4/m, 6, 6̄, 6/m, 23 and m3̄), repeated isotypic blocks and eigenspaces that mix representations are covered because only Lemmas C and D are used on $E_0^{nt}$. The same determinant argument, for the unstressed elastic tensor, appears in Franke, Sander and Garst.[9]

**S4.3 Prestress.** Under hydrostatic pressure the Gibbs potential depends on the Green strain E only, and $\partial\Pi/\partial E = \sigma + p\mathbf{1} = 0$ at the reference state, so $\delta^2\Pi[H] = e{:}C^G{:}e$ with e = sym H. For a rank-one gradient $H = a \otimes n$ the volume term $(\mathrm{tr}\, H)^2 - \mathrm{tr}\, H^2$ vanishes, so $a \cdot A(n)a = e{:}C^G{:}e$ exactly; this is Wallace's statement that wave propagation under hydrostatic stress is governed by the stress–strain coefficients, here $C^G$.[3] For the Helmholtz tangent of prescribed strain, $C^F$ (the second variation of U on symmetric gradients, including its $-p\, \mathrm{tr}\, e^2$ term), $a \cdot A(n)a = e{:}C^F{:}e - (p/2)\, |a \times n|^2$, so under prescribed strain ellipticity can be lost while $C^F$ is still positive definite; the classification therefore uses $C^G$ and A. The identity relies on the hydrostatic load. Under a non-hydrostatic nominal dead load it fails in general. For the local energy $W = S{:}E - tE_{12}^2 + |E|^4$ with S = diag(1, 1, 2) and load potential $-S{:}(F - \mathbf{1})$, the symmetric shear $E_{12}$ of this tetragonal state becomes a zero mode at t = 1, while every rank-one gradient $a \otimes n$ with unit a, n keeps a second variation of at least 1/2.

**S4.4 Totally symmetric sectors.** Cubic: $Sym_{inv} = \mathrm{span}\{\mathbf{1}\}$, no symmetrized rank-one strain,

so PD-only failure is pure dilation. Hexagonal, trigonal and tetragonal: $\mathrm{Sym}_{\mathrm{inv}} = \mathrm{span}\{e_{xx} + e_{yy}, e_{zz}\}$ and $\det e = \gamma_1^2\gamma_2$ for $e = \gamma_1(e_{xx} + e_{yy}) + \gamma_2 e_{zz}$; the only symmetrized rank-one strain is uniaxial $e_{zz}$. Orthorhombic, monoclinic and triclinic: symmetrized rank-one strains form a codimension-one subset. Larger totally symmetric sectors leave more room for PD-only failure. A totally symmetric failing strain on these symmetrized rank-one sets closes an acoustic branch and is not PD-only; the classification therefore uses the computed $\lambda_{\mathrm{LE}}$, not symmetry alone. The compatibility condition for ferroelastic domain walls[10] concerns the difference of two spontaneous strains and is a different question from a symmetrized rank-one strain inside an irreducible subspace.

**S4.5 Relation to the acoustic classification of elastic instabilities.** Cowley (Table I and Sec. II C)[11] assigned the repeated non-trivial representations 2B of the tetragonal classes 4, $\bar{4}$ and 4/m, 2E of the trigonal classes (3, $\bar{3}$ and 32, 3m, $\bar{3}$m), and 2B of the monoclinic classes to type 0. Sec. II C states that in general no acoustic mode need have zero velocity when a condition is violated. Terhune, Kushida and Ford[12] reported that these 2B and 2E entries are type I. The determinant step, proved without stress for all crystal classes,[9] and its hydrostatic extension in Theorem 5 supply the group-theoretic reason. Cowley's column counts waves along symmetry directions whose velocity is fixed by a single eigenvalue, whereas for a repeated representation the critical direction rotates with the elastic constants (with $c_{16}$, $c_{14}$, $c_{15}$ or $c_{46}$) and a zero-velocity wave exists along it. Supplementary Table 16 gives one elastic matrix on each of these stability boundaries, with its components in Supplementary Table 17; in every case the acoustic tensor along the critical direction is singular to below $5 \times 10^{-15}$. The $A_1$ and repeated $2A_1$ entries remain without acoustic softening for a generic critical strain, as Cowley states; a critical strain on the symmetrized rank-one sets of S4.4, such as uniaxial $e_{zz}$, is the non-generic exception.

## S5. Numerical models and verification

**S5.1 Beam model.** The consistent initial-stress tangent converges at the expected rate for a pin-ended Euler column and two portal frames (Supplementary Table 13). Element-level identities and the rotation invariance of self-stressed frames hold to machine precision (Supplementary Table 14). On six loaded nets the full tangent annihilates rigid rotations, its static condensation reproduces the relaxed macroscopic tangent, and the lowest Bloch eigenvalue converges quadratically to the acoustic-tensor prediction (Supplementary Table 15). On 138 atlas nets, refining from two to eight elements per strut changes no class (Supplementary Table 12).

**S5.2 Decorated Truchet frames.** The checkerboard is the weakest of the 35 periodic-cell representatives of the two-dimensional frame, which include repeated supercell descriptions of smaller cells (Supplementary Table 23). It is also the weakest of all 65,536 4 × 4 patterns: the 158 symmetry orbits without a checkerboard block (Supplementary Table 9) and the 511 orbits that contain one (Supplementary Table 20), the next-weakest of which fails at 1.054 times the checkerboard. Planted checkerboard domains approach its threshold, and the corrected excess falls as $R^{-2}$ (Supplementary Table 2). The three-dimensional frame has 22 periodic orbits on supercells up to 2 × 2 × 2 (Supplementary Table 21); its random samples and planted domains are listed in Supplementary Table 3, and the individual $10^3$ samples in Supplementary Table 24. Local rules that remove the weakest crystals raise the weakest allowed candidate and the random-sample thresholds (Supplementary Table 9). The no-checker rule forbids every 2 × 2 checkerboard block; the sparsity rule requires every translated 3 × 3 window to contain at most one tile in state 1. All 182 periodic tilings of the three rule catalogues (periods up to 3 × 3, the allowed 4 × 4 patterns and, for the sparsity rule, 3 × 4 cells) are drawn in Supplementary Fig. 1, counted up to translation and the symmetries of the square; 13 are blocked by the no-checker rule, and the eleven weakest are among them. Correlated samples (Supplementary Table 8) assign the staggered variable $\sigma = (2s - 1)(-1)^{x+y}$ to each tile state s, so that both checkerboard phases are uniform in σ, and draw σ from the two-dimensional Ising model with nearest-neighbour coupling J by checkerboard heat-bath updates (3,000 sweeps from a random start, eight independent samples per J, 64 × 64 tiles). Thresholds are those of the fixed periodic cell after a hydrostatic preload, with the consistent tangent of Methods; solving with the three cell strains free lowers them by less than 0.002% for all 108 configurations (Supplementary Table 7). At J = 0.38 and 0.42, chains of 30,000 sweeps from random and ordered starts give nearest-neighbour correlations whose time averages over sweeps 3,010–30,000 lie within 0.0008 of the infinite-lattice exact value for each six-chain mean. Final configurations deviate more: their six-chain means by up to 0.0053 (Supplementary Table 7) and single final configurations by up to 0.0293. Their mean thresholds lie within the scatter of the 3,000-sweep values; these checks do not certify equilibrium statistics of rare domains. The largest cluster is the largest connected region of uniform σ; clusters and squares wrap across the periodic boundary. The threshold excess of these samples falls with the side $R_{sq}$ of the largest σ-uniform square. For the 48 samples with $J \geq 0.32$ a pooled log–log fit gives a slope of −2.09, and the excess times $R_{sq}^2$ has a median of 3.4 (10–90%: 2.6–4.2). A fit with one slope and a separate intercept for each J, or the samples with $0.32 \leq J \leq 0.42$ alone, gives a slope of only about −1.45. For planted single domains the full-sample pressure excess times $R^2$ is 8.6–9.9 for $R \geq 16$ (Supplementary Table 22); the corrected compact-bound excess follows $12.3R^{-1.99}$ (Supplementary Table 2). Correlated configurations therefore lower the threshold more than one compact domain of the same inscribed size, on a different scaling curve.

**S5.3 Soft inclusions.** In all 24 inclusion tests of six nets classified PD-only on the $4^3$ Bloch grid (five on the $8^3$ grid; the sixth, ncj, is the control) the onset inside the inclusion exceeds its own PD threshold (Supplementary Table 1). The control, which fails by loss of ellipticity at its PD threshold, also survives, to $\rho_{PD}$ = 2.7–3.4, so at these sizes the stiffer matrix also confines localizable failure; the tests probe matrix confinement, not stabilization by disorder, and only cco and vne come close to their localizable thresholds. A soft pcu domain (R/L = 0.34, $s_P$ = 0.72) fails mostly outside itself by a macroscopic shear mode at L = 10 (22% of the mode energy in the domain, $\rho_{PD}$ = 2.21) but inside itself at L = 20 (98% of the mode energy, $\rho_{PD}$ = 1.98).

**S5.4 Domain-size scans.** Supplementary Table 22 lists the full and compact thresholds of planted two-dimensional domains, Supplementary Table 3 the random and planted three-dimensional samples of Fig. 2, and Supplementary Table 4 the compact bound of three-dimensional domains up to R = 10.

## S6. Atlas

Supplementary Table 10 lists the classes, thresholds and windows of all 876 nets.

## S7. TRUMOF-1 in the coarse-grained model

**S7.1 Rule.** We reconstructed the Truchet rule of TRUMOF-1 from the published average structure.[13] Each cluster takes one of 24 orientations of its maximally spread carboxylate decoration, and each 1,3-benzenedicarboxylate site is occupied by zero or two carboxylates. Among the 13,824 origin-fixed assignments of a 1 × 1 × 1 cell the rule has four zero-energy configurations in three symmetry classes, which correspond to the three symmetry-distinct 1 × 1 × 1 approximants of the published structure analysis.[13] The connectivity of all 30 published DFT approximants, ten 1 × 1 × 1 and twenty 2 × 2 × 2, satisfies the rule.

**S7.2 Potential and pressure convention.** The coarse-grained potential consists of harmonic node–node springs and three-body angle springs fitted to the elastic constants of MOF-5, in a soft and a hard parameter set.[14] The published pressure-dependent Born constants of MOF-5 are reproduced by $B_{44} = C_{44}(0)(\ell_0/\ell)^3 - 2.5p$, with $\ell/\ell_0$ the isotropic stretch, whereas the objective hydrostatic tangent and its small-k dynamics give $-p$. We use $-p$ throughout;

reference angles within $10^{-6}$ rad of π are set to π, so that the analytic straight-angle limit is used near collinearity. Applying the published form as an isotropic correction to the macroscopic tangent lowers the tabulated thresholds and ensemble means to 36–61% of their objective values. It keeps the order of the three 1 × 1 × 1 approximants and every random sample above the weakest approximant, but makes *Pc* and $P2_1$ fail first by loss of positive definiteness alone (Supplementary Table 25). That convention defines no finite-wavevector operator, so it does not fix a localizable bound. The bound below holds for the objective tangent only. Its analytic Hessian agrees with central finite differences of U + pV (one-node pcu cell of MOF-5 at 0.3 GPa, soft parameters; steps $10^{-3}$ to $3 \times 10^{-5}$). Energy second differences agree to $9 \times 10^{-10}$ (step $10^{-4}$) and gradient differences to $8 \times 10^{-11}$ (step $3 \times 10^{-5}$) relative; all steps agree to $10^{-7}$. This checks internal consistency of the objective tangent; it does not explain the origin of the published −2.5p form. Unlike the linear-preload beam frames, the thresholds of this section follow the relaxed equilibrium by nonlinear continuation in pressure. In the same model of MOF-5 (pcu) the shear constant $C_{44}$ is the first homogeneous stability constant to vanish, at 0.90 and 3.74 GPa with the soft and hard parameters (0.34 and 1.22 GPa under the published form); it is a symmetry-breaking strain, so this failure can localize.

**S7.3 Approximants.** The three 1 × 1 × 1 approximants fail by localizable modes, losing positive definiteness and ellipticity together at 0.75 (*Pc*), 0.92 ($P\bar{4}$) and 1.75 ($P2_1$) GPa with the soft parameters (Supplementary Table 26, Supplementary Fig. 2a). The $P\bar{4}$ approximant consists of two non-interacting interpenetrated frameworks, which Meekel et al.[14] identified as the origin of its instability at all pressures; the relative translation of the two frameworks is a load-independent zero mode, a three-dimensional kernel that the 1 × 1 × 1 cell also shows at **k** = (1/2, 1/2, 0). It violates (A0); we remove this kernel and report the first negative eigenvalue on its complement, so the $P\bar{4}$ result is a numerical illustration outside the literal (A0). The twenty 2 × 2 × 2 approximants have point group 1, in which every strain is totally symmetric, so the selection rule permits PD-only failure; with the soft parameters 18 of 20 fail PD-only and two at a finite wavevector, and with the hard parameters all 20 fail PD-only.

**S7.4 Large patches.** Each approximant tiles space. Inside a $4^3$ cell whose outer shell is frozen, a rule-preserving rearrangement of 14 to 26 clusters joins an approximant core to a disordered surrounding, so independent blocks embed patches of each approximant, of any size, at positive density in the geometric sense; whether synthesis produces them is not fixed by the rule.

**S7.5 Ensembles.** Random rule-satisfying samples reach mean thresholds of 1.67, 1.80 and 1.85 GPa for $2^3$, $3^3$ and $4^3$ cells, with no downward trend in this range. These are located onsets: in 29 of the 30 samples the relaxed homogeneous tangent loses positive definiteness at a root of the augmented equilibrium equations, and one sample first crosses zero at the wavevector (1/2, 1/2, 1/2), among Γ and the four supercell wavevectors searched (Supplementary Table 27, Supplementary Fig. 2b). Of the planted domains, only the largest, a cubic $P\bar{4}$ domain four cells on each side in a $6^3$ sample, concentrates the onset mode (62% of its energy) and lowers the threshold to 1.46 GPa; smaller domains in different backgrounds give mixed results. Under tangent-transfer assumptions corresponding to H2, which we have not verified for this model, and at $s_P = 1$ (planted domains give 0.77–1.24, Supplementary Table 27), the weakest approximant searched, *Pc*, would bound the cap of the rule from above. Because these thresholds come from nonlinear continuation of a spring and angle model, this section illustrates a related model and does not apply Theorem 1 directly.

**S7.6 Correlated configurations inside the rule.** We drew rule-satisfying $6^3$ samples by the Metropolis annealing used for the random samples of S7.5 (Supplementary Table 27) plus a fixed bias b on every cluster whose orientation differs from one approximant, either the weakest (*Pc*) or a strong one ($P2_1$). The rule term is annealed to zero violations, and b sets the fraction of clusters in the biased approximant (five samples per b, with the shared b = 0 samples evaluated against each target; Supplementary Table 28, Supplementary Fig. 2c). This is a generation protocol, not equilibrium sampling; b is not calibrated to synthesis. The samples are continued in pressure with the straight-angle kernel of S7.2, testing four supercell wavevectors at the last load. Five samples show a recorded sign change: two nearly or fully *Pc*-ordered samples at 0.737 and 0.746 GPa, and three fully $P2_1$-ordered samples at 1.746 GPa; the tiled crystals fail at 0.746 and 1.746 GPa. The other 80 end without a located onset: 76 continuation endpoints and four $P2_1$-biased stops at the iteration limit. Biased toward *Pc*, the mean endpoint falls from 1.82 GPa at fraction 0.04 to 1.34 GPa at 0.28 and 0.88 GPa at 0.64, and at fraction 0.79 or more every onset or endpoint lies at 0.68–0.76 GPa, around the tiled *Pc* value. Biased toward $P2_1$, the mean endpoints are 1.74 and 1.42 GPa at fractions 0.24 and 0.52; at 0.89 two endpoints lie at 1.13 and 1.33 GPa, and three samples reached the iteration limit at 1.26–1.30 GPa; a further endpoint at fraction 0.99 lies at 1.46 GPa, and a fourth stop, at fraction 0.995, at 1.37 GPa. Endpoints are not onsets; the trend is exploratory.

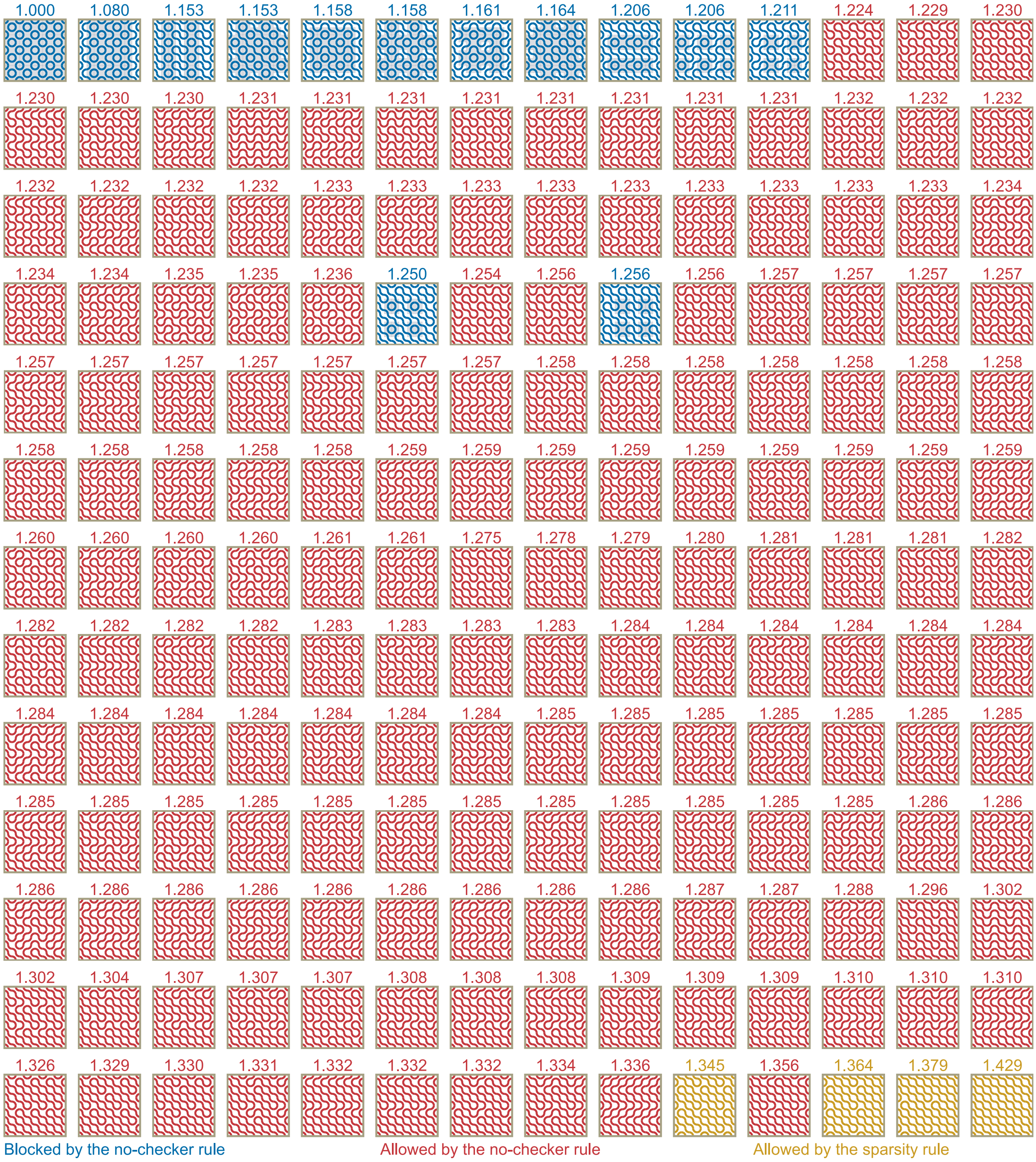

**Supplementary Fig. 1 | The tilings of Fig. 4 with their thresholds.** The 182 tilings of main-text Fig. 4 in the same order, each labelled with $p_c/p_0$ at fixed hydrostatic pressure. Symmetry-equivalent tilings are drawn once. Blue: blocked by the no-checker rule (13 tilings; checkerboard blocks shaded); red: allowed by it (165); ochre: also allowed by the sparsity rule (4). The weakest allowed tiling lies at 1.224. Other blocked $4 \times 4$ orbits are omitted; the weakest are listed in Supplementary Table 20.

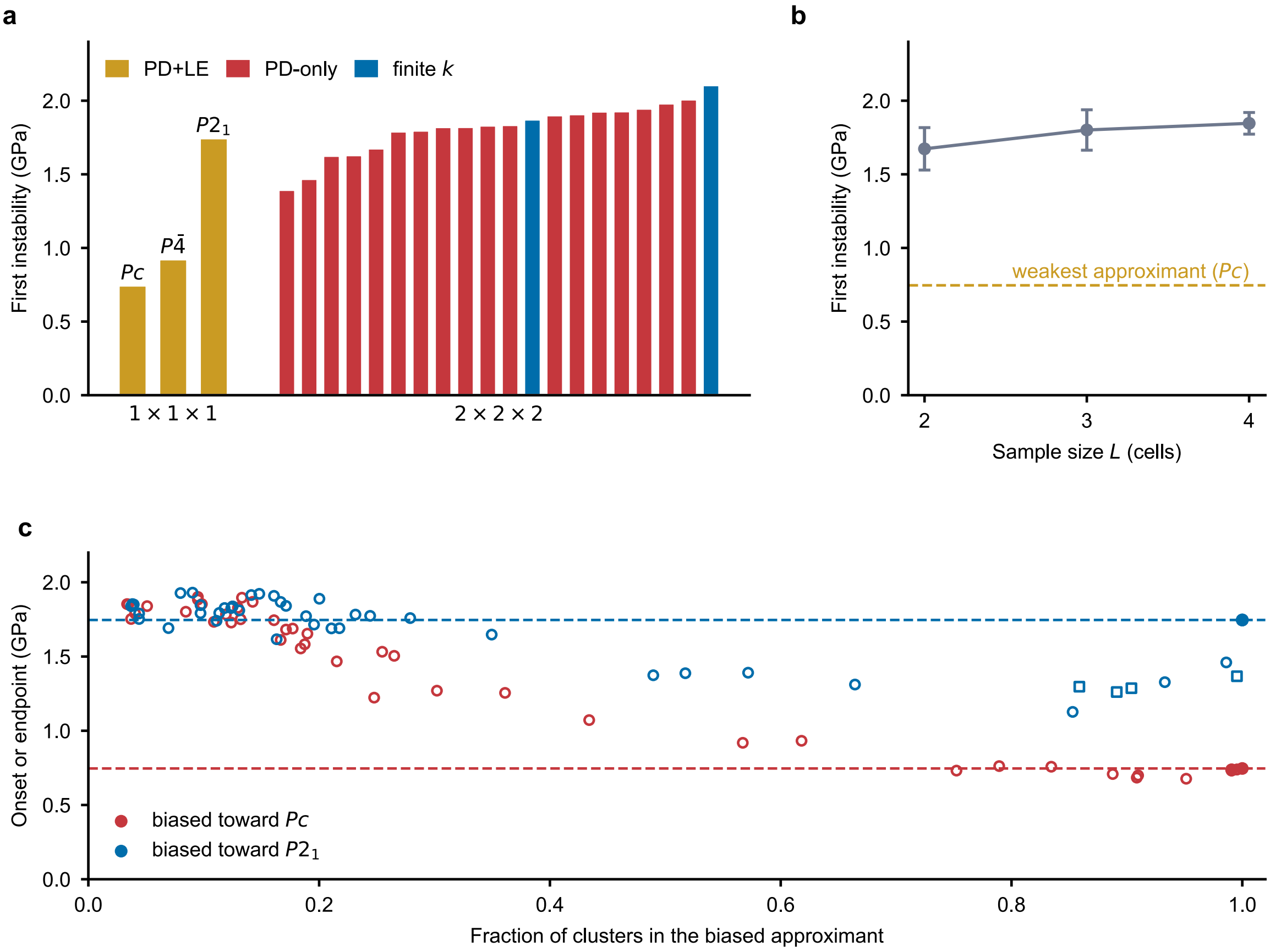

**Supplementary Fig. 2 | TRUMOF-1 within the coarse-grained model. a**, First-instability pressure of the three symmetry-distinct 1 × 1 × 1 approximants and the twenty published 2 × 2 × 2 approximants of the TRUMOF-1 rule, coloured by class (coarse-grained potential of Meekel et al.[14], soft parameters). **b**, Mean ± s.d. of ten random rule-satisfying samples versus size; the dashed line marks the weakest approximant searched (*Pc*). **c**, Correlated configurations of the rule biased toward *Pc* (red) or $P2_1$ (blue): load versus the fraction of clusters in the biased approximant; filled dots, located onsets; open circles, continuation endpoints, which are not onsets; open squares, endpoints at the iteration limit; dashed, the tiled approximant (Supplementary Table 28). The five unbiased configurations are shared by the two series, with order fractions evaluated for each target. The trend is exploratory.

## Supplementary Table 1 | Soft inclusions of nets classified PD-only on the $4^3$ Bloch grid

| Net | $L$ | $R/L$ | $s_P$ | $\rho_{PD}$ | $\rho_{loc}$ | Energy in inclusion | Energy in core |
|---|---|---|---|---|---|---|---|
| cco | 6 | 0.22 | 0.809 | 2.990 | 1.373 | 98.5% | 34.8% |
| cco | 6 | 0.34 | 0.769 | 2.543 | 1.168 | 99.0% | 70.1% |
| cco | 10 | 0.22 | 0.793 | 2.652 | 1.219 | 99.3% | 68.7% |
| cco | 10 | 0.34 | 0.773 | 2.411 | 1.108 | 99.8% | 92.9% |
| ncd | 6 | 0.22 | 0.819 | 3.229 | 2.034 | 97.5% | 36.0% |
| ncd | 6 | 0.34 | 0.788 | 2.769 | 1.744 | 99.2% | 69.1% |
| ncd | 8 | 0.22 | 0.724 | 2.701 | 1.701 | 98.5% | 57.0% |
| ncd | 8 | 0.34 | 0.810 | 2.733 | 1.722 | 99.7% | 83.1% |
| ncj (control) | 6 | 0.22 | 0.822 | 3.439 | 3.439 | 96.5% | 33.1% |
| ncj (control) | 6 | 0.34 | 0.799 | 2.833 | 2.833 | 98.6% | 65.2% |
| ncj (control) | 8 | 0.22 | 0.797 | 3.025 | 3.025 | 97.7% | 51.5% |
| ncj (control) | 8 | 0.34 | 0.816 | 2.713 | 2.713 | 99.3% | 74.8% |
| oof | 6 | 0.22 | 0.894 | 3.716 | 3.578 | 94.0% | 22.9% |
| oof | 6 | 0.34 | 0.804 | 2.806 | 2.702 | 98.2% | 49.4% |
| utp | 6 | 0.22 | 0.997 | 3.351 | 3.266 | 92.5% | 57.7% |
| utp | 6 | 0.34 | 0.997 | 2.664 | 2.597 | 95.1% | 75.2% |
| utp | 8 | 0.22 | 0.996 | 2.915 | 2.841 | 94.9% | 74.1% |
| utp | 8 | 0.34 | 0.997 | 2.361 | 2.301 | 97.7% | 89.5% |
| utp | 10 | 0.22 | 0.996 | 2.588 | 2.523 | 95.9% | 80.8% |
| utp | 10 | 0.34 | 0.996 | 2.221 | 2.164 | 98.9% | 95.2% |
| vne | 6 | 0.22 | 0.764 | 2.773 | 1.647 | 93.7% | 48.8% |
| vne | 6 | 0.34 | 0.791 | 2.285 | 1.357 | 98.3% | 45.3% |
| vne | 10 | 0.22 | 0.771 | 2.093 | 1.244 | 98.4% | 59.1% |
| vne | 10 | 0.34 | 0.773 | 1.953 | 1.160 | 99.5% | 81.6% |

A spherical region of radius $R$ in an $L^3$ supercell of the net has struts of $r/l$ = 0.02; the rest of the same net has $r/l$ = 0.03. $R/L$ is the radius in units of the supercell edge. $\rho_{PD}$ and $\rho_{loc}$ are the onset pressure in the inclusion, $s_P$ times the applied pressure, divided by the bulk $\lambda_{PD}$ and $\lambda_{loc}$ of the net; $s_P$ is the mean cell pressure inside the inclusion over the mean applied pressure. $\rho_{PD} > 1$ in every case: the inclusion survives past its own PD threshold, the PD+LE control ncj included ($\rho_{PD} = \rho_{loc}$ = 2.7–3.4), so at these sizes the matrix also confines localizable failure. For the largest cco and vne inclusions ($L$ = 10, $R/L$ = 0.34) the interior stress has a deviatoric fraction of 0.5% and 5.6%; the localizable threshold of the crystal loaded along that stress tensor gives $\rho_{loc}$ = 1.110 and 1.156, against 1.108 and 1.160 with the scalar correction. $\rho_{loc}$ uses $\lambda_{loc}$ of Supplementary Table 10. Energy in core: fraction of the mode energy in struts within $\max(R/2, R - l_{max})$ of the inclusion centre, $l_{max}$ the longest strut. ncj, PD-only on the $4^3$ grid, loses ellipticity together with positive definiteness on the $8^3$ grid and serves as the control.

## Supplementary Table 2 | Planted checkerboard domains in the two-dimensional frame

| Sample | Domain | Full sample, strain | Full sample, pressure | Energy in domain | Compact bound, strain | Compact bound, pressure | $s_P$ | Corrected excess |
|---|---|---|---|---|---|---|---|---|
| 64×64 | 16×16 | – | – | – | – | 1.0483 | 1.0006 | 0.0489 |
| 64×64 | 24×24 | – | – | – | – | 1.0212 | 1.0009 | 0.0221 |
| 64×64 | 32×32 | – | – | – | – | 1.0115 | 1.0006 | 0.0121 |
| 64×64 | 40×40 | – | – | – | – | 1.0072 | 1.0007 | 0.0079 |
| 64×64 | 48×48 | 1.0057 | 1.0043 | 99.8% | 1.0063 | 1.0049 | 1.0006 | 0.0055 |
| 96×96 | 64×64 | – | – | – | 1.0041 | 1.0024 | 1.0006 | 0.0031 |

Thresholds in units of the checkerboard threshold. “Full sample” is the first-instability threshold of the whole sample with the domain planted in an independent random background; “compact bound” restricts the trial field to the domain (an upper bound on the sample threshold). $s_P$ is the stress-concentration factor of the domain, and the corrected excess is the compact pressure ratio multiplied by $s_P$, minus one (the cap is $\lambda_{loc}/s_P$). A power-law fit of this corrected excess over the listed domains gives exponent 1.992. Full-sample values for the other domains are in Supplementary Table 22. Without a planted domain, 20 random 32×32 samples at fixed strain give 1.2193 (range 1.2060–1.2361).

**Supplementary Table 3 | Random samples and planted domains in the three-dimensional frame (Fig. 2c)**

| Sample | n | Pressure / weakest periodic | Participating cells | Energy in domain | $s_P$ |
|---|---|---|---|---|---|
| random, $L = 2$ | 10 | 1.4754 ± 0.1377 | 97–100% | – | – |
| random, $L = 4$ | 10 | 1.4277 ± 0.0329 | 97–98% | – | – |
| random, $L = 6$ | 10 | 1.4211 ± 0.0226 | 93–98% | – | – |
| random, $L = 8$ | 10 | 1.4245 ± 0.0149 | 91–97% | – | – |
| random, $L = 10$ | 10 | 1.4238 ± 0.0065 | 88–95% | – | – |
| $L = 4$, no domain | 1 | 1.4138 | 98.1% | – | – |
| $L = 4$, domain $R = 2$ | 1 | 1.3995 | 97.4% | 14.0% | 1.0000 |
| $L = 6$, no domain | 1 | 1.4078 | 95.6% | – | – |
| $L = 6$, domain $R = 2$ | 1 | 1.4039 | 95.8% | 3.9% | 0.9987 |
| $L = 6$, domain $R = 3$ | 1 | 1.3659 | 94.0% | 16.8% | 1.0004 |
| $L = 6$, domain $R = 4$ | 1 | 1.2894 | 84.1% | 42.1% | 1.0005 |
| $L = 8$, no domain | 1 | 1.4212 | 95.5% | – | – |
| $L = 8$, domain $R = 2$ | 1 | 1.4203 | 95.3% | 1.7% | 0.9999 |
| $L = 8$, domain $R = 3$ | 1 | 1.4071 | 87.3% | 9.6% | 1.0002 |
| $L = 8$, domain $R = 4$ | 1 | 1.3431 | 58.4% | 33.8% | 1.0005 |
| $L = 8$, domain $R = 5$ | 1 | 1.2699 | 51.1% | 55.6% | 1.0005 |
| $L = 8$, domain $R = 6$ | 1 | 1.2046 | 60.5% | 68.9% | 1.0004 |
| $L = 10$, no domain | 1 | 1.4108 | 92.1% | – | – |
| $L = 10$, domain $R = 2$ | 1 | 1.4092 | 91.3% | 1.2% | 0.9972 |
| $L = 10$, domain $R = 3$ | 1 | 1.4015 | 83.9% | 6.3% | 0.9996 |
| $L = 10$, domain $R = 4$ | 1 | 1.3684 | 42.6% | 29.2% | 0.9998 |
| $L = 10$, domain $R = 5$ | 1 | 1.2940 | 27.0% | 56.8% | 1.0004 |
| $L = 10$, domain $R = 6$ | 1 | 1.2316 | 30.3% | 70.2% | 1.0004 |
| $L = 10$, domain $R = 7$ | 1 | 1.1838 | 36.8% | 78.0% | 1.0005 |
| $L = 10$, domain $R = 8$ | 1 | 1.1377 | 49.7% | 81.6% | 1.0004 |

Random samples of $L^3$ cubes (mean ± s.d.) and samples with a central $R^3$ layer-uniform checkerboard domain in one random background per $L$. Participating cells: participation ratio of the cell-resolved critical-mode energy, as a fraction of all cells. $s_P$: mean interior pressure of the domain over the applied pressure (the pressure form of the force ratio defined in S3), computed on the identical configurations.

**Supplementary Table 4 | Compact-support bound of planted domains in the three-dimensional frame**

| Domain $R$ (cells) | Compact bound / weakest periodic | $s_P$ | Corrected excess | Local exponent to next $R$ |
|---|---|---|---|---|
| 2 | 3.0560 | 0.9999 | 2.0558 | 0.93 |
| 3 | 2.4093 | 1.0004 | 1.4103 | 1.50 |
| 4 | 1.9140 | 1.0005 | 0.9151 | 1.71 |
| 5 | 1.6239 | 1.0008 | 0.6253 | 1.83 |
| 6 | 1.4470 | 1.0005 | 0.4478 | 1.87 |
| 7 | 1.3348 | 1.0005 | 0.3355 | 1.91 |
| 8 | 1.2591 | 1.0006 | 0.2599 | 1.94 |
| 10 | 1.1679 | 1.0004 | 0.1684 | – |

Local exponent: $-\ln[\text{excess}(R')/\text{excess}(R)]/\ln(R'/R)$, with $R'$ the next listed size. One $14^3$ random background with a central $R^3$ domain of the weakest periodic motif, fixed hydrostatic pressure, consistent tangent. Corrected excess = compact ratio × $s_P$ − 1. A fit $\alpha_2 R^{-2} + \alpha_3 R^{-3}$ for $R \geq 4$ gives $\alpha_2 = 18.9$, $\alpha_3 = -16.9$ (relative rms 1.0%, maximum 2.2%).

**Supplementary Table 5 | Independent tiles: threshold and mode localization versus sample size**

| *L* | n | Threshold / $p_0$, mean ± s.d. | Range | Largest square, mean side | Mode energy in it, mean | PR | Participating tiles |
|---|---|---|---|---|---|---|---|
| 32 | 8 | 1.2145 ± 0.0123 | 1.1874–1.2290 | 3.0 | 8.5% | 33.88% | 347 |
| 64 | 8 | 1.2081 ± 0.0085 | 1.1958–1.2186 | 3.1 | 9.4% | 8.19% | 335 |
| 96 | 8 | 1.2020 ± 0.0078 | 1.1867–1.2130 | 3.2 | 15.5% | 1.89% | 174 |
| 128 | 8 | 1.1984 ± 0.0074 | 1.1877–1.2107 | 3.5 | 8.5% | 1.13% | 185 |
| 192 | 8 | 1.1971 ± 0.0086 | 1.1808–1.2039 | 3.4 | 7.9% | 0.46% | 170 |

Largest square: the largest checkerboard square of the sample; PR, participation ratio. Independent tile choices in $L \times L$ samples of the two-dimensional frame; fluctuations of the fixed periodic cell after hydrostatic preload, as in Supplementary Table 6; these eight samples per size are drawn independently of the 20 samples of Supplementary Table 2. The largest checkerboard square is searched across the periodic boundary; its mode energy is the larger value over tied squares. Participation ratio and participating tiles as in Supplementary Table 6.

**Supplementary Table 6 | Correlated configurations below the Ising transition: threshold and mode localization versus sample size**

| Coupling $J$ | Sample $L$ | n | Threshold / $p_0$, mean ± s.d. | Range | Largest uniform square side | Participation ratio | Participating tiles |
|---|---|---|---|---|---|---|---|
| 0.38 | 32 | 6 | 1.0873 ± 0.0106 | 1.0754–1.1060 | 6.0 | 25.27% | 259 |
| 0.38 | 64 | 6 | 1.0708 ± 0.0049 | 1.0645–1.0787 | 7.5 | 3.27% | 134 |
| 0.38 | 96 | 6 | 1.0694 ± 0.0114 | 1.0549–1.0888 | 7.7 | 1.95% | 180 |
| 0.38 | 128 | 6 | 1.0629 ± 0.0058 | 1.0561–1.0708 | 7.7 | 0.95% | 155 |
| 0.42 | 32 | 6 | 1.0593 ± 0.0119 | 1.0455–1.0724 | 8.0 | 28.43% | 291 |
| 0.42 | 64 | 6 | 1.0461 ± 0.0088 | 1.0387–1.0620 | 8.8 | 7.08% | 290 |
| 0.42 | 96 | 6 | 1.0443 ± 0.0064 | 1.0385–1.0556 | 9.7 | 2.32% | 213 |
| 0.42 | 128 | 6 | 1.0418 ± 0.0056 | 1.0363–1.0516 | 9.7 | 1.47% | 241 |

Tile states carry two-dimensional Ising correlations of the staggered variable (critical coupling $\ln(1+\sqrt{2})/2 \approx 0.4407$[15]), sampled by checkerboard heat-bath updates (3,000 sweeps from a random start; longer chains and an ordered start in Supplementary Table 7); hydrostatic preload, fluctuations of the fixed periodic cell (the free cell changes them by less than 0.002%, Supplementary Table 7). The largest uniform square is searched across the periodic boundary. Participation ratio: $1/(L^2\Sigma E_i^2)$ of the normalized tile energies of the critical mode; participating tiles = $L^2$ × participation ratio. Thresholds in units of the checkerboard threshold.

## Supplementary Table 7 | Correlated configurations: longer sampling, two starts and the free cell

| *L* | *J* | Start | NN, time avg. | NN, final | NN, exact | Threshold, 3k sweeps | Threshold, 30k sweeps | PR, 3k | PR, 30k |
|---|---|---|---|---|---|---|---|---|---|
| 64 | 0.38 | ordered | 0.5034 | 0.5007 | 0.5033 | 1.0718 ± 0.0079 | 1.0695 ± 0.0073 | 5.00% | 4.28% |
| 64 | 0.38 | random | 0.5034 | 0.4980 | 0.5033 | 1.0708 ± 0.0049 | 1.0663 ± 0.0061 | 3.27% | 3.55% |
| 64 | 0.42 | ordered | 0.6123 | 0.6140 | 0.6130 | 1.0486 ± 0.0074 | 1.0475 ± 0.0045 | 4.75% | 5.30% |
| 64 | 0.42 | random | 0.6134 | 0.6152 | 0.6130 | 1.0461 ± 0.0088 | 1.0480 ± 0.0052 | 7.08% | 7.31% |
| 128 | 0.38 | ordered | 0.5034 | 0.5013 | 0.5033 | 1.0660 ± 0.0034 | 1.0648 ± 0.0074 | 1.09% | 0.94% |
| 128 | 0.38 | random | 0.5034 | 0.5011 | 0.5033 | 1.0629 ± 0.0058 | 1.0629 ± 0.0067 | 0.95% | 0.89% |
| 128 | 0.42 | ordered | 0.6132 | 0.6141 | 0.6130 | 1.0447 ± 0.0060 | 1.0432 ± 0.0022 | 1.75% | 2.22% |
| 128 | 0.42 | random | 0.6129 | 0.6108 | 0.6130 | 1.0418 ± 0.0056 | 1.0415 ± 0.0027 | 1.47% | 1.61% |

*L*, sample size; *J*, coupling; NN, nearest-neighbour correlation; PR, participation ratio; 3k and 30k, 3,000 and 30,000 sweeps. Six heat-bath chains per row from a random or a fully ordered start; individual final configurations deviate from the exact NN value by up to 0.0293; thresholds of the fixed periodic cell after hydrostatic preload, in units of the checkerboard threshold. NN: nearest-neighbour correlation of the staggered variable, averaged over the six chains; time average over sweeps 3,010–30,000 of each chain, or the final configuration after 30,000 sweeps; exact, the infinite-lattice value[15], which finite size and sampling need not reach. Free cell: solving the coupled problem of all internal fluctuations and the three cell strains, with the pressure–volume term, for the 108 distinct configurations of Supplementary Tables 6 and 8 lowers the threshold by at most 0.0013%; the cell carries at most 0.003% of the mode energy.

## Supplementary Table 8 | Correlated samples of the two-dimensional frame

| Coupling $J$ | n | Threshold / $p_0$, mean | Range | Largest checkerboard cluster (tiles) | Mode energy in that cluster | \|staggered magnetization\| |
|---|---|---|---|---|---|---|
| 0.00 | 8 | 1.2076 | 1.1947–1.2139 | 209 | 21.6% | 0.01 |
| 0.15 | 8 | 1.1702 | 1.1643–1.1802 | 715 | 22.7% | 0.03 |
| 0.25 | 8 | 1.1324 | 1.1211–1.1415 | 1160 | 30.8% | 0.02 |
| 0.32 | 8 | 1.1057 | 1.0939–1.1168 | 1599 | 51.4% | 0.04 |
| 0.38 | 8 | 1.0731 | 1.0645–1.0861 | 1888 | 58.1% | 0.07 |
| 0.42 | 8 | 1.0463 | 1.0387–1.0620 | 2375 | 94.7% | 0.23 |
| 0.46 | 8 | 1.0258 | 1.0192–1.0299 | 3703 | 97.5% | 0.81 |
| 0.55 | 8 | 1.0110 | 1.0073–1.0150 | 3822 | 98.9% | 0.87 |
| 0.70 | 8 | 1.0040 | 1.0020–1.0075 | 3554 | 99.1% | 0.74 |

64×64 tiles, 3,000 heat-bath sweeps from a random start; fluctuations of the fixed periodic cell after hydrostatic preload; thresholds in units of the checkerboard threshold. Tile states carry two-dimensional Ising correlations of the staggered variable with coupling $J$ (critical value $\ln(1+\sqrt{2})/2 \approx 0.4407$); $J = 0$ is the independent-tile ensemble. Above $J_c$ the configurations still hold domains after 3,000 sweeps (|staggered magnetization| below the equilibrium value). Clusters are joined across the periodic boundary; mode energy is the positive elastic energy of the critical mode.

**Supplementary Table 9 | Local rules and upper estimates of the cap in the two-dimensional frame**

| Ensemble | Periodic candidates | Weakest allowed candidate | Class | Random 64×64 samples, mean (range) |
|---|---|---|---|---|
| Independent tiles | – | 1 (checkerboard) | Γ-optical | 1.2066 (1.1987–1.2147) |
| No 2×2 checkerboard block | 180 | 1.2243 (0000/0101/0000/0101) | Γ-optical | 1.2684 (1.2662–1.2706) |
| 3×3 sparsity rule | 7 | 1.3449 (100/000/000) | finite $k$ | – |

Pressures in units of the checkerboard threshold $p_0$. Class: first instability of the weakest allowed candidate. The no-checker candidates are the 22 allowed periodic-cell representatives with periods up to 3×3, including repeated supercell descriptions, and the 158 symmetry orbits of the 13,278 allowed 4×4 patterns; 12 samples per sampled ensemble. At $s_P = 1$ the weakest candidate is an upper estimate of the cap of each rule, not a certified infimum over all periods; a higher weakest candidate does not by itself show that the cap rises. All 152 zone-centre optical no-checker candidates have $\lambda_\Gamma$ within 0.11% of $\lambda_{LE}$, and the weakest no-checker candidate within $1.8\times10^{-7}$; that candidate is the 2×2 cell 00/01 written as a 4×4 cell (1.2316 at prescribed strain in Supplementary Table 23). The sparsity rule requires every translated 3×3 window to contain at most one tile in state 1; its seven candidates are the all-0 and single-1 cells with periods 1×1, 3×3, 3×4 and 4×4.

## Supplementary Table 10 | First pressure instability of 876 RCSR nets

| Net | Space group | Sys. | Class | $\lambda_{PD}$ | $\lambda_{LE}$ | $\lambda_{\mu}^{num}$ | $f_{inv}$ | Window | Class, *r*/*l* = 0.05 |
|---|---|---|---|---|---|---|---|---|---|
| abf | I$\bar{4}$m2 | tet | PD+LE | $2.11\times10^{-8}$ | $2.11\times10^{-8}$ | – | 0.00 | – | PD+LE |
| acc | P$6_3$/mmc | hex | $\mu$ | – | – | $1.02\times10^{-6}$ | – | – | $\mu$ |
| acg | P$\bar{6}$m2 | hex | $\mu$ | – | – | $1.42\times10^{-6}$ | – | – | $\mu$ |
| ach | P$6_3$/mmc | hex | $\mu$ | – | – | $1.65\times10^{-6}$ | – | – | $\mu$ |
| aci | P$\bar{6}$m2 | hex | $\mu$ | – | – | $8.14\times10^{-7}$ | – | – | $\mu$ |
| ack | P$6_3$/mmc | hex | $\mu$ | – | – | $2.11\times10^{-6}$ | – | – | $\mu$ |
| acl | P$\bar{6}$m2 | hex | $\mu$ | $4.81\times10^{-6}$ | $4.81\times10^{-6}$ | $2.83\times10^{-6}$ | 0.00 | – | $\mu$ |
| acn | P$6_3$/mmc | hex | $\mu$ | $1.29\times10^{-7}$ | $1.29\times10^{-7}$ | $7.26\times10^{-8}$ | 0.00 | – | $\mu$ |
| aco | P$6_3$/mmc | hex | $\mu$ | – | – | $3.34\times10^{-7}$ | – | – | $\mu$ |
| acs | P$6_3$/mmc | hex | $\mu$ | $6.84\times10^{-7}$ | $7.52\times10^{-7}$ | $5.74\times10^{-7}$ | 1.00 | – | $\mu$ |
| afw | R32 | trig | $\mu$ | $1.67\times10^{-7}$ | $1.67\times10^{-7}$ | $1.28\times10^{-7}$ | 0.00 | – | $\mu$ |
| ala | P$\bar{6}$2m | hex | $\mu$ | $4.87\times10^{-7}$ | $5.21\times10^{-7}$ | $3.99\times10^{-7}$ | 1.00 | – | $\mu$ |
| alb | P6/mmm | hex | PD+LE | $1.74\times10^{-6}$ | $1.74\times10^{-6}$ | – | 0.00 | – | PD+LE |
| ald | P6/mmm | hex | $\mu$ | – | – | $7.31\times10^{-7}$ | – | – | $\mu$ |
| ale | P6/mmm | hex | PD+LE | $2.86\times10^{-6}$ | $2.86\times10^{-6}$ | $2.86\times10^{-6}$ | 0.00 | – | PD+LE |
| alh | P6/mmm | hex | $\mu$ | $7.53\times10^{-6}$ | $7.53\times10^{-6}$ | $4.43\times10^{-6}$ | 0.00 | – | $\mu$ |
| ali | P6/mmm | hex | $\mu$ | – | – | $2.80\times10^{-6}$ | – | – | $\mu$ |
| alj | P6/mmm | hex | $\mu$ | – | – | $2.76\times10^{-6}$ | – | – | $\mu$ |
| alk | P6/mmm | hex | PD+LE | $2.74\times10^{-6}$ | $2.74\times10^{-6}$ | – | 0.00 | – | PD+LE |
| alm | P$4_2$/mnm | tet | PD+LE | $1.20\times10^{-7}$ | $1.20\times10^{-7}$ | $1.22\times10^{-7}$ | 0.00 | – | PD+LE |
| anh | P$6_2$22 | hex | $\mu$ | $1.40\times10^{-7}$ | $1.40\times10^{-7}$ | $1.33\times10^{-7}$ | 0.00 | – | $\mu$ |
| ant | I$4_1$/amd | tet | PD+LE | $8.77\times10^{-8}$ | $8.77\times10^{-8}$ | – | 0.00 | – | PD+LE |
| apo | Pbcn | orth | PD-only | $1.14\times10^{-7}$ | $1.23\times10^{-7}$ | – | 1.00 | 8.0% | PD-only |
| asc | P$\bar{6}$2m | hex | PD+LE | $1.37\times10^{-7}$ | $1.37\times10^{-7}$ | – | 0.00 | – | PD+LE |
| asd | P4/mmm | tet | PD+LE | $1.06\times10^{-7}$ | $1.06\times10^{-7}$ | – | 0.00 | – | PD+LE |
| ase | P6/mmm | hex | PD+LE | $1.63\times10^{-7}$ | $1.63\times10^{-7}$ | – | 0.00 | – | PD+LE |
| asv | P4/mmm | tet | $\mu$ | $4.22\times10^{-8}$ | $4.22\times10^{-8}$ | $3.39\times10^{-8}$ | 0.00 | – | $\mu$ |
| att | Pmma | orth | $\mu$ | – | – | $7.97\times10^{-8}$ | – | – | $\mu$ |
| baa | I$\bar{4}$2d | tet | $\mu$ | – | – | $1.61\times10^{-7}$ | – | – | $\mu$ |
| bab | P$\bar{6}$m2 | hex | $\mu$ | – | – | $3.63\times10^{-8}$ | – | – | $\mu$ |
| bal | P$\bar{3}$1m | trig | $\mu$ | $9.56\times10^{-8}$ | $9.56\times10^{-8}$ | $6.08\times10^{-8}$ | 0.00 | – | $\mu$ |
| bao | P6/mmm | hex | $\mu$ | $9.94\times10^{-8}$ | $9.94\times10^{-8}$ | $8.34\times10^{-8}$ | 0.00 | – | $\mu$ |
| baz | P$6_3$/mmc | hex | $\mu$ | – | – | $1.11\times10^{-8}$ | – | – | $\mu$ |
| bba | P6/mmm | hex | $\mu$ | – | – | $1.34\times10^{-7}$ | – | – | $\mu$ |
| bbc | P4/mmm | tet | $\mu$ | – | – | $9.08\times10^{-8}$ | – | – | $\mu$ |
| bbe | Imma | orth | $\mu$ | – | – | $6.94\times10^{-8}$ | – | – | $\mu$ |
| bbf | Pmna | orth | PD+LE | $1.71\times10^{-7}$ | $1.71\times10^{-7}$ | – | 0.00 | – | PD+LE |
| bbg | Pccm | orth | $\mu$ | $1.52\times10^{-7}$ | $1.52\times10^{-7}$ | $1.17\times10^{-7}$ | 0.00 | – | $\mu$ |
| bbh | Cmma | orth | $\mu$ | – | – | $1.35\times10^{-7}$ | – | – | $\mu$ |
| bbi | C$222_1$ | orth | $\mu$ | – | – | $1.22\times10^{-7}$ | – | – | $\mu$ |

| Net | Space group | Sys. | Class | $\lambda_{PD}$ | $\lambda_{LE}$ | $\lambda_\mu^{num}$ | $f_{inv}$ | Window | Class, *r*/*l* = 0.05 |
|---|---|---|---|---|---|---|---|---|---|
| bbj | Ama2 | orth | PD+LE | $1.15\times10^{-7}$ | $1.15\times10^{-7}$ | – | 0.00 | – | PD+LE |
| bbk | $P222_1$ | orth | PD+LE | $1.16\times10^{-7}$ | $1.16\times10^{-7}$ | – | 0.00 | – | PD+LE |
| bbl | $I2_12_12_1$ | orth | $\mu$ | – | – | $1.74\times10^{-7}$ | – | – | $\mu$ |
| bbm | C2/c | mono | PD+LE | $1.15\times10^{-7}$ | $1.15\times10^{-7}$ | – | 0.00 | – | PD+LE |
| bbr | $P4_122$ | tet | $\mu$ | $9.88\times10^{-8}$ | $9.88\times10^{-8}$ | $9.30\times10^{-8}$ | 0.00 | – | $\mu$ |
| bbs | $P6_122$ | hex | $\mu$ | $9.99\times10^{-8}$ | $9.99\times10^{-8}$ | $9.58\times10^{-8}$ | 0.00 | – | $\mu$ |
| bbw | $P6_3/mmc$ | hex | $\mu$ | – | – | $1.77\times10^{-8}$ | – | – | $\mu$ |
| bcg | Cccm | orth | PD+LE | $8.25\times10^{-7}$ | $8.25\times10^{-7}$ | – | 0.00 | – | PD+LE |
| bck | $P3_212$ | trig | $\mu$ | – | – | $6.11\times10^{-8}$ | – | – | $\mu$ |
| bcm | $P4_2/m$ | tet | $\mu$ | $2.26\times10^{-7}$ | $2.26\times10^{-7}$ | $1.49\times10^{-7}$ | 0.00 | – | $\mu$ |
| bcn | Pnn2 | orth | PD+LE | $1.38\times10^{-7}$ | $1.38\times10^{-7}$ | $1.38\times10^{-7}$ | 0.00 | – | PD+LE |
| bco | $P222_1$ | orth | PD+LE | $1.49\times10^{-7}$ | $1.49\times10^{-7}$ | – | 0.00 | – | PD+LE |
| bcp | Cmma | orth | $\mu$ | $1.32\times10^{-7}$ | $1.32\times10^{-7}$ | $1.03\times10^{-7}$ | 0.00 | – | $\mu$ |
| bct | I4/mmm | tet | PD+LE | $1.37\times10^{-6}$ | $1.37\times10^{-6}$ | – | 0.00 | – | PD+LE |
| bcu | $Im\bar{3}m$ | cub | PD+LE | $1.19\times10^{-6}$ | $1.19\times10^{-6}$ | – | 0.00 | – | PD+LE |
| bel | $P\bar{6}m2$ | hex | $\mu$ | $9.61\times10^{-8}$ | $9.61\times10^{-8}$ | $8.59\times10^{-8}$ | 0.00 | – | $\mu$ |
| ben | $P\bar{6}m2$ | hex | $\mu$ | $2.98\times10^{-7}$ | $2.98\times10^{-7}$ | $1.56\times10^{-7}$ | 0.00 | – | $\mu$ |
| bet | $Pm\bar{3}n$ | cub | $\mu$ | – | – | $7.27\times10^{-6}$ | – | – | $\mu$ |
| bik | Cmcm | orth | $\mu$ | – | – | $1.41\times10^{-7}$ | – | – | $\mu$ |
| bne | I2/m | mono | $\mu$ | – | – | $5.72\times10^{-8}$ | – | – | $\mu$ |
| bnn | P6/mmm | hex | $\mu$ | $4.51\times10^{-7}$ | $4.51\times10^{-7}$ | $2.65\times10^{-7}$ | 0.00 | – | $\mu$ |
| bob | $P\bar{4}3m$ | cub | $\mu$ | $6.12\times10^{-7}$ | $6.12\times10^{-7}$ | $4.41\times10^{-7}$ | 0.00 | – | $\mu$ |
| boc | $P\bar{4}3m$ | cub | $\mu$ | $1.44\times10^{-7}$ | $1.44\times10^{-7}$ | $1.04\times10^{-7}$ | 0.00 | – | $\mu$ |
| bod | $P\bar{4}3m$ | cub | $\mu$ | $1.72\times10^{-8}$ | $1.72\times10^{-8}$ | $1.18\times10^{-8}$ | 0.00 | – | $\mu$ |
| boe | P23 | cub | $\mu$ | $5.68\times10^{-8}$ | $5.68\times10^{-8}$ | $3.91\times10^{-8}$ | 0.00 | – | $\mu$ |
| boq | $P\bar{4}2m$ | tet | $\mu$ | $1.56\times10^{-7}$ | $1.56\times10^{-7}$ | $1.30\times10^{-7}$ | 0.00 | – | $\mu$ |
| bor | $P\bar{4}3m$ | cub | $\mu$ | $1.06\times10^{-7}$ | $1.06\times10^{-7}$ | $8.54\times10^{-8}$ | 0.00 | – | $\mu$ |
| bos | $P\bar{4}2m$ | tet | $\mu$ | $3.22\times10^{-7}$ | $3.22\times10^{-7}$ | $2.37\times10^{-7}$ | 0.00 | – | $\mu$ |
| bow | Immm | orth | $\mu$ | – | – | $7.60\times10^{-8}$ | – | – | $\mu$ |
| bpb | $P6_522$ | hex | $\mu$ | $3.47\times10^{-8}$ | $3.47\times10^{-8}$ | $3.08\times10^{-8}$ | 0.00 | – | $\mu$ |
| bpf | P2/c | mono | PD-only | $2.66\times10^{-8}$ | $2.92\times10^{-8}$ | – | 1.00 | 9.7% | PD-only |
| bpj | $P4_2/mmc$ | tet | $\mu$ | $1.78\times10^{-8}$ | $1.78\times10^{-8}$ | $1.19\times10^{-8}$ | 0.00 | – | $\mu$ |
| bpq | Imma | orth | PD+LE | $9.83\times10^{-8}$ | $9.83\times10^{-8}$ | – | 1.00 | – | PD+LE |
| bpt | Pnnn | orth | $\mu$ | – | – | $2.17\times10^{-8}$ | – | – | $\mu$ |
| brl | P6/mmm | hex | $\mu$ | $1.90\times10^{-7}$ | $1.90\times10^{-7}$ | $1.28\times10^{-7}$ | 0.00 | – | $\mu$ |
| brw | P23 | cub | $\mu$ | $4.58\times10^{-7}$ | $4.58\times10^{-7}$ | $4.53\times10^{-7}$ | 0.00 | – | $\mu$ |
| bsa | I4/mmm | tet | PD+LE | $2.43\times10^{-6}$ | $2.43\times10^{-6}$ | $3.06\times10^{-6}$ | 0.00 | – | PD+LE |
| bsd | P4/mmm | tet | PD+LE | $2.02\times10^{-6}$ | $2.02\times10^{-6}$ | $2.07\times10^{-6}$ | 0.00 | – | PD+LE |
| bsn | $I4_1/amd$ | tet | PD+LE | $6.84\times10^{-7}$ | $6.84\times10^{-7}$ | – | 0.00 | – | PD+LE |
| bto | $P6_222$ | hex | $\mu$ | $4.06\times10^{-8}$ | $4.06\times10^{-8}$ | $2.28\times10^{-8}$ | 0.00 | – | $\mu$ |
| btu | Pmmm | orth | $\mu$ | $4.90\times10^{-7}$ | $4.90\times10^{-7}$ | $3.61\times10^{-7}$ | 0.00 | – | $\mu$ |
| btv | Pmmm | orth | $\mu$ | $3.71\times10^{-7}$ | $3.71\times10^{-7}$ | $3.30\times10^{-7}$ | 0.00 | – | $\mu$ |
| cab | $Pm\bar{3}m$ | cub | $\mu$ | $9.96\times10^{-8}$ | $9.96\times10^{-8}$ | $4.53\times10^{-8}$ | 0.00 | – | $\mu$ |

| Net | Space group | Sys. | Class | $\lambda_{PD}$ | $\lambda_{LE}$ | $\lambda_{\mu}^{num}$ | $f_{inv}$ | Window | Class, *r/l* = 0.05 |
|---|---|---|---|---|---|---|---|---|---|
| cai | P4/mbm | tet | $\mu$ | 5.49×10$^{-7}$ | 5.49×10$^{-7}$ | 5.01×10$^{-7}$ | 0.00 | – | $\mu$ |
| cal | I4/mcm | tet | $\mu$ | – | – | 6.76×10$^{-6}$ | – | – | $\mu$ |
| can | P6$_3$/mmc | hex | $\mu$ | – | – | 1.29×10$^{-7}$ | – | – | $\mu$ |
| cap | I4$_1$/amd | tet | PD+LE | 6.85×10$^{-7}$ | 6.85×10$^{-7}$ | – | 0.00 | – | PD+LE |
| cat | P6$_3$/mmc | hex | $\mu$ | 2.84×10$^{-7}$ | 2.84×10$^{-7}$ | 2.70×10$^{-7}$ | 0.00 | – | $\mu$ |
| cbs | Pmma | orth | PD+LE | 5.69×10$^{-7}$ | 5.69×10$^{-7}$ | – | 0.00 | – | PD+LE |
| cco | Cmcm | orth | PD-only | 1.71×10$^{-6}$ | – | 3.73×10$^{-6}$ | 1.00 | 117.7% | PD-only |
| ccu | P6/mmm | hex | $\mu$ | – | – | 5.35×10$^{-6}$ | – | – | $\mu$ |
| cda | I4$_1$22 | tet | $\mu$ | – | – | 9.26×10$^{-8}$ | – | – | $\mu$ |
| cdh | Pm$\bar{3}$ | cub | PD+LE | 1.94×10$^{-7}$ | 1.94×10$^{-7}$ | – | 0.00 | – | PD+LE |
| cdl | P4$_2$/mmc | tet | PD+LE | 4.28×10$^{-8}$ | 4.28×10$^{-8}$ | 4.93×10$^{-8}$ | 0.00 | – | PD+LE |
| cdm | P4$_2$/mmc | tet | PD+LE | 2.87×10$^{-8}$ | 2.87×10$^{-8}$ | 3.10×10$^{-8}$ | 0.00 | – | PD+LE |
| cdn | P4$_2$/mmc | tet | PD+LE | 2.15×10$^{-8}$ | 2.15×10$^{-8}$ | 2.71×10$^{-8}$ | 0.00 | – | PD+LE |
| cdp | P4$_2$/ncm | tet | $\mu$ | 1.94×10$^{-7}$ | 1.94×10$^{-7}$ | 1.78×10$^{-7}$ | 0.00 | – | $\mu$ |
| cds | P4$_2$/mmc | tet | PD+LE | 7.66×10$^{-8}$ | 7.66×10$^{-8}$ | – | 0.00 | – | PD+LE |
| cdt | Pnna | orth | PD+LE | 3.11×10$^{-8}$ | 3.11×10$^{-8}$ | 6.05×10$^{-8}$ | 1.00 | – | PD+LE |
| cdz | Pmmm | orth | PD+LE | 5.39×10$^{-8}$ | 5.39×10$^{-8}$ | – | 0.00 | – | PD+LE |
| cec | P6/mmm | hex | $\mu$ | 3.49×10$^{-7}$ | 3.49×10$^{-7}$ | 1.70×10$^{-7}$ | 0.00 | – | $\mu$ |
| cee | P6/mmm | hex | $\mu$ | 1.18×10$^{-7}$ | 1.18×10$^{-7}$ | 6.27×10$^{-8}$ | 0.00 | – | $\mu$ |
| ceg | P$\bar{3}$m1 | trig | PD-only | 1.07×10$^{-7}$ | 1.30×10$^{-7}$ | 1.12×10$^{-7}$ | 1.00 | 4.2% | PD-only |
| cei | P$\bar{6}$m2 | hex | $\mu$ | 8.87×10$^{-8}$ | 8.87×10$^{-8}$ | 5.24×10$^{-8}$ | 0.00 | – | $\mu$ |
| cel | P6/mmm | hex | $\mu$ | 8.65×10$^{-7}$ | – | 8.37×10$^{-7}$ | 1.00 | – | $\mu$ |
| cen | P6/mmm | hex | PD+LE | 6.68×10$^{-7}$ | 6.68×10$^{-7}$ | – | 0.00 | – | PD+LE |
| ceo | P6$_3$/mcm | hex | $\mu$ | 5.32×10$^{-7}$ | 5.32×10$^{-7}$ | 3.93×10$^{-7}$ | 0.00 | – | $\mu$ |
| cep | P$\bar{3}$1m | trig | PD-only | 1.83×10$^{-7}$ | 1.99×10$^{-7}$ | 1.99×10$^{-7}$ | 1.00 | 8.4% | PD-only |
| ceq | P$\bar{6}$2m | hex | PD-only | 1.36×10$^{-7}$ | 2.07×10$^{-7}$ | 1.82×10$^{-7}$ | 1.00 | 33.9% | PD-only |
| ces | P$\bar{6}$2m | hex | PD-only | 1.79×10$^{-7}$ | 2.73×10$^{-7}$ | – | 1.00 | 52.8% | PD-only |
| cfc | P6$_3$/mmc | hex | $\mu$ | – | – | 2.29×10$^{-7}$ | – | – | $\mu$ |
| cfd | P6$_3$/mmc | hex | $\mu$ | – | – | 2.31×10$^{-7}$ | – | – | $\mu$ |
| chb | Cmcm | orth | $\mu$ | – | – | 1.53×10$^{-6}$ | – | – | $\mu$ |
| clh | P4$_2$/mmc | tet | PD+LE | 8.00×10$^{-9}$ | 8.00×10$^{-9}$ | 8.63×10$^{-9}$ | 0.00 | – | PD+LE |
| cml | Cmc2$_1$ | orth | PD-only | 1.16×10$^{-7}$ | 1.40×10$^{-7}$ | – | 1.00 | 20.8% | PD-only |
| cot | Pn$\bar{3}$m | cub | PD+LE | 8.80×10$^{-7}$ | 8.80×10$^{-7}$ | – | 0.00 | – | PD+LE |
| crb | I4/mmm | tet | $\mu$ | – | – | 1.34×10$^{-7}$ | – | – | $\mu$ |
| crg | C2/m | mono | PD-only | 2.16×10$^{-7}$ | 2.17×10$^{-7}$ | – | 1.00 | 0.5% | PD-only |
| crr | P$\bar{4}$3m | cub | $\mu$ | 1.48×10$^{-7}$ | 1.48×10$^{-7}$ | 1.16×10$^{-7}$ | 0.00 | – | $\mu$ |
| csb | P6/mmm | hex | $\mu$ | – | – | 2.23×10$^{-7}$ | – | – | $\mu$ |
| csg | P6/mmm | hex | $\mu$ | – | – | 5.07×10$^{-7}$ | – | – | $\mu$ |
| csm | P6/mmm | hex | $\mu$ | – | – | 6.53×10$^{-7}$ | – | – | $\mu$ |
| csq | P6/mmm | hex | $\mu$ | – | – | 1.50×10$^{-7}$ | – | – | $\mu$ |
| css | P$\bar{4}$n2 | tet | $\mu$ | 9.53×10$^{-7}$ | 9.53×10$^{-7}$ | 7.94×10$^{-7}$ | 0.00 | – | $\mu$ |
| ctp | P6/mmm | hex | $\mu$ | 3.50×10$^{-7}$ | 3.50×10$^{-7}$ | 2.73×10$^{-7}$ | 0.00 | – | $\mu$ |
| cut | P4/mmm | tet | PD+LE | 4.20×10$^{-8}$ | 4.20×10$^{-8}$ | – | 0.00 | – | PD+LE |

| Net | Space group | Sys. | Class | $\lambda_{PD}$ | $\lambda_{LE}$ | $\lambda_{\mu}^{num}$ | $f_{inv}$ | Window | Class, *r/l* = 0.05 |
|---|---|---|---|---|---|---|---|---|---|
| cyc | $P4_2/mmc$ | tet | PD+LE | $2.24\times10^{-7}$ | $2.24\times10^{-7}$ | – | 0.00 | – | PD+LE |
| cyf | $Pm\bar{3}m$ | cub | PD+LE | $3.33\times10^{-6}$ | $3.33\times10^{-6}$ | – | 0.00 | – | PD+LE |
| cyg | P4/mmm | tet | PD+LE | $3.82\times10^{-7}$ | $3.82\times10^{-7}$ | – | 0.00 | – | PD+LE |
| cys | $P4_332$ | cub | $\mu$ | – | – | $1.50\times10^{-7}$ | – | – | $\mu$ |
| cyt | $Pm\bar{3}m$ | cub | PD+LE | $3.50\times10^{-7}$ | $3.50\times10^{-7}$ | – | 0.00 | – | PD+LE |
| czg | P4/mmm | tet | PD+LE | $1.41\times10^{-7}$ | $1.41\times10^{-7}$ | – | 0.00 | – | PD+LE |
| czh | P4/mmm | tet | PD+LE | $2.73\times10^{-7}$ | $2.73\times10^{-7}$ | – | 0.00 | – | PD+LE |
| dag | $P4_2/mnm$ | tet | PD-only | $8.81\times10^{-8}$ | $1.09\times10^{-7}$ | – | 1.00 | 24.1% | PD-only |
| daz | C222 | orth | $\mu$ | $2.75\times10^{-8}$ | $2.75\times10^{-8}$ | $2.23\times10^{-8}$ | 0.00 | – | $\mu$ |
| ddi | $I4_1/amd$ | tet | PD+LE | $1.29\times10^{-6}$ | $1.29\times10^{-6}$ | – | 0.00 | – | PD+LE |
| dea | $Pna2_1$ | orth | $\mu$ | $9.08\times10^{-8}$ | $9.25\times10^{-8}$ | $9.03\times10^{-8}$ | 1.00 | – | $\mu$ |
| dfs | P6/mmm | hex | PD+LE | $3.67\times10^{-7}$ | $3.67\times10^{-7}$ | – | 0.00 | – | PD+LE |
| dft | $P4_2/mmc$ | tet | $\mu$ | $1.07\times10^{-7}$ | $1.07\times10^{-7}$ | $7.99\times10^{-8}$ | 0.00 | – | $\mu$ |
| dgo | P432 | cub | $\mu$ | $3.17\times10^{-6}$ | $3.17\times10^{-6}$ | $2.69\times10^{-6}$ | 0.00 | – | $\mu$ |
| dgp | I4/mmm | tet | PD+LE | $1.54\times10^{-6}$ | $1.54\times10^{-6}$ | $2.22\times10^{-6}$ | 0.00 | – | PD+LE |
| dia | $Fd\bar{3}m$ | cub | $\mu$ | – | – | $2.39\times10^{-7}$ | – | – | $\mu$ |
| dib | $I4_1/amd$ | tet | $\mu$ | – | – | $6.32\times10^{-7}$ | – | – | $\mu$ |
| did | $P\bar{4}3m$ | cub | PD+LE | $3.23\times10^{-7}$ | $3.23\times10^{-7}$ | – | 0.00 | – | PD+LE |
| dif | $F\bar{4}3m$ | cub | $\mu$ | – | – | $6.59\times10^{-6}$ | – | – | $\mu$ |
| dig | $P\bar{4}3m$ | cub | PD+LE | $3.92\times10^{-7}$ | $3.92\times10^{-7}$ | $3.93\times10^{-7}$ | 0.00 | – | PD+LE |
| diq | $I\bar{4}m2$ | tet | PD+LE | $6.00\times10^{-7}$ | $6.00\times10^{-7}$ | – | 0.00 | – | PD+LE |
| dmc | $P2_1/c$ | mono | PD-only | $6.45\times10^{-8}$ | $6.48\times10^{-8}$ | – | 1.00 | 0.5% | PD-only |
| dmd | $P4_2/mmc$ | tet | PD+LE | $2.10\times10^{-8}$ | $2.10\times10^{-8}$ | – | 0.00 | – | PD+LE |
| dme | C2/m | mono | $\mu$ | $5.13\times10^{-8}$ | $5.16\times10^{-8}$ | $4.35\times10^{-8}$ | 1.00 | – | $\mu$ |
| dmf | C222 | orth | $\mu$ | $5.93\times10^{-8}$ | $5.96\times10^{-8}$ | $5.78\times10^{-8}$ | 1.00 | – | $\mu$ |
| dmg | Pmmm | orth | PD+LE | $3.13\times10^{-8}$ | $3.13\times10^{-8}$ | – | 0.00 | – | PD+LE |
| dmh | Pmna | orth | $\mu$ | $4.16\times10^{-8}$ | $4.17\times10^{-8}$ | $4.15\times10^{-8}$ | 1.00 | – | PD+$\mu$ |
| dmj | $P6_522$ | hex | $\mu$ | – | – | $4.90\times10^{-8}$ | – | – | $\mu$ |
| dmp | Pnna | orth | PD+LE | $2.35\times10^{-7}$ | $2.35\times10^{-7}$ | – | 0.00 | – | PD+LE |
| dnm | $P\bar{6}m2$ | hex | $\mu$ | $6.32\times10^{-7}$ | $6.32\times10^{-7}$ | $4.27\times10^{-7}$ | 0.00 | – | $\mu$ |
| dnq | $P\bar{4}m2$ | tet | $\mu$ | $2.36\times10^{-7}$ | $2.36\times10^{-7}$ | $2.09\times10^{-7}$ | 0.00 | – | $\mu$ |
| dum | P6/mmm | hex | $\mu$ | – | – | $4.53\times10^{-9}$ | – | – | $\mu$ |
| eca | $P6_3/mmc$ | hex | $\mu$ | $1.02\times10^{-6}$ | $1.02\times10^{-6}$ | $6.22\times10^{-7}$ | 0.00 | – | $\mu$ |
| ece | $P6_122$ | hex | $\mu$ | – | – | $2.09\times10^{-6}$ | – | – | $\mu$ |
| ecf | $P6_3/mmc$ | hex | $\mu$ | $1.06\times10^{-6}$ | $1.06\times10^{-6}$ | $9.54\times10^{-7}$ | 0.00 | – | $\mu$ |
| ecg | $P6_3/mmc$ | hex | PD+LE | $9.50\times10^{-7}$ | $9.50\times10^{-7}$ | – | 0.00 | – | PD+LE |
| eci | $P4_12_12$ | tet | $\mu$ | $1.35\times10^{-6}$ | $1.35\times10^{-6}$ | $1.34\times10^{-6}$ | 0.00 | – | $\mu$ |
| eck | I4/mcm | tet | $\mu$ | – | – | $2.60\times10^{-7}$ | – | – | $\mu$ |
| ecl | I4/mmm | tet | PD+LE | $8.53\times10^{-7}$ | $8.53\times10^{-7}$ | – | 0.00 | – | PD+LE |
| ecm | $I\bar{4}2d$ | tet | $\mu$ | – | – | $1.32\times10^{-6}$ | – | – | $\mu$ |
| ecn | R3m | trig | PD+LE | $2.03\times10^{-6}$ | $2.03\times10^{-6}$ | $2.47\times10^{-6}$ | 0.00 | – | PD+LE |
| ecp | $P6_3/m$ | hex | PD-only | $1.10\times10^{-6}$ | – | $2.39\times10^{-6}$ | 1.00 | 116.7% | PD-only |
| ecq | P6/mcc | hex | $\mu$ | $5.04\times10^{-7}$ | $5.04\times10^{-7}$ | $3.29\times10^{-7}$ | 0.00 | – | $\mu$ |

| Net | Space group | Sys. | Class | $\lambda_{PD}$ | $\lambda_{LE}$ | $\lambda_{\mu}^{num}$ | $f_{inv}$ | Window | Class, *r*/*l* = 0.05 |
|---|---|---|---|---|---|---|---|---|---|
| ecr | $P3_221$ | trig | $\mu$ | $1.93\times10^{-6}$ | $1.93\times10^{-6}$ | $1.91\times10^{-6}$ | 0.00 | – | $\mu$ |
| ecs | $P\bar{3}c1$ | trig | $\mu$ | – | – | $1.30\times10^{-6}$ | – | – | $\mu$ |
| ect | $P\bar{3}c1$ | trig | $\mu$ | $1.97\times10^{-6}$ | $1.97\times10^{-6}$ | $1.71\times10^{-6}$ | 0.00 | – | $\mu$ |
| ecu | Cmcm | orth | PD+LE | $1.06\times10^{-6}$ | $1.06\times10^{-6}$ | $1.11\times10^{-6}$ | 1.00 | – | PD+LE |
| ecz | $R\bar{3}m$ | trig | $\mu$ | – | – | $1.15\times10^{-6}$ | – | – | $\mu$ |
| edc | $P6_3/mmc$ | hex | PD+LE | $8.76\times10^{-7}$ | $8.76\times10^{-7}$ | – | 0.00 | – | PD+LE |
| edi | $P\bar{4}m2$ | tet | $\mu$ | $9.78\times10^{-8}$ | $9.78\times10^{-8}$ | $6.10\times10^{-8}$ | 0.00 | – | $\mu$ |
| eft | P4/mmm | tet | PD+LE | $1.02\times10^{-7}$ | $1.02\times10^{-7}$ | – | 0.00 | – | PD+LE |
| ela | $P4_2/mnm$ | tet | PD+LE | $2.73\times10^{-6}$ | $2.73\times10^{-6}$ | – | 0.00 | – | PD+LE |
| elb | Cmca | orth | $\mu$ | – | – | $1.87\times10^{-6}$ | – | – | $\mu$ |
| elc | C2/m | mono | $\mu$ | $2.05\times10^{-6}$ | $2.05\times10^{-6}$ | $1.87\times10^{-6}$ | 1.00 | – | $\mu$ |
| eld | $P3_121$ | trig | $\mu$ | – | – | $1.92\times10^{-6}$ | – | – | $\mu$ |
| elf | $P6_122$ | hex | $\mu$ | – | – | $1.92\times10^{-6}$ | – | – | $\mu$ |
| eoj | $I4_132$ | cub | $\mu$ | – | – | $1.92\times10^{-5}$ | – | – | $\mu$ |
| eok | $I4_132$ | cub | $\mu$ | – | – | $7.92\times10^{-6}$ | – | – | $\mu$ |
| epb | $P4_2/nnm$ | tet | $\mu$ | $2.42\times10^{-7}$ | $2.42\times10^{-7}$ | $2.14\times10^{-7}$ | 0.00 | – | $\mu$ |
| epx | $I\bar{4}m2$ | tet | $\mu$ | – | – | $4.76\times10^{-7}$ | – | – | $\mu$ |
| esg | C2/c | mono | PD+LE | $1.62\times10^{-7}$ | $1.63\times10^{-7}$ | – | 1.00 | – | PD+LE |
| est | $P4_2/n$ | tet | $\mu$ | $5.25\times10^{-7}$ | $5.25\times10^{-7}$ | $5.00\times10^{-7}$ | 0.00 | – | $\mu$ |
| eta | $P6_222$ | hex | $\mu$ | $3.93\times10^{-8}$ | $3.93\times10^{-8}$ | $2.03\times10^{-8}$ | 0.00 | – | $\mu$ |
| etc | $P6_3/mmc$ | hex | $\mu$ | $1.78\times10^{-8}$ | $1.78\times10^{-8}$ | $9.33\times10^{-9}$ | 0.00 | – | $\mu$ |
| etd | $P6_122$ | hex | $\mu$ | $3.51\times10^{-8}$ | $3.51\times10^{-8}$ | $2.22\times10^{-8}$ | 0.00 | – | $\mu$ |
| ete | $P6_422$ | hex | $\mu$ | – | – | $3.66\times10^{-8}$ | – | – | $\mu$ |
| eth | Pmna | orth | PD+LE | $2.02\times10^{-8}$ | $2.02\times10^{-8}$ | – | 0.00 | – | PD+LE |
| eye | P6/mmm | hex | $\mu$ | – | – | $4.91\times10^{-7}$ | – | – | $\mu$ |
| fcu | $Fm\bar{3}m$ | cub | $\mu$ | – | – | $5.66\times10^{-6}$ | – | – | $\mu$ |
| fdu | $P4_2/mmc$ | tet | PD+LE | $1.49\times10^{-7}$ | $1.49\times10^{-7}$ | – | 0.00 | – | PD+LE |
| feb | Pnma | orth | PD+LE | $2.03\times10^{-6}$ | $2.03\times10^{-6}$ | $3.68\times10^{-6}$ | 1.00 | – | PD+LE |
| fee | P4/mmm | tet | $\mu$ | $1.53\times10^{-7}$ | $1.53\times10^{-7}$ | $1.06\times10^{-7}$ | 0.00 | – | $\mu$ |
| fet | P4/mbm | tet | $\mu$ | – | – | $1.06\times10^{-7}$ | – | – | $\mu$ |
| ffa | Cmmm | orth | $\mu$ | – | – | $1.67\times10^{-7}$ | – | – | $\mu$ |
| ffb | P6/mmm | hex | $\mu$ | $6.36\times10^{-8}$ | $6.36\times10^{-8}$ | $3.31\times10^{-8}$ | 0.00 | – | $\mu$ |
| ffc | P4/mmm | tet | PD+LE | $1.23\times10^{-8}$ | $1.23\times10^{-8}$ | – | 0.00 | – | PD+LE |
| ffh | $P\bar{4}3m$ | cub | $\mu$ | – | – | $2.90\times10^{-6}$ | – | – | $\mu$ |
| fit | I4/mmm | tet | PD+LE | $1.03\times10^{-6}$ | $1.03\times10^{-6}$ | $1.03\times10^{-6}$ | 0.00 | – | PD+LE |
| fkm | $R\bar{3}m$ | trig | $\mu$ | – | – | $1.63\times10^{-6}$ | – | – | $\mu$ |
| flb | I4/mmm | tet | $\mu$ | – | – | $1.10\times10^{-6}$ | – | – | $\mu$ |
| fld | $Pm\bar{3}m$ | cub | $\mu$ | – | – | $9.66\times10^{-7}$ | – | – | $\mu$ |
| fll | $R\bar{3}m$ | trig | PD+LE | $1.45\times10^{-6}$ | $1.45\times10^{-6}$ | – | 0.00 | – | PD+LE |
| flp | $Fm\bar{3}m$ | cub | $\mu$ | – | – | $3.16\times10^{-6}$ | – | – | $\mu$ |
| flq | I4/mmm | tet | PD+LE | $8.27\times10^{-7}$ | $8.27\times10^{-7}$ | – | 0.00 | – | PD+LE |
| flu | $Fm\bar{3}m$ | cub | PD+LE | $5.93\times10^{-7}$ | $5.93\times10^{-7}$ | – | 0.00 | – | PD+LE |
| flw | $Pm\bar{3}$ | cub | $\mu$ | – | – | $1.95\times10^{-6}$ | – | – | $\mu$ |

| Net | Space group | Sys. | Class | $\lambda_{PD}$ | $\lambda_{LE}$ | $\lambda_{\mu}^{num}$ | $f_{inv}$ | Window | Class, *r*/*l* = 0.05 |
|---|---|---|---|---|---|---|---|---|---|
| fly | Imma | orth | PD+LE | $8.54\times10^{-8}$ | $8.54\times10^{-8}$ | – | 1.00 | – | PD+LE |
| fna | P6$_1$22 | hex | PD-only | $1.98\times10^{-7}$ | $2.03\times10^{-7}$ | $2.03\times10^{-7}$ | 1.00 | 2.7% | PD-only |
| fnb | Im$\bar{3}$ | cub | PD+LE | $2.41\times10^{-7}$ | $2.41\times10^{-7}$ | – | 0.00 | – | PD+LE |
| fne | P6$_2$22 | hex | $\mu$ | $4.35\times10^{-7}$ | $4.35\times10^{-7}$ | $2.73\times10^{-7}$ | 0.00 | – | $\mu$ |
| fni | I4$_1$22 | tet | $\mu$ | – | – | $1.10\times10^{-7}$ | – | – | $\mu$ |
| fnr | P6$_3$/mmc | hex | PD-only | $3.24\times10^{-7}$ | $3.31\times10^{-7}$ | – | 1.00 | 2.2% | PD-only |
| fns | P6$_2$22 | hex | $\mu$ | – | – | $4.16\times10^{-8}$ | – | – | $\mu$ |
| fnt | P6$_2$22 | hex | $\mu$ | $4.48\times10^{-7}$ | $4.48\times10^{-7}$ | $3.88\times10^{-7}$ | 0.00 | – | $\mu$ |
| fnu | R$\bar{3}$m | trig | $\mu$ | $6.83\times10^{-7}$ | $6.83\times10^{-7}$ | $5.97\times10^{-7}$ | 0.00 | – | $\mu$ |
| fnv | P6$_2$ | hex | $\mu$ | $2.30\times10^{-7}$ | $3.41\times10^{-7}$ | $1.53\times10^{-7}$ | 1.00 | – | $\mu$ |
| fnw | P6$_3$/mcm | hex | PD+LE | $2.25\times10^{-7}$ | $2.25\times10^{-7}$ | – | 0.00 | – | PD+LE |
| fnx | P4$_1$22 | tet | $\mu$ | $5.58\times10^{-7}$ | $5.61\times10^{-7}$ | $5.53\times10^{-7}$ | 1.00 | – | $\mu$ |
| fny | P6$_3$/mcm | hex | $\mu$ | $2.55\times10^{-7}$ | $2.55\times10^{-7}$ | $1.85\times10^{-7}$ | 0.00 | – | $\mu$ |
| fnz | P6/mcc | hex | PD+LE | $4.65\times10^{-7}$ | $4.65\times10^{-7}$ | – | 0.00 | – | PD+LE |
| foa | P6/mcc | hex | $\mu$ | – | – | $8.15\times10^{-8}$ | – | – | $\mu$ |
| fon | P4$_2$/nmc | tet | PD+LE | $6.97\times10^{-7}$ | $6.97\times10^{-7}$ | $8.50\times10^{-7}$ | 0.00 | – | PD+LE |
| fqr | P4/mmm | tet | $\mu$ | $1.18\times10^{-7}$ | $1.18\times10^{-7}$ | $6.33\times10^{-8}$ | 0.00 | – | $\mu$ |
| frl | Cmmm | orth | $\mu$ | – | – | $1.01\times10^{-7}$ | – | – | $\mu$ |
| fsb | P6/mmm | hex | PD+LE | $2.53\times10^{-7}$ | $2.53\times10^{-7}$ | – | 0.00 | – | PD+LE |
| fsc | P4/mmm | tet | PD+LE | $2.94\times10^{-7}$ | $2.94\times10^{-7}$ | – | 0.00 | – | PD+LE |
| fsd | P4/mmm | tet | PD+LE | $8.54\times10^{-8}$ | $8.54\times10^{-8}$ | – | 0.00 | – | PD+LE |
| fse | I$\bar{4}$m2 | tet | $\mu$ | $5.15\times10^{-7}$ | $5.15\times10^{-7}$ | $4.71\times10^{-7}$ | 0.00 | – | $\mu$ |
| fsf | Imm2 | orth | $\mu$ | – | – | $3.27\times10^{-7}$ | – | – | $\mu$ |
| fsg | Pmmm | orth | PD+LE | $2.25\times10^{-7}$ | $2.25\times10^{-7}$ | – | 0.00 | – | PD+LE |
| fsh | R$\bar{3}$m | trig | PD+LE | $3.14\times10^{-7}$ | $3.14\times10^{-7}$ | – | 0.00 | – | PD+LE |
| fsj | P4/mmm | tet | PD+LE | $1.02\times10^{-7}$ | $1.02\times10^{-7}$ | – | 0.00 | – | PD+LE |
| fsk | Pnma | orth | PD-only | $3.45\times10^{-7}$ | $3.48\times10^{-7}$ | – | 1.00 | 0.9% | PD-only |
| fsl | P2$_1$/c | mono | PD+LE | $4.24\times10^{-7}$ | $4.24\times10^{-7}$ | – | 0.00 | – | PD+LE |
| fsm | Pnma | orth | PD+LE | $4.00\times10^{-7}$ | $4.00\times10^{-7}$ | – | 0.00 | – | PD+LE |
| fsn | Pa$\bar{3}$ | cub | $\mu$ | – | – | $4.85\times10^{-7}$ | – | – | $\mu$ |
| fso | P2$_1$ | mono | PD-only | $2.29\times10^{-7}$ | $2.30\times10^{-7}$ | – | 1.00 | 0.4% | PD-only |
| fst | P6/mmm | hex | $\mu$ | $2.24\times10^{-7}$ | $2.24\times10^{-7}$ | $1.57\times10^{-7}$ | 0.00 | – | $\mu$ |
| fsx | I4/mmm | tet | PD+LE | $4.25\times10^{-7}$ | $4.25\times10^{-7}$ | – | 0.00 | – | PD+LE |
| fsy | Cmmm | orth | PD+LE | $1.22\times10^{-7}$ | $1.22\times10^{-7}$ | – | 0.00 | – | PD+LE |
| fta | Amm2 | orth | PD+LE | $1.92\times10^{-7}$ | $1.92\times10^{-7}$ | – | 0.00 | – | PD+LE |
| ftb | F222 | orth | PD+LE | $3.78\times10^{-7}$ | $3.78\times10^{-7}$ | – | 0.00 | – | PD+LE |
| ftw | Pm$\bar{3}$m | cub | $\mu$ | – | – | $1.16\times10^{-6}$ | – | – | $\mu$ |
| fuq | I4/mmm | tet | $\mu$ | – | – | $5.50\times10^{-6}$ | – | – | $\mu$ |
| fwk | R$\bar{3}$m | trig | $\mu$ | $3.24\times10^{-6}$ | $3.24\times10^{-6}$ | $2.29\times10^{-6}$ | 0.00 | – | $\mu$ |
| fwq | P4/mmm | tet | $\mu$ | – | – | $1.49\times10^{-6}$ | – | – | $\mu$ |
| fws | P4/mmm | tet | $\mu$ | $3.01\times10^{-6}$ | $3.01\times10^{-6}$ | $1.98\times10^{-6}$ | 0.00 | – | $\mu$ |
| fwt | Pm$\bar{3}$m | cub | $\mu$ | $2.17\times10^{-6}$ | $2.17\times10^{-6}$ | $1.79\times10^{-6}$ | 0.00 | – | $\mu$ |
| fwv | P$\bar{4}$3m | cub | $\mu$ | $1.97\times10^{-6}$ | $1.97\times10^{-6}$ | $1.48\times10^{-6}$ | 0.00 | – | $\mu$ |

| Net | Space group | Sys. | Class | $\lambda_{PD}$ | $\lambda_{LE}$ | $\lambda_{\mu}^{num}$ | $f_{inv}$ | Window | Class, $r/l$ = 0.05 |
|---|---|---|---|---|---|---|---|---|---|
| gaf | Cmmm | orth | $\mu$ | – | – | $1.00\times10^{-7}$ | – | – | $\mu$ |
| gfc | $P6_3/mmc$ | hex | $\mu$ | $1.91\times10^{-7}$ | $1.91\times10^{-7}$ | $1.87\times10^{-7}$ | 0.00 | – | $\mu$ |
| gfy | $P6_3/mmc$ | hex | $\mu$ | $1.53\times10^{-7}$ | $1.53\times10^{-7}$ | $8.03\times10^{-8}$ | 0.00 | – | $\mu$ |
| gpu | Fddd | orth | PD+LE | $1.71\times10^{-6}$ | $1.71\times10^{-6}$ | $1.71\times10^{-6}$ | 0.00 | – | PD+LE |
| gra | $P6_3/mmc$ | hex | PD+LE | $1.89\times10^{-7}$ | $1.89\times10^{-7}$ | – | 0.00 | – | PD+LE |
| gwg | $P2_1/n$ | mono | PD+LE | $3.74\times10^{-8}$ | $3.74\times10^{-8}$ | – | 1.00 | – | PD+LE |
| hbm | P4/mmm | tet | PD+LE | $5.44\times10^{-8}$ | $5.44\times10^{-8}$ | – | 0.00 | – | PD+LE |
| hbr | P6/mmm | hex | $\mu$ | – | – | $1.27\times10^{-6}$ | – | – | $\mu$ |
| hcd | $P3_112$ | trig | $\mu$ | $7.92\times10^{-7}$ | $7.92\times10^{-7}$ | $6.52\times10^{-7}$ | 0.00 | – | $\mu$ |
| hcj | C2/m | mono | $\mu$ | – | – | $5.30\times10^{-7}$ | – | – | $\mu$ |
| hcp | $P6_3/mmc$ | hex | $\mu$ | – | – | $5.66\times10^{-6}$ | – | – | $\mu$ |
| hex | P6/mmm | hex | PD+LE | $9.44\times10^{-7}$ | $9.44\times10^{-7}$ | – | 0.00 | – | PD+LE |
| hjm | C2/c | mono | $\mu$ | $8.79\times10^{-7}$ | $8.79\times10^{-7}$ | $7.91\times10^{-7}$ | 0.00 | – | $\mu$ |
| hms | $P\bar{6}m2$ | hex | PD+LE | $1.89\times10^{-7}$ | $1.89\times10^{-7}$ | – | 0.00 | – | PD+LE |
| hof | I4/mmm | tet | PD+LE | $2.40\times10^{-8}$ | $2.40\times10^{-8}$ | – | 0.00 | – | PD+LE |
| hst | $P6_422$ | hex | $\mu$ | $5.38\times10^{-8}$ | $5.38\times10^{-8}$ | $4.82\times10^{-8}$ | 0.00 | – | $\mu$ |
| htp | Imma | orth | PD-only | $2.72\times10^{-7}$ | $2.91\times10^{-7}$ | – | 1.00 | 7.0% | PD-only |
| hww | $P4_2/mnm$ | tet | PD+LE | $2.00\times10^{-7}$ | $2.00\times10^{-7}$ | $3.72\times10^{-7}$ | 0.00 | – | PD+LE |
| hwx | $P4_2/mnm$ | tet | PD+LE | $1.02\times10^{-7}$ | $1.02\times10^{-7}$ | – | 0.00 | – | PD+LE |
| hxc | $P\bar{4}3m$ | cub | PD+LE | $9.43\times10^{-7}$ | $9.43\times10^{-7}$ | – | 0.00 | – | PD+LE |
| hxg | $Pn\bar{3}m$ | cub | $\mu$ | – | – | $1.11\times10^{-6}$ | – | – | $\mu$ |
| hxp | $Pn\bar{3}m$ | cub | $\mu$ | – | – | $3.02\times10^{-6}$ | – | – | $\mu$ |
| hyb | $P6_3/mmc$ | hex | $\mu$ | $6.38\times10^{-8}$ | $6.38\times10^{-8}$ | $6.23\times10^{-8}$ | 0.00 | – | $\mu$ |
| ilc | $R\bar{3}m$ | trig | PD+LE | $1.31\times10^{-6}$ | $1.31\times10^{-6}$ | – | 0.00 | – | PD+LE |
| ild | $R\bar{3}m$ | trig | $\mu$ | – | – | $5.42\times10^{-6}$ | – | – | $\mu$ |
| ile | Fmmm | orth | PD+LE | $1.46\times10^{-6}$ | $1.46\times10^{-6}$ | – | 0.00 | – | PD+LE |
| ins | Pnnm | orth | $\mu$ | $6.49\times10^{-8}$ | $6.49\times10^{-8}$ | $5.88\times10^{-8}$ | 1.00 | – | $\mu$ |
| irl | Cccm | orth | $\mu$ | – | – | $1.01\times10^{-7}$ | – | – | $\mu$ |
| isq | $P4_2/mnm$ | tet | $\mu$ | $1.88\times10^{-7}$ | $1.88\times10^{-7}$ | $1.83\times10^{-7}$ | 0.00 | – | $\mu$ |
| itb | $Pm\bar{3}n$ | cub | $\mu$ | $5.40\times10^{-6}$ | $5.40\times10^{-6}$ | $3.50\times10^{-6}$ | 0.00 | – | $\mu$ |
| ith | $Pm\bar{3}n$ | cub | PD+LE | $1.12\times10^{-6}$ | $1.12\times10^{-6}$ | – | 0.00 | – | PD+LE |
| itv | $I\bar{4}2d$ | tet | $\mu$ | – | – | $1.98\times10^{-7}$ | – | – | $\mu$ |
| itw | $Pm\bar{3}n$ | cub | PD+LE | $1.42\times10^{-6}$ | $1.42\times10^{-6}$ | – | 0.00 | – | PD+LE |
| jbw | Pmma | orth | PD+LE | $1.24\times10^{-7}$ | $1.24\times10^{-7}$ | – | 0.00 | – | PD+LE |
| jea | Imm2 | orth | $\mu$ | – | – | $5.87\times10^{-8}$ | – | – | $\mu$ |
| jeb | Cmmm | orth | $\mu$ | – | – | $5.24\times10^{-8}$ | – | – | $\mu$ |
| jem | Cmmm | orth | PD+LE | $1.10\times10^{-7}$ | $1.10\times10^{-7}$ | – | 0.00 | – | PD+LE |
| jhw | P4/mnc | tet | PD-only | $1.68\times10^{-7}$ | $2.12\times10^{-7}$ | – | 1.00 | 26.3% | PD-only |
| jjt | $P\bar{6}2c$ | hex | PD-only | $1.12\times10^{-7}$ | $1.55\times10^{-7}$ | – | 1.00 | 38.4% | PD-only |
| jkz | Fmmm | orth | $\mu$ | – | – | $1.65\times10^{-6}$ | – | – | $\mu$ |
| jsa | $P\bar{4}3m$ | cub | $\mu$ | – | – | $8.59\times10^{-8}$ | – | – | $\mu$ |
| jsm | $I4_122$ | tet | PD+LE | $4.95\times10^{-7}$ | $4.95\times10^{-7}$ | $1.26\times10^{-6}$ | 0.00 | – | PD+LE |
| jus | Cmcm | orth | PD-only | $2.78\times10^{-7}$ | $3.02\times10^{-7}$ | – | 1.00 | 8.6% | PD-only |

| Net | Space group | Sys. | Class | $\lambda_{PD}$ | $\lambda_{LE}$ | $\lambda_{\mu}^{num}$ | $f_{inv}$ | Window | Class, *r/l* = 0.05 |
|---|---|---|---|---|---|---|---|---|---|
| jxj | $P4_2/ncm$ | tet | PD+LE | $7.14\times10^{-8}$ | $7.14\times10^{-8}$ | – | 0.00 | – | PD+LE |
| kag | P6/mmm | hex | $\mu$ | $5.47\times10^{-7}$ | $5.47\times10^{-7}$ | $5.33\times10^{-7}$ | 0.00 | – | $\mu$ |
| kdd | $Cmc2_1$ | orth | PD+LE | $2.60\times10^{-7}$ | $2.60\times10^{-7}$ | – | 0.00 | – | PD+LE |
| kea | $P4_32_12$ | tet | PD+LE | $1.79\times10^{-7}$ | $1.79\times10^{-7}$ | $1.79\times10^{-7}$ | 0.00 | – | PD+LE |
| ken | $P6_3/mmc$ | hex | $\mu$ | $6.67\times10^{-7}$ | $6.67\times10^{-7}$ | $4.22\times10^{-7}$ | 0.00 | – | $\mu$ |
| kew | P6/mmm | hex | PD+LE | $1.53\times10^{-7}$ | $1.53\times10^{-7}$ | – | 0.00 | – | PD+LE |
| key | P622 | hex | $\mu$ | – | – | $2.87\times10^{-7}$ | – | – | $\mu$ |
| kkm | Pbam | orth | PD+LE | $2.50\times10^{-7}$ | $2.50\times10^{-7}$ | – | 0.00 | – | PD+LE |
| knb | P6/mmm | hex | PD-only | $7.35\times10^{-7}$ | $1.01\times10^{-6}$ | $8.87\times10^{-7}$ | 1.00 | 20.7% | PD-only |
| knc | $P4_2/mmc$ | tet | PD-only | $5.49\times10^{-7}$ | $6.65\times10^{-7}$ | $5.89\times10^{-7}$ | 1.00 | 7.3% | PD-only |
| kty | P4/mmm | tet | $\mu$ | $7.93\times10^{-7}$ | $7.93\times10^{-7}$ | $7.24\times10^{-7}$ | 0.00 | – | $\mu$ |
| kwh | $R\bar{3}m$ | trig | PD+LE | $9.98\times10^{-7}$ | $9.98\times10^{-7}$ | – | 0.00 | – | PD+LE |
| laf | $P4_2/mnm$ | tet | PD+LE | $7.12\times10^{-8}$ | $7.12\times10^{-8}$ | $2.27\times10^{-7}$ | 0.00 | – | PD+LE |
| lcq | $P4_12_12$ | tet | $\mu$ | – | – | $2.90\times10^{-6}$ | – | – | PD+LE |
| lcu | $P4_332$ | cub | $\mu$ | $1.40\times10^{-5}$ | $1.40\times10^{-5}$ | $1.18\times10^{-5}$ | 0.00 | – | $\mu$ |
| lcv | $I4_132$ | cub | $\mu$ | – | – | $3.49\times10^{-8}$ | – | – | $\mu$ |
| lcx | $Pm\bar{3}n$ | cub | PD+LE | $1.62\times10^{-6}$ | $1.62\times10^{-6}$ | – | 0.00 | – | PD+LE |
| lcy | $P4_132$ | cub | $\mu$ | – | – | $6.48\times10^{-7}$ | – | – | $\mu$ |
| lfm | Imma | orth | PD+LE | $6.18\times10^{-8}$ | $6.18\times10^{-8}$ | – | 0.00 | – | PD+LE |
| lhb | Fdd2 | orth | $\mu$ | $1.57\times10^{-6}$ | $1.57\times10^{-6}$ | $1.52\times10^{-6}$ | 0.00 | – | $\mu$ |
| lhh | $P6_122$ | hex | $\mu$ | $1.81\times10^{-7}$ | $1.81\times10^{-7}$ | $1.24\times10^{-7}$ | 0.00 | – | $\mu$ |
| lib | P4/mmm | tet | $\mu$ | $4.42\times10^{-7}$ | $4.42\times10^{-7}$ | $2.13\times10^{-7}$ | 0.00 | – | $\mu$ |
| lil | Imma | orth | PD-only | $4.48\times10^{-8}$ | $5.01\times10^{-8}$ | – | 1.00 | 11.9% | PD-only |
| lim | Imma | orth | PD-only | $1.93\times10^{-8}$ | $2.05\times10^{-8}$ | – | 1.00 | 6.1% | PD-only |
| llj | I4/mmm | tet | PD+LE | $2.04\times10^{-7}$ | $2.04\times10^{-7}$ | – | 0.00 | – | PD+LE |
| loh | Cmca | orth | $\mu$ | – | – | $5.96\times10^{-7}$ | – | – | $\mu$ |
| lok | $I4_132$ | cub | $\mu$ | – | – | $9.41\times10^{-7}$ | – | – | $\mu$ |
| lom | Cmca | orth | PD-only | $9.69\times10^{-7}$ | $9.75\times10^{-7}$ | – | 1.00 | 0.5% | PD-only |
| lon | $P6_3/mmc$ | hex | $\mu$ | – | – | $2.05\times10^{-7}$ | – | – | $\mu$ |
| lsz | $P4_2/nnm$ | tet | $\mu$ | $1.48\times10^{-6}$ | $1.48\times10^{-6}$ | $1.31\times10^{-6}$ | 0.00 | – | $\mu$ |
| lvt | $I4_1/amd$ | tet | PD-only | $6.91\times10^{-8}$ | $8.85\times10^{-8}$ | – | 1.00 | 28.1% | PD-only |
| lys | $P4_332$ | cub | $\mu$ | – | – | $6.56\times10^{-7}$ | – | – | $\mu$ |
| mab | Pmna | orth | PD-only | $5.00\times10^{-7}$ | $5.34\times10^{-7}$ | – | 1.00 | 6.7% | PD-only |
| may | P6/mmm | hex | PD+LE | $3.69\times10^{-7}$ | $3.69\times10^{-7}$ | – | 0.00 | – | PD+LE |
| mbc | Cmcm | orth | $\mu$ | – | – | $1.64\times10^{-6}$ | – | – | $\mu$ |
| mcd | Cmmm | orth | PD+LE | $9.39\times10^{-8}$ | $9.39\times10^{-8}$ | – | 0.00 | – | PD+LE |
| mcf | Cmmm | orth | PD+LE | $1.77\times10^{-7}$ | $1.77\times10^{-7}$ | – | 0.00 | – | PD+LE |
| mcn | P4/mbm | tet | PD+LE | $1.62\times10^{-7}$ | $1.62\times10^{-7}$ | – | 0.00 | – | PD+LE |
| mco | $P\bar{6}2m$ | hex | $\mu$ | $8.95\times10^{-8}$ | $8.95\times10^{-8}$ | $8.81\times10^{-8}$ | 0.00 | – | $\mu$ |
| mdf | $I4_122$ | tet | $\mu$ | – | – | $4.05\times10^{-7}$ | – | – | $\mu$ |
| mgq | $I4_1/amd$ | tet | $\mu$ | $7.83\times10^{-6}$ | $7.83\times10^{-6}$ | $5.96\times10^{-6}$ | 0.00 | – | $\mu$ |
| mgz | $P6_3/mmc$ | hex | PD-only | $2.90\times10^{-6}$ | $3.46\times10^{-6}$ | – | 1.00 | 19.2% | PD-only |
| mhg | P4/mbm | tet | PD+LE | $6.21\times10^{-7}$ | $6.21\times10^{-7}$ | – | 0.00 | – | PD+LE |

| Net | Space group | Sys. | Class | $\lambda_{PD}$ | $\lambda_{LE}$ | $\lambda_{\mu}^{num}$ | $f_{inv}$ | Window | Class, *r/l* = 0.05 |
|---|---|---|---|---|---|---|---|---|---|
| mjb | P4/mmm | tet | *μ* | 1.85×10⁻⁸ | 1.85×10⁻⁸ | 1.75×10⁻⁸ | 0.00 | – | *μ* |
| mjd | $P6_422$ | hex | *μ* | – | – | 5.39×10⁻⁸ | – | – | *μ* |
| mjh | Fmmm | orth | *μ* | – | – | 1.53×10⁻⁶ | – | – | *μ* |
| mmg | Imma | orth | PD+LE | 4.09×10⁻⁷ | 4.09×10⁻⁷ | – | 0.00 | – | PD+LE |
| mmo | R32 | trig | *μ* | – | – | 1.29×10⁻⁶ | – | – | *μ* |
| mmt | Pccn | orth | *μ* | 2.41×10⁻⁷ | 2.41×10⁻⁷ | 2.10×10⁻⁷ | 0.00 | – | *μ* |
| moa | Cmmm | orth | *μ* | – | – | 3.43×10⁻⁸ | – | – | *μ* |
| mob | $I4_1$/amd | tet | *μ* | – | – | 1.63×10⁻⁶ | – | – | *μ* |
| moc | Cmmm | orth | PD+LE | 5.99×10⁻⁸ | 5.99×10⁻⁸ | – | 0.00 | – | PD+LE |
| mog | Cmmm | orth | *μ* | – | – | 1.20×10⁻⁷ | – | – | *μ* |
| mok | Cccm | orth | *μ* | 3.28×10⁻⁷ | 3.28×10⁻⁷ | 2.53×10⁻⁷ | 0.00 | – | *μ* |
| mot | P4/mmm | tet | PD+LE | 8.54×10⁻⁸ | 8.54×10⁻⁸ | – | 0.00 | – | PD+LE |
| mou | P4/mmm | tet | PD+LE | 2.53×10⁻⁸ | 2.53×10⁻⁸ | – | 0.00 | – | PD+LE |
| mrc | Pnnm | orth | PD+LE | 3.97×10⁻⁷ | 3.97×10⁻⁷ | – | 0.00 | – | PD+LE |
| msp | C2/c | mono | PD-only | 1.36×10⁻⁷ | 1.39×10⁻⁷ | – | 1.00 | 2.8% | PD-only |
| msq | P4/mmm | tet | *μ* | 8.44×10⁻⁷ | 8.44×10⁻⁷ | 3.69×10⁻⁷ | 0.00 | – | *μ* |
| msw | $P4_2$/nnm | tet | PD+LE | 6.84×10⁻⁷ | 6.84×10⁻⁷ | – | 0.00 | – | PD+LE |
| mvy | Pnnm | orth | PD+LE | 7.26×10⁻⁸ | 7.26×10⁻⁸ | – | 0.00 | – | PD+LE |
| mys | Pmna | orth | PD+LE | 3.86×10⁻⁸ | 3.86×10⁻⁸ | – | 0.00 | – | PD+LE |
| nab | $I\bar{4}m2$ | tet | *μ* | – | – | 6.98×10⁻⁸ | – | – | *μ* |
| nbo | $Im\bar{3}m$ | cub | PD+LE | 1.61×10⁻⁷ | 1.61×10⁻⁷ | – | 0.00 | – | PD+LE |
| ncb | $I\bar{4}3m$ | cub | PD+LE | 3.15×10⁻⁶ | 3.15×10⁻⁶ | 3.15×10⁻⁶ | 0.00 | – | PD+LE |
| ncd | $P4_2$/mnm | tet | PD-only | 1.41×10⁻⁶ | 2.24×10⁻⁶ | – | 1.00 | 58.8% | PD-only |
| nce | I4/mmm | tet | *μ* | – | – | 9.99×10⁻⁷ | – | – | *μ* |
| ncf | I4/m | tet | *μ* | – | – | 2.94×10⁻⁶ | – | – | *μ* |
| ncg | $P6_222$ | hex | *μ* | 1.10×10⁻⁶ | 1.10×10⁻⁶ | 1.08×10⁻⁶ | 0.00 | – | *μ* |
| nch | $P\bar{1}$ | tric | PD+LE | 1.25×10⁻⁶ | 1.25×10⁻⁶ | – | 1.00 | – | PD+LE |
| nci | Fmmm | orth | PD+LE | 1.05×10⁻⁶ | 1.05×10⁻⁶ | – | 0.00 | – | PD+LE |
| ncj | C2/m | mono | PD+LE | 1.28×10⁻⁶ | 1.28×10⁻⁶ | – | 1.00 | – | PD+LE |
| nia | $P6_3$/mmc | hex | PD-only | 6.84×10⁻⁷ | 7.01×10⁻⁷ | – | 1.00 | 2.4% | PD-only |
| nic | $P6_3$/mmc | hex | *μ* | – | – | 2.01×10⁻⁶ | – | – | *μ* |
| nid | $P6_3$/mmc | hex | *μ* | – | – | 1.24×10⁻⁶ | – | – | *μ* |
| nih | $P6_3$/mmc | hex | *μ* | 4.81×10⁻⁶ | 4.81×10⁻⁶ | 2.48×10⁻⁶ | 0.00 | – | *μ* |
| nik | $P6_3$/mmc | hex | *μ* | – | – | 1.42×10⁻⁶ | – | – | *μ* |
| nin | P6/mmm | hex | *μ* | – | – | 5.36×10⁻⁶ | – | – | *μ* |
| nix | $P6_3$/mmc | hex | *μ* | 4.81×10⁻⁶ | 4.81×10⁻⁶ | 2.75×10⁻⁶ | 0.00 | – | *μ* |
| nkc | $P4_2$/mmc | tet | PD+LE | 1.23×10⁻⁸ | 1.23×10⁻⁸ | – | 0.00 | – | PD+LE |
| nkd | $P4_2$/mmc | tet | PD+LE | 1.94×10⁻⁸ | 1.94×10⁻⁸ | – | 0.00 | – | PD+LE |
| nkn | Pmmm | orth | PD+LE | 1.46×10⁻⁸ | 1.46×10⁻⁸ | – | 0.00 | – | PD+LE |
| nko | Pmmm | orth | PD+LE | 1.80×10⁻⁸ | 1.80×10⁻⁸ | – | 0.00 | – | PD+LE |
| nkp | Pmmm | orth | PD+LE | 1.13×10⁻⁸ | 1.13×10⁻⁸ | – | 0.00 | – | PD+LE |
| nkq | Pmmm | orth | PD+LE | 2.35×10⁻⁸ | 2.35×10⁻⁸ | – | 0.00 | – | PD+LE |
| nob | Pmna | orth | PD-only | 1.88×10⁻⁸ | 1.90×10⁻⁸ | – | 1.00 | 1.0% | PD-only |

| Net | Space group | Sys. | Class | $\lambda_{PD}$ | $\lambda_{LE}$ | $\lambda_{\mu}^{num}$ | $f_{inv}$ | Window | Class, *r/l* = 0.05 |
|---|---|---|---|---|---|---|---|---|---|
| noj | $P4_322$ | tet | PD+LE | $1.25\times10^{-8}$ | $1.25\times10^{-8}$ | – | 0.00 | – | PD+LE |
| nom | $P6_122$ | hex | $\mu$ | $9.56\times10^{-8}$ | $9.56\times10^{-8}$ | $6.14\times10^{-8}$ | 0.00 | – | $\mu$ |
| noo | P4/mnc | tet | $\mu$ | – | – | $4.48\times10^{-7}$ | – | – | $\mu$ |
| noq | C2/c | mono | $\mu$ | $1.67\times10^{-7}$ | $1.68\times10^{-7}$ | $1.62\times10^{-7}$ | 1.00 | – | $\mu$ |
| nor | C2/m | mono | PD+LE | $1.61\times10^{-7}$ | $1.61\times10^{-7}$ | – | 0.00 | – | PD+LE |
| not | $P2_1/c$ | mono | PD-only | $1.18\times10^{-7}$ | $1.24\times10^{-7}$ | – | 1.00 | 5.1% | PD-only |
| nou | P4/mbm | tet | PD+LE | $1.47\times10^{-7}$ | $1.47\times10^{-7}$ | – | 0.00 | – | PD+LE |
| nov | Cmca | orth | PD+LE | $4.38\times10^{-7}$ | $4.38\times10^{-7}$ | – | 0.00 | – | PD+LE |
| noy | Cmcm | orth | $\mu$ | – | – | $1.91\times10^{-7}$ | – | – | $\mu$ |
| noz | Cmca | orth | $\mu$ | – | – | $3.91\times10^{-7}$ | – | – | $\mu$ |
| nsp | $P6_122$ | hex | $\mu$ | – | – | $2.41\times10^{-6}$ | – | – | $\mu$ |
| nuc | $P4_32_12$ | tet | PD+LE | $2.78\times10^{-7}$ | $2.78\times10^{-7}$ | $2.89\times10^{-7}$ | 0.00 | – | PD+LE |
| oab | Cmmm | orth | PD+LE | $2.42\times10^{-7}$ | $2.42\times10^{-7}$ | – | 0.00 | – | PD+LE |
| och | P6/mmm | hex | PD+LE | $2.94\times10^{-7}$ | $2.94\times10^{-7}$ | – | 0.00 | – | PD+LE |
| ooa | Pnna | orth | PD+LE | $1.09\times10^{-7}$ | $1.09\times10^{-7}$ | $1.09\times10^{-7}$ | 0.00 | – | PD+LE |
| oob | C2/m | mono | PD+LE | $8.24\times10^{-7}$ | $8.24\times10^{-7}$ | – | 0.00 | – | PD+LE |
| ooc | C2/m | mono | PD-only | $9.33\times10^{-7}$ | $1.00\times10^{-6}$ | $9.56\times10^{-7}$ | 1.00 | 2.4% | PD-only |
| oof | C2/c | mono | PD-only | $1.15\times10^{-6}$ | $1.19\times10^{-6}$ | – | 1.00 | 3.8% | PD-only |
| ook | $I4_132$ | cub | $\mu$ | – | – | $1.02\times10^{-6}$ | – | – | $\mu$ |
| osa | Cmcm | orth | $\mu$ | – | – | $3.45\times10^{-7}$ | – | – | $\mu$ |
| osb | Cmcm | orth | $\mu$ | – | – | $9.56\times10^{-7}$ | – | – | $\mu$ |
| osc | Pnma | orth | $\mu$ | – | – | $4.03\times10^{-7}$ | – | – | $\mu$ |
| osd | Pnma | orth | $\mu$ | – | – | $9.80\times10^{-7}$ | – | – | $\mu$ |
| ose | Immm | orth | $\mu$ | – | – | $4.60\times10^{-7}$ | – | – | $\mu$ |
| osf | Cmcm | orth | $\mu$ | – | – | $5.49\times10^{-7}$ | – | – | $\mu$ |
| oso | $P6_222$ | hex | $\mu$ | – | – | $2.24\times10^{-8}$ | – | – | $\mu$ |
| pca | $P\bar{4}3m$ | cub | $\mu$ | – | – | $8.03\times10^{-7}$ | – | – | $\mu$ |
| pcc | R3m | trig | $\mu$ | $1.90\times10^{-6}$ | $1.90\times10^{-6}$ | $1.38\times10^{-6}$ | 0.00 | – | $\mu$ |
| pce | $P\bar{4}3m$ | cub | $\mu$ | $2.00\times10^{-6}$ | $2.00\times10^{-6}$ | $1.93\times10^{-6}$ | 0.00 | – | $\mu$ |
| pch | $R\bar{3}m$ | trig | $\mu$ | $4.81\times10^{-6}$ | $4.81\times10^{-6}$ | $2.83\times10^{-6}$ | 0.00 | – | $\mu$ |
| pck | $R\bar{3}m$ | trig | $\mu$ | – | – | $1.27\times10^{-6}$ | – | – | $\mu$ |
| pco | $P\bar{4}3m$ | cub | $\mu$ | $2.67\times10^{-7}$ | $2.67\times10^{-7}$ | $1.66\times10^{-7}$ | 0.00 | – | $\mu$ |
| pcs | P4/mmm | tet | PD+LE | $6.81\times10^{-7}$ | $6.81\times10^{-7}$ | – | 0.00 | – | PD+LE |
| pcu | $Pm\bar{3}m$ | cub | PD+LE | $6.84\times10^{-7}$ | $6.84\times10^{-7}$ | – | 0.00 | – | PD+LE |
| pfm | I4/mmm | tet | PD+LE | $1.14\times10^{-6}$ | $1.14\times10^{-6}$ | – | 0.00 | – | PD+LE |
| pft | $P4_232$ | cub | $\mu$ | $1.34\times10^{-5}$ | $1.34\times10^{-5}$ | $6.28\times10^{-6}$ | 0.00 | – | $\mu$ |
| pha | $P\bar{4}3m$ | cub | $\mu$ | $1.84\times10^{-7}$ | $1.84\times10^{-7}$ | $9.89\times10^{-8}$ | 0.00 | – | $\mu$ |
| phq | $P6_222$ | hex | $\mu$ | – | – | $4.90\times10^{-7}$ | – | – | $\mu$ |
| pht | $P6_222$ | hex | PD-only | $6.84\times10^{-7}$ | $7.39\times10^{-7}$ | – | 1.00 | 8.0% | PD-only |
| phw | $P6_222$ | hex | $\mu$ | – | – | $7.93\times10^{-8}$ | – | – | $\mu$ |
| phx | $P6_222$ | hex | $\mu$ | $2.73\times10^{-8}$ | $2.73\times10^{-8}$ | $2.01\times10^{-8}$ | 0.00 | – | $\mu$ |
| pll | $P4_122$ | tet | PD+LE | $2.55\times10^{-7}$ | $2.55\times10^{-7}$ | – | 0.00 | – | PD+LE |
| pok | $I4_132$ | cub | $\mu$ | – | – | $2.02\times10^{-6}$ | – | – | $\mu$ |

| Net | Space group | Sys. | Class | $\lambda_{PD}$ | $\lambda_{LE}$ | $\lambda_{\mu}^{num}$ | $f_{inv}$ | Window | Class, $r/l$ = 0.05 |
|---|---|---|---|---|---|---|---|---|---|
| psb | $P4_2/mmc$ | tet | $\mu$ | $1.31\times10^{-6}$ | $1.31\times10^{-6}$ | $1.19\times10^{-6}$ | 0.00 | – | $\mu$ |
| psp | $P4_2/mmc$ | tet | PD+LE | $2.71\times10^{-7}$ | $2.71\times10^{-7}$ | – | 0.00 | – | PD+LE |
| pss | $P4_2/mmc$ | tet | PD+LE | $5.80\times10^{-7}$ | $5.80\times10^{-7}$ | – | 0.00 | – | PD+LE |
| pst | $P4_2/mmc$ | tet | PD+LE | $5.07\times10^{-7}$ | $5.07\times10^{-7}$ | – | 0.00 | – | PD+LE |
| pte | Cmmm | orth | PD+LE | $3.76\times10^{-7}$ | $3.76\times10^{-7}$ | – | 0.00 | – | PD+LE |
| ptg | $P6_222$ | hex | $\mu$ | $3.34\times10^{-7}$ | $3.34\times10^{-7}$ | $3.28\times10^{-7}$ | 0.00 | – | $\mu$ |
| pth | $P6_222$ | hex | $\mu$ | $1.45\times10^{-7}$ | $1.45\times10^{-7}$ | $1.36\times10^{-7}$ | 0.00 | – | $\mu$ |
| ptk | $P6_222$ | hex | $\mu$ | $2.38\times10^{-7}$ | $2.38\times10^{-7}$ | $2.30\times10^{-7}$ | 0.00 | – | $\mu$ |
| ptl | $P6_222$ | hex | $\mu$ | $9.51\times10^{-7}$ | $9.51\times10^{-7}$ | $9.24\times10^{-7}$ | 0.00 | – | $\mu$ |
| pts | $P4_2/mmc$ | tet | PD+LE | $9.44\times10^{-8}$ | $9.44\times10^{-8}$ | – | 0.00 | – | PD+LE |
| ptx | $P6_222$ | hex | $\mu$ | $5.95\times10^{-7}$ | $5.95\times10^{-7}$ | $5.74\times10^{-7}$ | 0.00 | – | $\mu$ |
| pyp | $Pa\bar{3}$ | cub | PD+LE | $2.55\times10^{-6}$ | $2.55\times10^{-6}$ | $2.55\times10^{-6}$ | 0.00 | – | PD+LE |
| pyr | $Pa\bar{3}$ | cub | PD+LE | $2.34\times10^{-7}$ | $2.34\times10^{-7}$ | $2.34\times10^{-7}$ | 0.00 | – | PD+LE |
| pys | Pbca | orth | PD+LE | $6.92\times10^{-7}$ | $6.92\times10^{-7}$ | – | 1.00 | – | PD+LE |
| pyu | $Pa\bar{3}$ | cub | PD+LE | $3.81\times10^{-6}$ | $3.81\times10^{-6}$ | $3.81\times10^{-6}$ | 1.00 | – | $\mu$ |
| qdl | $P6_222$ | hex | $\mu$ | $9.42\times10^{-8}$ | $9.42\times10^{-8}$ | $6.74\times10^{-8}$ | 0.00 | – | $\mu$ |
| qok | $P4_332$ | cub | PD+LE | $2.48\times10^{-6}$ | $2.48\times10^{-6}$ | – | 0.00 | – | PD+LE |
| qrf | Pmna | orth | PD+LE | $6.08\times10^{-8}$ | $6.08\times10^{-8}$ | $6.08\times10^{-8}$ | 0.00 | – | PD+LE |
| qtq | $P6_222$ | hex | $\mu$ | – | – | $6.15\times10^{-7}$ | – | – | $\mu$ |
| qtz | $P6_222$ | hex | $\mu$ | $3.01\times10^{-7}$ | $3.01\times10^{-7}$ | $1.55\times10^{-7}$ | 0.00 | – | $\mu$ |
| qza | $P6_422$ | hex | $\mu$ | – | – | $6.14\times10^{-8}$ | – | – | $\mu$ |
| qzd | $P6_222$ | hex | $\mu$ | $2.05\times10^{-7}$ | $2.05\times10^{-7}$ | $1.59\times10^{-7}$ | 0.00 | – | $\mu$ |
| rac | Pnna | orth | $\mu$ | – | – | $1.05\times10^{-7}$ | – | – | $\mu$ |
| reb | $P\bar{4}3m$ | cub | $\mu$ | $1.98\times10^{-6}$ | $1.98\times10^{-6}$ | $1.31\times10^{-6}$ | 0.00 | – | $\mu$ |
| ren | $Pm\bar{3}m$ | cub | $\mu$ | – | – | $1.92\times10^{-6}$ | – | – | $\mu$ |
| reo | $Pm\bar{3}m$ | cub | $\mu$ | $2.54\times10^{-6}$ | $2.54\times10^{-6}$ | $1.22\times10^{-6}$ | 0.00 | – | $\mu$ |
| rep | $Pm\bar{3}m$ | cub | $\mu$ | $3.12\times10^{-6}$ | $3.12\times10^{-6}$ | $1.30\times10^{-6}$ | 0.00 | – | $\mu$ |
| rgd | R3m | trig | $\mu$ | $5.72\times10^{-6}$ | $5.72\times10^{-6}$ | $5.39\times10^{-6}$ | 0.00 | – | $\mu$ |
| rkf | $P6_222$ | hex | $\mu$ | $3.92\times10^{-7}$ | $3.92\times10^{-7}$ | $2.53\times10^{-7}$ | 0.00 | – | $\mu$ |
| rna | Imma | orth | PD+LE | $2.58\times10^{-8}$ | $2.58\times10^{-8}$ | – | 1.00 | – | PD+LE |
| roa | Cccm | orth | PD+LE | $9.60\times10^{-7}$ | $9.60\times10^{-7}$ | – | 0.00 | – | PD+LE |
| rob | Cccm | orth | PD+LE | $5.75\times10^{-7}$ | $5.75\times10^{-7}$ | – | 0.00 | – | PD+LE |
| rok | $P4_332$ | cub | PD+LE | $2.48\times10^{-6}$ | $2.48\times10^{-6}$ | – | 0.00 | – | PD+LE |
| rtl | $P4_2/mnm$ | tet | PD+LE | $1.01\times10^{-7}$ | $1.01\times10^{-7}$ | – | 0.00 | – | PD+LE |
| rtw | P4/mbm | tet | PD+LE | $2.67\times10^{-7}$ | $2.67\times10^{-7}$ | – | 0.00 | – | PD+LE |
| scg | A2/n | mono | $\mu$ | – | – | $6.38\times10^{-7}$ | – | – | $\mu$ |
| scn | Immm | orth | PD+LE | $1.42\times10^{-7}$ | $1.42\times10^{-7}$ | – | 0.00 | – | PD+LE |
| scu | P4/mmm | tet | PD+LE | $1.89\times10^{-7}$ | $1.89\times10^{-7}$ | – | 0.00 | – | PD+LE |
| sda | $I4_1md$ | tet | PD+LE | $3.58\times10^{-7}$ | $3.58\times10^{-7}$ | – | 0.00 | – | PD+LE |
| sea | Cmmm | orth | PD+LE | $6.16\times10^{-7}$ | $6.16\times10^{-7}$ | – | 1.00 | – | PD+LE |
| seb | Pmmm | orth | PD+LE | $7.46\times10^{-7}$ | $7.46\times10^{-7}$ | – | 0.00 | – | PD+LE |
| seh | $P6_3/mmc$ | hex | PD+LE | $9.05\times10^{-7}$ | $9.05\times10^{-7}$ | – | 0.00 | – | PD+LE |
| sep | $P6_3mc$ | hex | PD+LE | $5.00\times10^{-7}$ | $5.00\times10^{-7}$ | $5.91\times10^{-7}$ | 0.00 | – | PD+LE |

| Net | Space group | Sys. | Class | $\lambda_{PD}$ | $\lambda_{LE}$ | $\lambda_{\mu}^{num}$ | $f_{inv}$ | Window | Class, *r*/*l* = 0.05 |
|---|---|---|---|---|---|---|---|---|---|
| sha | $P\bar{6}m2$ | hex | $\mu$ | $1.05\times10^{-6}$ | $1.05\times10^{-6}$ | $7.76\times10^{-7}$ | 0.00 | – | $\mu$ |
| shg | P6/mmm | hex | PD+LE | $1.65\times10^{-6}$ | $1.65\times10^{-6}$ | $1.65\times10^{-6}$ | 0.00 | – | PD+LE |
| shh | P6/mmm | hex | $\mu$ | – | – | $3.13\times10^{-6}$ | – | – | $\mu$ |
| shk | P6/mmm | hex | $\mu$ | – | – | $1.36\times10^{-6}$ | – | – | $\mu$ |
| shm | P6/mmm | hex | PD+LE | $1.05\times10^{-6}$ | $1.05\times10^{-6}$ | $1.05\times10^{-6}$ | 0.00 | – | PD+LE |
| shn | $P6_3/mmc$ | hex | $\mu$ | $9.32\times10^{-7}$ | $9.32\times10^{-7}$ | $7.40\times10^{-7}$ | 0.00 | – | $\mu$ |
| shp | P6/mmm | hex | $\mu$ | $1.03\times10^{-6}$ | – | $8.59\times10^{-7}$ | 1.00 | – | $\mu$ |
| shu | P6/mmm | hex | PD+LE | $2.56\times10^{-6}$ | $2.56\times10^{-6}$ | $2.56\times10^{-6}$ | 0.00 | – | PD+LE |
| shx | P6/mmm | hex | $\mu$ | $1.98\times10^{-6}$ | $1.98\times10^{-6}$ | $1.23\times10^{-6}$ | 0.00 | – | $\mu$ |
| shy | $Pm\bar{3}$ | cub | PD+LE | $3.37\times10^{-6}$ | $3.37\times10^{-6}$ | $3.37\times10^{-6}$ | 0.00 | – | PD+LE |
| sit | Imm2 | orth | PD+LE | $1.13\times10^{-7}$ | $1.13\times10^{-7}$ | – | 1.00 | – | PD+LE |
| ska | Cmca | orth | $\mu$ | – | – | $4.65\times10^{-7}$ | – | – | $\mu$ |
| skc | Ibam | orth | $\mu$ | – | – | $2.38\times10^{-7}$ | – | – | $\mu$ |
| skd | Imma | orth | $\mu$ | – | – | $1.19\times10^{-7}$ | – | – | $\mu$ |
| ske | Imma | orth | PD+LE | $6.77\times10^{-7}$ | $6.77\times10^{-7}$ | – | 0.00 | – | PD+LE |
| skf | Cmcm | orth | $\mu$ | – | – | $2.25\times10^{-7}$ | – | – | $\mu$ |
| skl | $P4_2/nmc$ | tet | PD+LE | $1.05\times10^{-6}$ | $1.05\times10^{-6}$ | $1.48\times10^{-6}$ | 0.00 | – | PD+LE |
| sma | $P6_222$ | hex | $\mu$ | $6.87\times10^{-7}$ | $6.87\times10^{-7}$ | $4.56\times10^{-7}$ | 0.00 | – | $\mu$ |
| smb | R3m | trig | PD-only | $4.45\times10^{-7}$ | $6.19\times10^{-7}$ | $5.38\times10^{-7}$ | 1.00 | 20.9% | PD-only |
| smc | R32 | trig | $\mu$ | $7.02\times10^{-7}$ | $7.02\times10^{-7}$ | $6.98\times10^{-7}$ | 0.00 | – | $\mu$ |
| smh | $P6_122$ | hex | $\mu$ | – | – | $7.21\times10^{-8}$ | – | – | $\mu$ |
| smi | $P6_122$ | hex | $\mu$ | $6.66\times10^{-7}$ | $6.79\times10^{-7}$ | $4.30\times10^{-7}$ | 1.00 | – | $\mu$ |
| smj | $P6_122$ | hex | $\mu$ | $5.46\times10^{-7}$ | $5.46\times10^{-7}$ | $3.43\times10^{-7}$ | 0.00 | – | $\mu$ |
| smk | $P6_122$ | hex | PD+LE | $5.31\times10^{-7}$ | $5.31\times10^{-7}$ | – | 0.00 | – | PD+LE |
| sml | $P6_222$ | hex | PD+LE | $5.15\times10^{-7}$ | $5.15\times10^{-7}$ | – | 0.00 | – | PD+LE |
| smm | R32 | trig | $\mu$ | $5.70\times10^{-7}$ | $5.70\times10^{-7}$ | $5.06\times10^{-7}$ | 0.00 | – | $\mu$ |
| smt | $P6_1$ | hex | $\mu$ | $8.04\times10^{-7}$ | $8.04\times10^{-7}$ | $7.30\times10^{-7}$ | 0.00 | – | $\mu$ |
| smu | $P\bar{6}2c$ | hex | PD+LE | $6.44\times10^{-7}$ | $6.44\times10^{-7}$ | – | 0.00 | – | PD+LE |
| smv | $P6_3/mmc$ | hex | $\mu$ | $2.89\times10^{-7}$ | $2.89\times10^{-7}$ | $1.92\times10^{-7}$ | 0.00 | – | $\mu$ |
| smw | P6/mcc | hex | $\mu$ | $4.55\times10^{-7}$ | $4.55\times10^{-7}$ | $2.91\times10^{-7}$ | 0.00 | – | $\mu$ |
| sne | $I\bar{4}2d$ | tet | $\mu$ | – | – | $6.21\times10^{-7}$ | – | – | $\mu$ |
| snf | $Pm\bar{3}$ | cub | PD+LE | $1.91\times10^{-7}$ | $1.91\times10^{-7}$ | – | 0.00 | – | PD+LE |
| sni | I4/mcm | tet | $\mu$ | – | – | $2.45\times10^{-7}$ | – | – | $\mu$ |
| snk | $I4_1/amd$ | tet | PD+LE | $6.54\times10^{-7}$ | $6.54\times10^{-7}$ | – | 0.00 | – | PD+LE |
| snm | $P4_122$ | tet | $\mu$ | $2.10\times10^{-7}$ | $2.10\times10^{-7}$ | $2.03\times10^{-7}$ | 0.00 | – | $\mu$ |
| snp | $P4_2/mmc$ | tet | $\mu$ | $2.24\times10^{-7}$ | $2.24\times10^{-7}$ | $1.24\times10^{-7}$ | 0.00 | – | $\mu$ |
| snq | I4/mmm | tet | $\mu$ | – | – | $1.97\times10^{-7}$ | – | – | $\mu$ |
| snw | $I4_1/amd$ | tet | $\mu$ | – | – | $6.95\times10^{-7}$ | – | – | $\mu$ |
| snx | $P6_3/mmc$ | hex | PD-only | $7.07\times10^{-7}$ | $7.15\times10^{-7}$ | – | 1.00 | 1.1% | PD-only |
| sny | $P6_3/mmc$ | hex | $\mu$ | – | – | $7.20\times10^{-8}$ | – | – | $\mu$ |
| snz | $R\bar{3}m$ | trig | PD+LE | $4.99\times10^{-7}$ | $4.99\times10^{-7}$ | – | 0.00 | – | PD+LE |
| sod | $Im\bar{3}m$ | cub | $\mu$ | $1.91\times10^{-7}$ | $1.91\times10^{-7}$ | $1.81\times10^{-7}$ | 0.00 | – | $\mu$ |
| soe | Pnna | orth | $\mu$ | – | – | $2.69\times10^{-7}$ | – | – | $\mu$ |

| Net | Space group | Sys. | Class | $\lambda_{PD}$ | $\lambda_{LE}$ | $\lambda_{\mu}^{num}$ | $f_{inv}$ | Window | Class, *r/l* = 0.05 |
|---|---|---|---|---|---|---|---|---|---|
| sof | Pnna | orth | $\mu$ | – | – | $1.11\times10^{-7}$ | – | – | $\mu$ |
| soj | Pbcn | orth | $\mu$ | $4.75\times10^{-7}$ | $4.75\times10^{-7}$ | $4.64\times10^{-7}$ | 0.00 | – | $\mu$ |
| sol | Pbcn | orth | PD-only | $7.29\times10^{-7}$ | $7.42\times10^{-7}$ | – | 1.00 | 1.8% | PD-only |
| sos | Pccn | orth | $\mu$ | – | – | $9.72\times10^{-7}$ | – | – | $\mu$ |
| sow | Pbcn | orth | PD+LE | $9.66\times10^{-8}$ | $9.66\times10^{-8}$ | $9.77\times10^{-8}$ | 1.00 | – | PD+LE |
| spp | P6/mmm | hex | PD+LE | $1.82\times10^{-6}$ | $1.82\times10^{-6}$ | $1.82\times10^{-6}$ | 0.00 | – | PD+LE |
| sqc | $I4_1$/amd | tet | PD-only | $5.49\times10^{-7}$ | $6.65\times10^{-7}$ | $6.34\times10^{-7}$ | 1.00 | 15.5% | PD-only |
| sqd | $I4_1$/amd | tet | $\mu$ | – | – | $1.65\times10^{-6}$ | – | – | $\mu$ |
| sqp | I4/mmm | tet | $\mu$ | – | – | $3.88\times10^{-7}$ | – | – | $\mu$ |
| sqq | $I4_1$/amd | tet | $\mu$ | – | – | $1.64\times10^{-6}$ | – | – | $\mu$ |
| sqs | P6/mmm | hex | $\mu$ | $1.98\times10^{-7}$ | $1.98\times10^{-7}$ | $9.80\times10^{-8}$ | 0.00 | – | $\mu$ |
| sqv | $I4_1$/amd | tet | PD+LE | $1.53\times10^{-6}$ | $1.53\times10^{-6}$ | $2.20\times10^{-6}$ | 0.00 | – | PD+LE |
| sra | Imma | orth | $\mu$ | – | – | $9.06\times10^{-8}$ | – | – | $\mu$ |
| srb | Imma | orth | $\mu$ | – | – | $4.78\times10^{-8}$ | – | – | $\mu$ |
| srd | $P4_232$ | cub | $\mu$ | $2.47\times10^{-7}$ | $2.47\times10^{-7}$ | $1.41\times10^{-7}$ | 0.00 | – | $\mu$ |
| srs | $I4_132$ | cub | $\mu$ | – | – | $3.37\times10^{-8}$ | – | – | $\mu$ |
| ssa | $P6_3$/mmc | hex | $\mu$ | $7.44\times10^{-8}$ | – | $7.37\times10^{-8}$ | 1.00 | – | $\mu$ |
| sse | P4/mmm | tet | PD+LE | $1.25\times10^{-7}$ | $1.25\times10^{-7}$ | – | 0.00 | – | PD+LE |
| ssg | $P6_3$/mmc | hex | $\mu$ | – | – | $4.04\times10^{-7}$ | – | – | $\mu$ |
| sss | P6/mmm | hex | $\mu$ | – | – | $5.96\times10^{-7}$ | – | – | $\mu$ |
| sta | $P6_3$/mmc | hex | PD-only | $6.84\times10^{-7}$ | $6.98\times10^{-7}$ | – | 1.00 | 2.0% | PD-only |
| stb | $R\bar{3}m$ | trig | PD-only | $6.84\times10^{-7}$ | $7.14\times10^{-7}$ | $7.18\times10^{-7}$ | 1.00 | 4.3% | PD-only |
| stc | $R\bar{3}m$ | trig | PD-only | $6.84\times10^{-7}$ | $7.10\times10^{-7}$ | – | 1.00 | 3.7% | PD-only |
| std | $P6_3$/mmc | hex | PD-only | $6.84\times10^{-7}$ | $7.23\times10^{-7}$ | – | 1.00 | 5.6% | PD-only |
| ste | $P6_3$/mmc | hex | PD-only | $6.84\times10^{-7}$ | $6.94\times10^{-7}$ | – | 1.00 | 1.4% | PD-only |
| sti | P6/mmm | hex | $\mu$ | $4.60\times10^{-7}$ | $4.60\times10^{-7}$ | $2.66\times10^{-7}$ | 0.00 | – | $\mu$ |
| stm | P6/mmm | hex | $\mu$ | $2.16\times10^{-7}$ | $2.16\times10^{-7}$ | $1.16\times10^{-7}$ | 0.00 | – | $\mu$ |
| stn | P6/mmm | hex | $\mu$ | – | – | $8.01\times10^{-7}$ | – | – | $\mu$ |
| sto | $P6_422$ | hex | $\mu$ | – | – | $8.79\times10^{-8}$ | – | – | $\mu$ |
| stp | P6/mmm | hex | $\mu$ | $1.41\times10^{-7}$ | $1.41\times10^{-7}$ | $7.72\times10^{-8}$ | 0.00 | – | $\mu$ |
| sub | P4/mmm | tet | PD+LE | $2.59\times10^{-6}$ | $2.59\times10^{-6}$ | $2.59\times10^{-6}$ | 0.00 | – | PD+LE |
| sur | Cccm | orth | PD-only | $5.94\times10^{-8}$ | – | $6.09\times10^{-8}$ | 1.00 | 2.5% | PD-only |
| sus | P4/mmm | tet | PD+LE | $6.29\times10^{-7}$ | $6.29\times10^{-7}$ | – | 0.00 | – | PD+LE |
| sva | $I4_1$md | tet | $\mu$ | – | – | $1.29\times10^{-6}$ | – | – | $\mu$ |
| svd | $P4_2$/mmc | tet | $\mu$ | $2.37\times10^{-7}$ | $2.37\times10^{-7}$ | $1.34\times10^{-7}$ | 0.00 | – | $\mu$ |
| sve | P4/mbm | tet | PD+LE | $8.03\times10^{-7}$ | $8.03\times10^{-7}$ | – | 0.00 | – | PD+LE |
| svf | I4/mcm | tet | $\mu$ | – | – | $2.60\times10^{-7}$ | – | – | $\mu$ |
| svg | I4/mmm | tet | PD+LE | $6.98\times10^{-7}$ | $6.98\times10^{-7}$ | $7.02\times10^{-7}$ | 0.00 | – | PD+LE |
| svh | $I4_1$/amd | tet | PD+LE | $8.16\times10^{-7}$ | $8.16\times10^{-7}$ | – | 0.00 | – | PD+LE |
| svi | I4/mcm | tet | PD+LE | $8.61\times10^{-7}$ | $8.61\times10^{-7}$ | – | 0.00 | – | PD+LE |
| svj | P6/m | hex | PD+LE | $6.69\times10^{-7}$ | $6.69\times10^{-7}$ | – | 0.00 | – | PD+LE |
| svk | Cmmm | orth | PD+LE | $7.65\times10^{-7}$ | $7.65\times10^{-7}$ | – | 0.00 | – | PD+LE |
| svn | $Pa\bar{3}$ | cub | PD+LE | $1.39\times10^{-6}$ | $1.39\times10^{-6}$ | – | 0.00 | – | PD+LE |

| Net | Space group | Sys. | Class | $\lambda_{PD}$ | $\lambda_{LE}$ | $\lambda_{\mu}^{num}$ | $f_{inv}$ | Window | Class, *r/l* = 0.05 |
|---|---|---|---|---|---|---|---|---|---|
| svs | P4/mcc | tet | $\mu$ | $4.65\times10^{-7}$ | $4.65\times10^{-7}$ | $1.92\times10^{-7}$ | 0.00 | – | $\mu$ |
| swe | $P6_2$ | hex | $\mu$ | $1.04\times10^{-6}$ | $1.04\times10^{-6}$ | $9.03\times10^{-7}$ | 0.00 | – | $\mu$ |
| swf | P6/mcc | hex | $\mu$ | – | – | $2.92\times10^{-7}$ | – | – | $\mu$ |
| swg | P6/mcc | hex | $\mu$ | – | – | $9.47\times10^{-8}$ | – | – | $\mu$ |
| swh | P6/mcc | hex | PD+LE | $6.13\times10^{-7}$ | $6.13\times10^{-7}$ | – | 0.00 | – | PD+LE |
| swi | P6/mcc | hex | PD+LE | $7.25\times10^{-7}$ | $7.25\times10^{-7}$ | – | 0.00 | – | PD+LE |
| swj | $P6_3$/mcm | hex | $\mu$ | $3.42\times10^{-7}$ | $3.42\times10^{-7}$ | $2.59\times10^{-7}$ | 0.00 | – | $\mu$ |
| swk | $P6_3$/mcm | hex | PD+LE | $2.68\times10^{-7}$ | $2.68\times10^{-7}$ | $2.73\times10^{-7}$ | 0.00 | – | PD+LE |
| sww | Pm$\bar{3}$n | cub | $\mu$ | $2.96\times10^{-7}$ | $2.96\times10^{-7}$ | $2.89\times10^{-7}$ | 0.00 | – | $\mu$ |
| sxa | Cmma | orth | $\mu$ | – | – | $6.84\times10^{-7}$ | – | – | $\mu$ |
| sxb | Cccm | orth | PD+LE | $6.13\times10^{-7}$ | $6.13\times10^{-7}$ | – | 0.00 | – | PD+LE |
| sxc | $P6_222$ | hex | $\mu$ | $6.83\times10^{-7}$ | $6.83\times10^{-7}$ | $6.74\times10^{-7}$ | 0.00 | – | $\mu$ |
| sxd | Imma | orth | $\mu$ | – | – | $3.28\times10^{-7}$ | – | – | $\mu$ |
| tav | C2/c | mono | PD-only | $2.58\times10^{-7}$ | $3.11\times10^{-7}$ | – | 1.00 | 20.6% | PD-only |
| tca | $P6_3$/mmc | hex | $\mu$ | – | – | $1.20\times10^{-6}$ | – | – | $\mu$ |
| tcb | Pnna | orth | PD-only | $1.89\times10^{-7}$ | $2.24\times10^{-7}$ | – | 1.00 | 18.9% | PD-only |
| tcc | $P6_222$ | hex | $\mu$ | $1.78\times10^{-6}$ | $1.78\times10^{-6}$ | $1.58\times10^{-6}$ | 0.00 | – | $\mu$ |
| tcd | R$\bar{3}$m | trig | $\mu$ | – | – | $1.20\times10^{-6}$ | – | – | $\mu$ |
| tce | I4/m | tet | PD+LE | $2.01\times10^{-6}$ | $2.01\times10^{-6}$ | – | 0.00 | – | PD+LE |
| tcg | Pnma | orth | PD+LE | $2.27\times10^{-6}$ | $2.27\times10^{-6}$ | $3.68\times10^{-6}$ | 1.00 | – | PD+LE |
| tch | $P6_3$/m | hex | $\mu$ | – | – | $3.68\times10^{-6}$ | – | – | $\mu$ |
| tcj | $P6_3$/mmc | hex | $\mu$ | – | – | $5.50\times10^{-6}$ | – | – | $\mu$ |
| tck | $P6_3$/mmc | hex | $\mu$ | – | – | $5.61\times10^{-6}$ | – | – | $\mu$ |
| tcl | R$\bar{3}$m | trig | $\mu$ | – | – | $5.57\times10^{-6}$ | – | – | $\mu$ |
| tcm | R$\bar{3}$m | trig | $\mu$ | – | – | $5.61\times10^{-6}$ | – | – | $\mu$ |
| tcn | $P6_422$ | hex | $\mu$ | $1.54\times10^{-6}$ | $1.54\times10^{-6}$ | $1.51\times10^{-6}$ | 0.00 | – | $\mu$ |
| tco | Cccm | orth | PD+LE | $1.50\times10^{-6}$ | $1.50\times10^{-6}$ | – | 0.00 | – | PD+LE |
| tcs | I4/mmm | tet | $\mu$ | – | – | $2.39\times10^{-7}$ | – | – | $\mu$ |
| tec | Pm$\bar{3}$m | cub | $\mu$ | – | – | $2.53\times10^{-7}$ | – | – | $\mu$ |
| ter | Pm$\bar{3}$m | cub | $\mu$ | $2.20\times10^{-6}$ | $2.20\times10^{-6}$ | $1.41\times10^{-6}$ | 0.00 | – | $\mu$ |
| tfa | I$\bar{4}$m2 | tet | PD+LE | $7.77\times10^{-8}$ | $7.77\times10^{-8}$ | – | 0.00 | – | PD+LE |
| tfc | Cmmm | orth | PD+LE | $3.64\times10^{-8}$ | $3.64\times10^{-8}$ | – | 0.00 | – | PD+LE |
| tff | Cm | mono | PD+LE | $1.22\times10^{-7}$ | $1.22\times10^{-7}$ | – | 0.00 | – | PD+LE |
| tfi | $P4_2$/mmc | tet | PD+LE | $5.18\times10^{-8}$ | $5.18\times10^{-8}$ | $5.76\times10^{-8}$ | 0.00 | – | PD+LE |
| tfk | $P4_2$/mmc | tet | PD+LE | $1.48\times10^{-8}$ | $1.48\times10^{-8}$ | – | 0.00 | – | PD+LE |
| tfn | $P4_2$/mnm | tet | PD+LE | $8.80\times10^{-8}$ | $8.80\times10^{-8}$ | – | 0.00 | – | PD+LE |
| tfp | P6/mmm | hex | $\mu$ | $9.16\times10^{-9}$ | $9.16\times10^{-9}$ | $5.05\times10^{-9}$ | 0.00 | – | $\mu$ |
| tfq | $P6_422$ | hex | $\mu$ | – | – | $5.51\times10^{-8}$ | – | – | $\mu$ |
| tfs | P6/mmm | hex | PD+LE | $5.30\times10^{-7}$ | $5.30\times10^{-7}$ | – | 0.00 | – | PD+LE |
| tfv | P$\bar{4}$n2 | tet | PD-only | $8.38\times10^{-8}$ | $1.04\times10^{-7}$ | $9.46\times10^{-8}$ | 1.00 | 12.9% | PD-only |
| tfx | Imma | orth | $\mu$ | – | – | $1.06\times10^{-7}$ | – | – | $\mu$ |
| tfz | P6/mmm | hex | $\mu$ | $6.36\times10^{-8}$ | $6.36\times10^{-8}$ | $3.31\times10^{-8}$ | 0.00 | – | $\mu$ |
| the | Pm$\bar{3}$m | cub | $\mu$ | $3.48\times10^{-7}$ | $3.48\times10^{-7}$ | $1.73\times10^{-7}$ | 0.00 | – | $\mu$ |

| Net | Space group | Sys. | Class | $\lambda_{PD}$ | $\lambda_{LE}$ | $\lambda_{\mu}^{num}$ | $f_{inv}$ | Window | Class, *r/l* = 0.05 |
|---|---|---|---|---|---|---|---|---|---|
| thi | Pm3̄m | cub | $\mu$ | – | – | $2.06\times10^{-7}$ | – | – | $\mu$ |
| thj | $I2_12_12_1$ | orth | $\mu$ | – | – | $1.14\times10^{-7}$ | – | – | $\mu$ |
| thl | $P2_1/c$ | mono | PD-only | $1.34\times10^{-7}$ | $1.57\times10^{-7}$ | – | 1.00 | 17.2% | PD-only |
| thn | Pa3̄ | cub | $\mu$ | – | – | $6.62\times10^{-6}$ | – | – | $\mu$ |
| thp | I4̄3d | cub | $\mu$ | – | – | $2.80\times10^{-6}$ | – | – | $\mu$ |
| ths | $I4_1/amd$ | tet | PD+LE | $1.78\times10^{-8}$ | $1.78\times10^{-8}$ | – | 0.00 | – | PD+LE |
| tnn | P6̄m2 | hex | $\mu$ | – | – | $9.75\times10^{-8}$ | – | – | $\mu$ |
| toc | Pn3̄m | cub | PD+LE | $3.87\times10^{-7}$ | $3.87\times10^{-7}$ | – | 0.00 | – | PD+LE |
| tok | $I4_132$ | cub | $\mu$ | – | – | $6.02\times10^{-7}$ | – | – | $\mu$ |
| ton | Pn3̄m | cub | $\mu$ | – | – | $7.47\times10^{-7}$ | – | – | $\mu$ |
| tow | $P4_2/nnm$ | tet | PD+LE | $6.87\times10^{-7}$ | $6.87\times10^{-7}$ | – | 0.00 | – | PD+LE |
| tpr | P6̄m2 | hex | $\mu$ | $4.93\times10^{-8}$ | $4.93\times10^{-8}$ | $4.08\times10^{-8}$ | 0.00 | – | $\mu$ |
| tpt | P6/mcc | hex | $\mu$ | $2.40\times10^{-6}$ | $2.51\times10^{-6}$ | $1.93\times10^{-6}$ | 1.00 | – | $\mu$ |
| trn | $P4_2/mmc$ | tet | $\mu$ | $1.77\times10^{-7}$ | $1.77\times10^{-7}$ | $1.20\times10^{-7}$ | 0.00 | – | $\mu$ |
| tsa | $P6_3/mmc$ | hex | PD+LE | $7.39\times10^{-8}$ | $7.39\times10^{-8}$ | – | 0.00 | – | PD+LE |
| tsb | $P6_322$ | hex | $\mu$ | $1.61\times10^{-7}$ | $1.61\times10^{-7}$ | $1.55\times10^{-7}$ | 0.00 | – | $\mu$ |
| tsd | Pmmm | orth | PD+LE | $4.07\times10^{-8}$ | $4.07\times10^{-8}$ | – | 0.00 | – | PD+LE |
| tsf | Pmmm | orth | PD+LE | $1.24\times10^{-7}$ | $1.24\times10^{-7}$ | – | 0.00 | – | PD+LE |
| tsg | P6̄m2 | hex | $\mu$ | $7.64\times10^{-8}$ | $7.64\times10^{-8}$ | $4.01\times10^{-8}$ | 0.00 | – | $\mu$ |
| tsi | $I4_1/amd$ | tet | PD+LE | $9.88\times10^{-7}$ | $9.88\times10^{-7}$ | – | 0.00 | – | PD+LE |
| tsl | $I4_1/amd$ | tet | $\mu$ | – | – | $7.56\times10^{-6}$ | – | – | $\mu$ |
| tsq | I4/mmm | tet | PD+LE | $3.91\times10^{-7}$ | $3.91\times10^{-7}$ | – | 0.00 | – | PD+LE |
| tsx | I4̄2m | tet | PD+LE | $1.75\times10^{-7}$ | $1.75\times10^{-7}$ | – | 0.00 | – | PD+LE |
| ttg | P4̄n2 | tet | $\mu$ | $4.06\times10^{-8}$ | $4.06\times10^{-8}$ | $3.94\times10^{-8}$ | 0.00 | – | $\mu$ |
| ttm | I4̄m2 | tet | PD+LE | $1.23\times10^{-6}$ | $1.23\times10^{-6}$ | – | 0.00 | – | PD+LE |
| ttn | I4̄m2 | tet | $\mu$ | – | – | $1.42\times10^{-6}$ | – | – | $\mu$ |
| ttp | P6/mmm | hex | $\mu$ | $3.14\times10^{-8}$ | $3.14\times10^{-8}$ | $1.49\times10^{-8}$ | 0.00 | – | $\mu$ |
| ttv | Pm3̄ | cub | PD+LE | $1.13\times10^{-7}$ | $1.13\times10^{-7}$ | – | 0.00 | – | PD+LE |
| ttw | P6/mmm | hex | $\mu$ | – | – | $4.08\times10^{-8}$ | – | – | $\mu$ |
| ttx | P6/mmm | hex | $\mu$ | $7.74\times10^{-8}$ | $7.74\times10^{-8}$ | $4.07\times10^{-8}$ | 0.00 | – | $\mu$ |
| tty | P4/mmm | tet | PD+LE | $1.04\times10^{-7}$ | $1.04\times10^{-7}$ | – | 0.00 | – | PD+LE |
| twt | $P6_222$ | hex | $\mu$ | $4.37\times10^{-8}$ | $5.41\times10^{-8}$ | $3.14\times10^{-8}$ | 1.00 | – | $\mu$ |
| txs | P6̄2c | hex | PD-only | $3.40\times10^{-7}$ | $3.84\times10^{-7}$ | $3.44\times10^{-7}$ | 1.00 | 1.2% | PD-only |
| ucn | $P6_3/mmc$ | hex | $\mu$ | $9.62\times10^{-8}$ | $9.62\times10^{-8}$ | $7.71\times10^{-8}$ | 0.00 | – | $\mu$ |
| ukf | $P6_222$ | hex | $\mu$ | $1.72\times10^{-7}$ | $1.72\times10^{-7}$ | $8.64\times10^{-8}$ | 0.00 | – | $\mu$ |
| ukg | $P4_2/n$ | tet | PD+LE | $1.75\times10^{-7}$ | $1.75\times10^{-7}$ | $1.93\times10^{-7}$ | 0.00 | – | PD+LE |
| ukh | $P6_222$ | hex | $\mu$ | $2.45\times10^{-7}$ | $2.66\times10^{-7}$ | $1.85\times10^{-7}$ | 1.00 | – | $\mu$ |
| uku | P23 | cub | $\mu$ | $1.17\times10^{-7}$ | $1.17\times10^{-7}$ | $8.10\times10^{-8}$ | 0.00 | – | $\mu$ |
| umb | $P6_222$ | hex | $\mu$ | – | – | $3.56\times10^{-8}$ | – | – | $\mu$ |
| umf | $P4_222$ | tet | $\mu$ | $7.30\times10^{-8}$ | $7.30\times10^{-8}$ | $6.31\times10^{-8}$ | 0.00 | – | $\mu$ |
| umg | $P6_222$ | hex | $\mu$ | – | – | $1.00\times10^{-7}$ | – | – | $\mu$ |
| umj | $P6_222$ | hex | $\mu$ | – | – | $1.03\times10^{-7}$ | – | – | $\mu$ |
| umm | $P6_222$ | hex | $\mu$ | – | – | $3.43\times10^{-8}$ | – | – | $\mu$ |

| Net | Space group | Sys. | Class | $\lambda_{PD}$ | $\lambda_{LE}$ | $\lambda_{\mu}^{num}$ | $f_{inv}$ | Window | Class, *r/l* = 0.05 |
|---|---|---|---|---|---|---|---|---|---|
| umo | $P6_122$ | hex | $\mu$ | – | – | $3.88\times10^{-8}$ | – | – | $\mu$ |
| umw | $P4_122$ | tet | PD+LE | $8.70\times10^{-8}$ | $8.70\times10^{-8}$ | – | 0.00 | – | PD+LE |
| umx | $P4_2/mcm$ | tet | $\mu$ | $9.08\times10^{-8}$ | $9.08\times10^{-8}$ | $8.32\times10^{-8}$ | 0.00 | – | $\mu$ |
| umy | $P6_122$ | hex | $\mu$ | $1.42\times10^{-7}$ | $1.53\times10^{-7}$ | $1.40\times10^{-7}$ | 1.00 | – | $\mu$ |
| una | $P6_122$ | hex | $\mu$ | – | – | $1.09\times10^{-7}$ | – | – | $\mu$ |
| unc | $P4_122$ | tet | $\mu$ | $1.72\times10^{-7}$ | $1.72\times10^{-7}$ | $1.51\times10^{-7}$ | 0.00 | – | $\mu$ |
| unf | $P\bar{3}1c$ | trig | $\mu$ | – | – | $2.65\times10^{-7}$ | – | – | $\mu$ |
| unh | $P6_122$ | hex | $\mu$ | $1.03\times10^{-7}$ | $1.03\times10^{-7}$ | $5.68\times10^{-8}$ | 0.00 | – | $\mu$ |
| uni | $P6_122$ | hex | $\mu$ | $2.21\times10^{-7}$ | $2.21\times10^{-7}$ | $2.18\times10^{-7}$ | 0.00 | – | $\mu$ |
| unj | $P6_122$ | hex | $\mu$ | $2.40\times10^{-7}$ | $2.40\times10^{-7}$ | $1.56\times10^{-7}$ | 0.00 | – | $\mu$ |
| unm | $P4_22_12$ | tet | $\mu$ | $2.62\times10^{-7}$ | $2.62\times10^{-7}$ | $1.88\times10^{-7}$ | 0.00 | – | $\mu$ |
| uoc | $I4_122$ | tet | PD+LE | $1.20\times10^{-7}$ | $1.20\times10^{-7}$ | – | 0.00 | – | PD+LE |
| uof | $P6_222$ | hex | PD+LE | $1.21\times10^{-7}$ | $1.21\times10^{-7}$ | – | 0.00 | – | PD+LE |
| uog | $P6_222$ | hex | $\mu$ | $2.41\times10^{-7}$ | $2.41\times10^{-7}$ | $1.16\times10^{-7}$ | 0.00 | – | $\mu$ |
| uoj | $P\bar{4}n2$ | tet | $\mu$ | $1.69\times10^{-7}$ | $1.69\times10^{-7}$ | $1.50\times10^{-7}$ | 0.00 | – | $\mu$ |
| uol | $P6_122$ | hex | $\mu$ | $1.75\times10^{-7}$ | $1.82\times10^{-7}$ | $1.51\times10^{-7}$ | 1.00 | – | $\mu$ |
| uom | $P6_222$ | hex | $\mu$ | $2.67\times10^{-7}$ | $2.67\times10^{-7}$ | $1.65\times10^{-7}$ | 0.00 | – | $\mu$ |
| uon | $P6_222$ | hex | PD-only | $1.06\times10^{-7}$ | $1.75\times10^{-7}$ | $1.10\times10^{-7}$ | 1.00 | 3.4% | PD-only |
| uoo | $P6_222$ | hex | $\mu$ | $2.30\times10^{-7}$ | $2.30\times10^{-7}$ | $8.10\times10^{-8}$ | 0.00 | – | $\mu$ |
| uop | $P6_222$ | hex | $\mu$ | – | – | $1.19\times10^{-7}$ | – | – | $\mu$ |
| uoq | $P6_222$ | hex | $\mu$ | – | – | $1.51\times10^{-7}$ | – | – | $\mu$ |
| uov | $P\bar{4}c2$ | tet | $\mu$ | $1.68\times10^{-7}$ | $1.68\times10^{-7}$ | $1.09\times10^{-7}$ | 0.00 | – | $\mu$ |
| upa | $P4_22_12$ | tet | $\mu$ | $2.25\times10^{-7}$ | $2.25\times10^{-7}$ | $2.06\times10^{-7}$ | 0.00 | – | $\mu$ |
| urk | $Pm\bar{3}m$ | cub | PD+LE | $4.05\times10^{-7}$ | $4.05\times10^{-7}$ | – | 0.00 | – | PD+LE |
| urs | $P4/mmm$ | tet | $\mu$ | $6.96\times10^{-7}$ | $6.96\times10^{-7}$ | $4.26\times10^{-7}$ | 1.00 | – | $\mu$ |
| ury | $P6/mmm$ | hex | PD+LE | $1.53\times10^{-7}$ | $1.53\times10^{-7}$ | – | 0.00 | – | PD+LE |
| utk | $P4_122$ | tet | $\mu$ | $4.73\times10^{-8}$ | $4.87\times10^{-8}$ | $4.00\times10^{-8}$ | 1.00 | – | $\mu$ |
| utp | Pnna | orth | PD-only | $2.43\times10^{-8}$ | $2.50\times10^{-8}$ | – | 1.00 | 2.6% | PD-only |
| utx | I432 | cub | $\mu$ | – | – | $2.23\times10^{-6}$ | – | – | $\mu$ |
| vbn | Pbcn | orth | PD+LE | $5.68\times10^{-7}$ | $5.68\times10^{-7}$ | – | 0.00 | – | PD+LE |
| vbo | $C222_1$ | orth | $\mu$ | – | – | $4.81\times10^{-7}$ | – | – | $\mu$ |
| vbt | Pccn | orth | $\mu$ | – | – | $3.51\times10^{-7}$ | – | – | $\mu$ |
| vbu | Pccn | orth | $\mu$ | – | – | $3.97\times10^{-7}$ | – | – | $\mu$ |
| vby | Cccm | orth | PD+LE | $6.67\times10^{-7}$ | $6.67\times10^{-7}$ | – | 0.00 | – | PD+LE |
| vcg | Ibam | orth | $\mu$ | – | – | $5.22\times10^{-7}$ | – | – | $\mu$ |
| vcj | Pbca | orth | PD+LE | $8.19\times10^{-7}$ | $8.19\times10^{-7}$ | $1.00\times10^{-6}$ | 1.00 | – | PD+LE |
| vcm | Cmca | orth | PD+LE | $8.08\times10^{-7}$ | $8.08\times10^{-7}$ | – | 0.00 | – | PD+LE |
| vcn | Cmca | orth | PD+LE | $7.50\times10^{-7}$ | $7.50\times10^{-7}$ | – | 0.00 | – | PD+LE |
| vco | Cmca | orth | $\mu$ | – | – | $6.21\times10^{-7}$ | – | – | $\mu$ |
| vcr | Pnna | orth | PD+LE | $8.85\times10^{-7}$ | $8.85\times10^{-7}$ | $9.02\times10^{-7}$ | 0.00 | – | PD+LE |
| vcs | Pnma | orth | PD+LE | $1.05\times10^{-6}$ | $1.05\times10^{-6}$ | $2.24\times10^{-6}$ | 1.00 | – | PD+LE |
| vct | Pbcn | orth | $\mu$ | $1.71\times10^{-6}$ | $1.73\times10^{-6}$ | $1.67\times10^{-6}$ | 1.00 | – | $\mu$ |
| vcu | Pbcn | orth | $\mu$ | – | – | $9.67\times10^{-7}$ | – | – | $\mu$ |

| Net | Space group | Sys. | Class | $\lambda_{PD}$ | $\lambda_{LE}$ | $\lambda_{\mu}^{num}$ | $f_{inv}$ | Window | Class, $r/l$ = 0.05 |
|---|---|---|---|---|---|---|---|---|---|
| vcv | Pbcn | orth | PD-only | $1.05\times10^{-6}$ | – | $2.02\times10^{-6}$ | 1.00 | 93.0% | PD-only |
| vcw | Pbca | orth | $\mu$ | – | – | $1.48\times10^{-6}$ | – | – | $\mu$ |
| vcx | Cmca | orth | PD-only | $1.35\times10^{-6}$ | $1.49\times10^{-6}$ | $1.59\times10^{-6}$ | 1.00 | 10.5% | PD-only |
| vcy | Pbca | orth | PD-only | $2.03\times10^{-6}$ | – | $2.86\times10^{-6}$ | 1.00 | 41.1% | PD-only |
| vcz | Pbcn | orth | PD+LE | $2.49\times10^{-7}$ | $2.49\times10^{-7}$ | $2.49\times10^{-7}$ | 0.00 | – | PD+LE |
| vma | C2/c | mono | PD-only | $4.08\times10^{-7}$ | $4.25\times10^{-7}$ | $4.22\times10^{-7}$ | 1.00 | 3.4% | PD-only |
| vmb | C2/c | mono | $\mu$ | – | – | $3.25\times10^{-7}$ | – | – | $\mu$ |
| vmc | C2/c | mono | $\mu$ | – | – | $1.09\times10^{-7}$ | – | – | $\mu$ |
| vmd | C2/c | mono | $\mu$ | – | – | $4.03\times10^{-7}$ | – | – | $\mu$ |
| vmf | C2/m | mono | $\mu$ | – | – | $4.23\times10^{-7}$ | – | – | $\mu$ |
| vmg | C2/c | mono | $\mu$ | – | – | $5.46\times10^{-7}$ | – | – | $\mu$ |
| vmh | A2/n | mono | $\mu$ | – | – | $4.43\times10^{-7}$ | – | – | $\mu$ |
| vmi | I2/a | mono | $\mu$ | – | – | $3.88\times10^{-7}$ | – | – | $\mu$ |
| vmj | I2/a | mono | $\mu$ | $5.72\times10^{-7}$ | – | $3.92\times10^{-7}$ | 1.00 | – | $\mu$ |
| vmk | I2/a | mono | $\mu$ | – | – | $2.49\times10^{-7}$ | – | – | $\mu$ |
| vml | I2/a | mono | $\mu$ | $7.16\times10^{-7}$ | $7.26\times10^{-7}$ | $5.36\times10^{-7}$ | 1.00 | – | $\mu$ |
| vmn | I2/a | mono | PD+LE | $8.47\times10^{-7}$ | $8.47\times10^{-7}$ | – | 1.00 | – | PD+LE |
| vmo | C2/c | mono | $\mu$ | – | – | $7.41\times10^{-7}$ | – | – | $\mu$ |
| vmp | I2/m | mono | PD+LE | $8.05\times10^{-7}$ | $8.05\times10^{-7}$ | – | 0.00 | – | PD+LE |
| vmq | C2/c | mono | PD+LE | $8.85\times10^{-7}$ | $8.85\times10^{-7}$ | $9.01\times10^{-7}$ | 0.00 | – | PD+LE |
| vmr | $P2_1/c$ | mono | PD+LE | $8.17\times10^{-7}$ | $8.17\times10^{-7}$ | $1.00\times10^{-6}$ | 1.00 | – | PD+LE |
| vms | I2/a | mono | $\mu$ | $1.26\times10^{-6}$ | $1.26\times10^{-6}$ | $1.10\times10^{-6}$ | 0.00 | – | $\mu$ |
| vmt | C2/c | mono | PD+LE | $1.08\times10^{-6}$ | $1.08\times10^{-6}$ | – | 0.00 | – | PD+LE |
| vmu | C2/c | mono | $\mu$ | – | – | $9.80\times10^{-7}$ | – | – | $\mu$ |
| vmw | I2/a | mono | $\mu$ | – | – | $9.72\times10^{-7}$ | – | – | $\mu$ |
| vmx | C2/c | mono | PD+LE | $9.21\times10^{-7}$ | $9.21\times10^{-7}$ | – | 0.00 | – | PD+LE |
| vmy | I2/a | mono | PD-only | $1.05\times10^{-6}$ | – | $2.01\times10^{-6}$ | 1.00 | 91.7% | PD-only |
| vmz | $P2_1/c$ | mono | $\mu$ | $1.04\times10^{-6}$ | $1.04\times10^{-6}$ | $9.67\times10^{-7}$ | 1.00 | – | $\mu$ |
| vna | $P2_1/n$ | mono | PD+LE | $1.32\times10^{-6}$ | $1.32\times10^{-6}$ | – | 0.00 | – | PD+LE |
| vnb | C2/c | mono | PD-only | $1.22\times10^{-6}$ | $1.24\times10^{-6}$ | – | 1.00 | 1.6% | PD-only |
| vnc | C2/c | mono | PD+LE | $1.78\times10^{-6}$ | $1.78\times10^{-6}$ | $2.81\times10^{-6}$ | 0.00 | – | PD+LE |
| vnd | I2/a | mono | PD+LE | $1.80\times10^{-6}$ | $1.80\times10^{-6}$ | $2.79\times10^{-6}$ | 0.00 | – | PD+LE |
| vne | $P2_1/c$ | mono | PD-only | $1.67\times10^{-6}$ | – | $2.81\times10^{-6}$ | 1.00 | 68.3% | PD-only |
| vnf | $P2_1/c$ | mono | PD+LE | $1.41\times10^{-6}$ | $1.41\times10^{-6}$ | – | 0.00 | – | PD+LE |
| vng | C2/m | mono | $\mu$ | – | – | $4.03\times10^{-7}$ | – | – | $\mu$ |
| vnh | C2/c | mono | $\mu$ | $7.07\times10^{-7}$ | $7.15\times10^{-7}$ | $5.38\times10^{-7}$ | 1.00 | – | $\mu$ |
| vni | C2/c | mono | PD+LE | $8.29\times10^{-7}$ | $8.29\times10^{-7}$ | – | 1.00 | – | PD+LE |
| vtx | $I4_1/amd$ | tet | PD+LE | $5.49\times10^{-7}$ | $5.49\times10^{-7}$ | – | 0.00 | – | PD+LE |
| wfb | $P4_2/n$ | tet | PD+LE | $1.66\times10^{-6}$ | $1.66\times10^{-6}$ | $1.66\times10^{-6}$ | 1.00 | – | PD+LE |
| wfd | I4/m | tet | PD-only | $1.30\times10^{-6}$ | $1.37\times10^{-6}$ | – | 1.00 | 5.5% | PD-only |
| wfe | $P4_2/n$ | tet | PD+LE | $1.35\times10^{-6}$ | $1.35\times10^{-6}$ | $1.35\times10^{-6}$ | 1.00 | – | PD+LE |
| wff | $I\bar{4}2m$ | tet | PD-only | $1.88\times10^{-6}$ | $1.90\times10^{-6}$ | – | 1.00 | 1.0% | PD+LE |
| wfi | $P4_12_12$ | tet | PD+LE | $1.08\times10^{-6}$ | $1.08\times10^{-6}$ | $2.08\times10^{-6}$ | 0.00 | – | PD+LE |

| Net | Space group | Sys. | Class | $\lambda_{PD}$ | $\lambda_{LE}$ | $\lambda_{\mu}^{num}$ | $f_{inv}$ | Window | Class, *r*/*l* = 0.05 |
|---|---|---|---|---|---|---|---|---|---|
| wfs | I$\bar{4}$2m | tet | PD-only | $9.18\times10^{-7}$ | $1.90\times10^{-6}$ | $1.89\times10^{-6}$ | 1.00 | 106.3% | PD-only |
| wft | P$\bar{4}2_1$c | tet | PD-only | $1.51\times10^{-6}$ | $1.84\times10^{-6}$ | – | 1.00 | 21.7% | PD+LE |
| wfu | P$\bar{4}2_1$c | tet | PD-only | $9.48\times10^{-7}$ | $1.94\times10^{-6}$ | $1.83\times10^{-6}$ | 1.00 | 93.2% | PD-only |
| wfy | I$\bar{4}$ | tet | PD-only | $8.55\times10^{-7}$ | $1.68\times10^{-6}$ | – | 1.00 | 96.9% | PD-only |
| wfz | P$4_12_12$ | tet | PD+LE | $8.19\times10^{-7}$ | $8.19\times10^{-7}$ | $8.23\times10^{-7}$ | 0.00 | – | PD+LE |
| wgu | P$4_2$/mbc | tet | $\mu$ | $5.87\times10^{-7}$ | $5.87\times10^{-7}$ | $5.82\times10^{-7}$ | 0.00 | – | $\mu$ |
| wgv | P$4_2$/mnm | tet | PD+LE | $4.73\times10^{-7}$ | $4.73\times10^{-7}$ | – | 0.00 | – | PD+LE |
| wgw | P$4_2$/ncm | tet | $\mu$ | $4.99\times10^{-7}$ | $4.99\times10^{-7}$ | $4.96\times10^{-7}$ | 0.00 | – | $\mu$ |
| wha | P$\bar{4}2_1$c | tet | PD-only | $8.89\times10^{-7}$ | $8.94\times10^{-7}$ | – | 1.00 | 0.6% | PD+LE |
| whb | P$4_2$/nmc | tet | $\mu$ | – | – | $8.58\times10^{-7}$ | – | – | $\mu$ |
| whc | P$\bar{4}$c2 | tet | $\mu$ | $4.40\times10^{-7}$ | $4.40\times10^{-7}$ | $1.26\times10^{-7}$ | 0.00 | – | $\mu$ |
| whs | P$4_1$22 | tet | PD+LE | $1.19\times10^{-7}$ | $1.19\times10^{-7}$ | – | 0.00 | – | PD+LE |
| wht | P$4_12_12$ | tet | $\mu$ | $7.09\times10^{-7}$ | $7.09\times10^{-7}$ | $6.79\times10^{-7}$ | 0.00 | – | $\mu$ |
| wix | P$6_1$22 | hex | $\mu$ | $3.78\times10^{-8}$ | $3.78\times10^{-8}$ | $3.21\times10^{-8}$ | 0.00 | – | $\mu$ |
| wiy | P$6_1$22 | hex | $\mu$ | $3.15\times10^{-8}$ | $3.72\times10^{-8}$ | $2.37\times10^{-8}$ | 1.00 | – | $\mu$ |
| wiz | P$6_2$22 | hex | $\mu$ | $4.20\times10^{-8}$ | $4.20\times10^{-8}$ | $2.05\times10^{-8}$ | 0.00 | – | $\mu$ |
| wja | P$6_3$/m | hex | $\mu$ | – | – | $4.70\times10^{-7}$ | – | – | $\mu$ |
| wjd | P$\bar{6}$2c | hex | PD+LE | $6.47\times10^{-7}$ | $6.47\times10^{-7}$ | – | 0.00 | – | PD+LE |
| wje | P$6_2$22 | hex | $\mu$ | – | – | $2.23\times10^{-7}$ | – | – | $\mu$ |
| wjf | P$6_2$22 | hex | $\mu$ | – | – | $7.93\times10^{-8}$ | – | – | $\mu$ |
| wjg | P$6_2$22 | hex | $\mu$ | $2.92\times10^{-7}$ | $2.92\times10^{-7}$ | $2.47\times10^{-7}$ | 0.00 | – | $\mu$ |
| wjh | P$6_1$22 | hex | $\mu$ | $1.11\times10^{-7}$ | $1.11\times10^{-7}$ | $6.23\times10^{-8}$ | 0.00 | – | $\mu$ |
| wji | P$6_1$22 | hex | $\mu$ | $1.37\times10^{-7}$ | $1.37\times10^{-7}$ | $1.17\times10^{-7}$ | 0.00 | – | $\mu$ |
| wjk | P$6_2$22 | hex | $\mu$ | $1.67\times10^{-7}$ | $1.67\times10^{-7}$ | $1.66\times10^{-7}$ | 0.00 | – | $\mu$ |
| wjl | P$6_1$22 | hex | PD-only | $1.45\times10^{-7}$ | $1.51\times10^{-7}$ | – | 1.00 | 4.2% | PD-only |
| wjm | P$6_2$22 | hex | $\mu$ | – | – | $9.35\times10^{-8}$ | – | – | $\mu$ |
| wjn | P$6_2$22 | hex | $\mu$ | $2.19\times10^{-7}$ | $2.19\times10^{-7}$ | $1.19\times10^{-7}$ | 0.00 | – | $\mu$ |
| wjo | P$6_1$22 | hex | PD-only | $2.03\times10^{-7}$ | $2.60\times10^{-7}$ | – | 1.00 | 27.9% | PD-only |
| wjp | P$6_1$22 | hex | $\mu$ | $2.29\times10^{-7}$ | $2.29\times10^{-7}$ | $1.96\times10^{-7}$ | 0.00 | – | $\mu$ |
| wjq | P$6_2$22 | hex | $\mu$ | $2.97\times10^{-7}$ | $2.97\times10^{-7}$ | $2.21\times10^{-7}$ | 0.00 | – | $\mu$ |
| wjs | P$6_1$22 | hex | $\mu$ | $4.79\times10^{-7}$ | $5.07\times10^{-7}$ | $4.67\times10^{-7}$ | 1.00 | – | $\mu$ |
| wjt | P$6_2$22 | hex | $\mu$ | $2.08\times10^{-7}$ | $2.08\times10^{-7}$ | $1.31\times10^{-7}$ | 0.00 | – | $\mu$ |
| wju | P$6_2$22 | hex | $\mu$ | $4.36\times10^{-7}$ | $4.36\times10^{-7}$ | $3.44\times10^{-7}$ | 0.00 | – | $\mu$ |
| wjv | P$6_2$22 | hex | $\mu$ | $3.87\times10^{-7}$ | $3.87\times10^{-7}$ | $3.12\times10^{-7}$ | 0.00 | – | $\mu$ |
| wjw | P$6_2$22 | hex | $\mu$ | – | – | $5.50\times10^{-7}$ | – | – | $\mu$ |
| wjx | P$6_2$22 | hex | $\mu$ | $2.23\times10^{-7}$ | $2.23\times10^{-7}$ | $1.49\times10^{-7}$ | 0.00 | – | $\mu$ |
| wjy | P$6_1$22 | hex | $\mu$ | $2.25\times10^{-7}$ | $2.25\times10^{-7}$ | $1.35\times10^{-7}$ | 0.00 | – | $\mu$ |
| wjz | P$6_1$22 | hex | $\mu$ | $2.15\times10^{-7}$ | $2.15\times10^{-7}$ | $1.24\times10^{-7}$ | 0.00 | – | $\mu$ |
| wka | P$6_1$22 | hex | $\mu$ | – | – | $1.30\times10^{-7}$ | – | – | $\mu$ |
| wkb | P$6_1$22 | hex | $\mu$ | $5.13\times10^{-7}$ | $5.13\times10^{-7}$ | $4.21\times10^{-7}$ | 0.00 | – | $\mu$ |
| wkc | P$6_1$22 | hex | $\mu$ | $4.66\times10^{-7}$ | $4.66\times10^{-7}$ | $3.29\times10^{-7}$ | 0.00 | – | $\mu$ |
| wkd | P$6_1$22 | hex | $\mu$ | – | – | $4.66\times10^{-7}$ | – | – | $\mu$ |
| wke | P$6_3$22 | hex | $\mu$ | $2.27\times10^{-7}$ | $2.27\times10^{-7}$ | $1.36\times10^{-7}$ | 0.00 | – | $\mu$ |

| **Net** | **Space group** | **Sys.** | **Class** | $\lambda_{PD}$ | $\lambda_{LE}$ | $\lambda_{\mu}^{num}$ | $f_{inv}$ | **Window** | **Class, *r/l* = 0.05** |
|---|---|---|---|---|---|---|---|---|---|
| wkf | $P6_122$ | hex | PD-only | $6.42\times10^{-7}$ | $8.33\times10^{-7}$ | $7.93\times10^{-7}$ | 1.00 | 23.5% | PD-only |
| wkg | $P6_122$ | hex | PD+LE | $7.56\times10^{-7}$ | $7.56\times10^{-7}$ | $7.54\times10^{-7}$ | 0.00 | – | $\mu$ |
| wkh | $P6_122$ | hex | $\mu$ | $7.25\times10^{-7}$ | $7.25\times10^{-7}$ | $6.46\times10^{-7}$ | 0.00 | – | $\mu$ |
| wki | $P6_122$ | hex | $\mu$ | $6.91\times10^{-7}$ | $6.91\times10^{-7}$ | $5.80\times10^{-7}$ | 0.00 | – | $\mu$ |
| wkj | $P6_222$ | hex | $\mu$ | $2.59\times10^{-7}$ | $2.59\times10^{-7}$ | $1.56\times10^{-7}$ | 0.00 | – | $\mu$ |
| wkk | $P6_222$ | hex | $\mu$ | $3.17\times10^{-7}$ | $3.17\times10^{-7}$ | $2.21\times10^{-7}$ | 0.00 | – | $\mu$ |
| wkl | $P6_122$ | hex | $\mu$ | $2.44\times10^{-7}$ | $2.44\times10^{-7}$ | $1.68\times10^{-7}$ | 0.00 | – | $\mu$ |
| wkm | $P6_122$ | hex | $\mu$ | $2.88\times10^{-7}$ | $2.88\times10^{-7}$ | $1.85\times10^{-7}$ | 0.00 | – | $\mu$ |
| wkn | $P6_122$ | hex | $\mu$ | – | – | $1.97\times10^{-7}$ | – | – | $\mu$ |
| wko | $P6_122$ | hex | $\mu$ | $5.92\times10^{-7}$ | $5.92\times10^{-7}$ | $4.89\times10^{-7}$ | 0.00 | – | $\mu$ |
| wkp | $P6_122$ | hex | $\mu$ | $5.67\times10^{-7}$ | $5.67\times10^{-7}$ | $5.06\times10^{-7}$ | 0.00 | – | $\mu$ |
| wkq | $P6_122$ | hex | $\mu$ | $6.60\times10^{-7}$ | $6.60\times10^{-7}$ | $5.45\times10^{-7}$ | 0.00 | – | $\mu$ |
| wkt | $P6_122$ | hex | $\mu$ | – | – | $9.72\times10^{-7}$ | – | – | $\mu$ |
| wku | $P6_122$ | hex | PD+LE | $1.70\times10^{-6}$ | $1.70\times10^{-6}$ | $1.70\times10^{-6}$ | 1.00 | – | $\mu$ |
| wkv | $P6_122$ | hex | PD+LE | $8.49\times10^{-7}$ | $8.49\times10^{-7}$ | $1.11\times10^{-6}$ | 0.00 | – | PD+LE |
| wkw | $P6_1$ | hex | $\mu$ | $1.35\times10^{-6}$ | $1.35\times10^{-6}$ | $1.16\times10^{-6}$ | 0.00 | – | $\mu$ |
| wkx | $P6_122$ | hex | $\mu$ | $3.30\times10^{-7}$ | $3.30\times10^{-7}$ | $2.15\times10^{-7}$ | 0.00 | – | $\mu$ |
| wlc | $P\bar{3}1c$ | trig | $\mu$ | $1.81\times10^{-7}$ | $1.81\times10^{-7}$ | $1.12\times10^{-7}$ | 0.00 | – | $\mu$ |
| wld | $P\bar{3}c1$ | trig | $\mu$ | $4.26\times10^{-7}$ | $4.41\times10^{-7}$ | $4.12\times10^{-7}$ | 1.00 | – | $\mu$ |
| wli | $P3_221$ | trig | PD+LE | $4.70\times10^{-7}$ | $4.70\times10^{-7}$ | – | 0.00 | – | PD+LE |
| wlj | $P3_221$ | trig | PD+LE | $4.64\times10^{-7}$ | $4.64\times10^{-7}$ | $4.64\times10^{-7}$ | 0.00 | – | PD+LE |
| wma | $P3_221$ | trig | PD-only | $7.23\times10^{-7}$ | $8.59\times10^{-7}$ | $7.33\times10^{-7}$ | 1.00 | 1.4% | PD-only |
| wmb | $P3_221$ | trig | $\mu$ | $6.23\times10^{-7}$ | $6.23\times10^{-7}$ | $4.15\times10^{-7}$ | 0.00 | – | $\mu$ |
| wmd | $P6_122$ | hex | $\mu$ | $8.91\times10^{-7}$ | $8.91\times10^{-7}$ | $4.83\times10^{-7}$ | 0.00 | – | $\mu$ |
| wme | $P6_322$ | hex | $\mu$ | $7.29\times10^{-7}$ | $7.29\times10^{-7}$ | $6.60\times10^{-7}$ | 0.00 | – | $\mu$ |
| wmj | $P\bar{3}1c$ | trig | $\mu$ | $5.48\times10^{-7}$ | $5.48\times10^{-7}$ | $3.79\times10^{-7}$ | 0.00 | – | $\mu$ |
| wmk | $P\bar{3}1c$ | trig | $\mu$ | $5.47\times10^{-7}$ | $5.47\times10^{-7}$ | $5.44\times10^{-7}$ | 0.00 | – | $\mu$ |
| wml | $P\bar{3}c1$ | trig | $\mu$ | – | – | $2.44\times10^{-7}$ | – | – | $\mu$ |
| wmm | $P\bar{3}c1$ | trig | $\mu$ | $8.24\times10^{-7}$ | $9.77\times10^{-7}$ | $7.07\times10^{-7}$ | 1.00 | – | $\mu$ |
| wmn | $P\bar{3}c1$ | trig | $\mu$ | $5.61\times10^{-7}$ | $5.61\times10^{-7}$ | $5.34\times10^{-7}$ | 0.00 | – | $\mu$ |
| wmo | $P\bar{3}c1$ | trig | $\mu$ | $7.09\times10^{-7}$ | $7.09\times10^{-7}$ | $6.84\times10^{-7}$ | 0.00 | – | $\mu$ |
| wni | $P3_221$ | trig | $\mu$ | $1.70\times10^{-6}$ | $1.70\times10^{-6}$ | $1.24\times10^{-6}$ | 0.00 | – | $\mu$ |
| wnj | $P3_221$ | trig | PD-only | $1.08\times10^{-6}$ | $1.22\times10^{-6}$ | – | 1.00 | 12.9% | $\mu$ |
| wnk | $P3_221$ | trig | PD+LE | $8.88\times10^{-7}$ | $8.88\times10^{-7}$ | $1.11\times10^{-6}$ | 0.00 | – | PD+LE |
| wnl | $P6_222$ | hex | $\mu$ | $3.16\times10^{-7}$ | $3.16\times10^{-7}$ | $2.02\times10^{-7}$ | 0.00 | – | $\mu$ |
| wnr | $P3_221$ | trig | PD+LE | $6.93\times10^{-7}$ | $6.93\times10^{-7}$ | – | 0.00 | – | PD+LE |
| wns | $P\bar{3}c1$ | trig | PD+LE | $9.78\times10^{-7}$ | $9.78\times10^{-7}$ | $1.17\times10^{-6}$ | 0.00 | – | PD+LE |
| wnt | $P\bar{3}c1$ | trig | PD+LE | $8.05\times10^{-7}$ | $8.05\times10^{-7}$ | – | 0.00 | – | PD+LE |
| wys | R32 | trig | PD-only | $3.62\times10^{-7}$ | $4.48\times10^{-7}$ | $3.95\times10^{-7}$ | 1.00 | 9.1% | PD-only |
| wyt | R3m | trig | $\mu$ | $5.27\times10^{-7}$ | $5.27\times10^{-7}$ | $4.37\times10^{-7}$ | 0.00 | – | $\mu$ |
| xac | $Pn\bar{3}m$ | cub | PD+LE | $3.25\times10^{-7}$ | $3.25\times10^{-7}$ | – | 0.00 | – | PD+LE |
| xah | P4/mmm | tet | PD+LE | $1.04\times10^{-7}$ | $1.04\times10^{-7}$ | – | 0.00 | – | PD+LE |
| xai | P6/mmm | hex | $\mu$ | $7.55\times10^{-8}$ | $7.55\times10^{-8}$ | $4.11\times10^{-8}$ | 0.00 | – | $\mu$ |

| Net | Space group | Sys. | Class | $\lambda_{PD}$ | $\lambda_{LE}$ | $\lambda_{\mu}^{num}$ | $f_{inv}$ | Window | Class, *r*/*l* = 0.05 |
|---|---|---|---|---|---|---|---|---|---|
| xak | I4/mmm | tet | PD+LE | 1.33×10$^{-6}$ | 1.33×10$^{-6}$ | 1.33×10$^{-6}$ | 0.00 | – | PD+LE |
| xal | P4/mmm | tet | $\mu$ | 4.79×10$^{-6}$ | 4.79×10$^{-6}$ | 3.09×10$^{-6}$ | 0.00 | – | $\mu$ |
| xan | I4/mmm | tet | $\mu$ | – | – | 5.65×10$^{-7}$ | – | – | $\mu$ |
| xao | I4/mmm | tet | PD+LE | 1.04×10$^{-7}$ | 1.04×10$^{-7}$ | – | 0.00 | – | PD+LE |
| xaq | P$\bar{4}$3m | cub | $\mu$ | 2.52×10$^{-8}$ | 2.52×10$^{-8}$ | 1.31×10$^{-8}$ | 0.00 | – | $\mu$ |
| xat | P6$_3$/mmc | hex | PD+LE | 2.87×10$^{-6}$ | 2.87×10$^{-6}$ | 2.87×10$^{-6}$ | 0.00 | – | PD+LE |
| xax | P6$_3$/mmc | hex | $\mu$ | 3.96×10$^{-7}$ | 3.96×10$^{-7}$ | 3.63×10$^{-7}$ | 0.00 | – | $\mu$ |
| xay | P$\bar{4}$3m | cub | $\mu$ | – | – | 3.23×10$^{-6}$ | – | – | $\mu$ |
| xbe | I4/mmm | tet | PD+LE | 4.47×10$^{-8}$ | 4.47×10$^{-8}$ | – | 0.00 | – | PD+LE |
| xbg | Pm$\bar{3}$m | cub | $\mu$ | – | – | 8.68×10$^{-7}$ | – | – | $\mu$ |
| xbj | I4$_1$/amd | tet | $\mu$ | 3.28×10$^{-7}$ | 3.28×10$^{-7}$ | 3.26×10$^{-7}$ | 0.00 | – | $\mu$ |
| xbm | Pm$\bar{3}$m | cub | PD+LE | 2.61×10$^{-7}$ | 2.61×10$^{-7}$ | – | 0.00 | – | PD+LE |
| xbn | Pm$\bar{3}$m | cub | $\mu$ | – | – | 1.53×10$^{-6}$ | – | – | $\mu$ |
| xbo | Pm$\bar{3}$m | cub | $\mu$ | – | – | 1.84×10$^{-6}$ | – | – | $\mu$ |
| xbp | F$\bar{4}$3m | cub | PD+LE | 1.18×10$^{-6}$ | 1.18×10$^{-6}$ | 1.18×10$^{-6}$ | 0.00 | – | PD+LE |
| xbs | P$\bar{4}$3m | cub | PD+LE | 1.63×10$^{-6}$ | 1.63×10$^{-6}$ | – | 0.00 | – | PD+LE |
| xbt | Pn$\bar{3}$m | cub | $\mu$ | – | – | 1.56×10$^{-6}$ | – | – | $\mu$ |
| xby | P6$_3$/mmc | hex | PD+LE | 3.43×10$^{-6}$ | 3.43×10$^{-6}$ | – | 0.00 | – | PD+LE |
| xca | Immm | orth | $\mu$ | – | – | 8.07×10$^{-8}$ | – | – | $\mu$ |
| xci | P6/mmm | hex | $\mu$ | 5.06×10$^{-8}$ | 5.06×10$^{-8}$ | 2.68×10$^{-8}$ | 0.00 | – | $\mu$ |
| xda | Immm | orth | $\mu$ | – | – | 1.01×10$^{-7}$ | – | – | $\mu$ |
| xfa | P4$_1$2$_1$2 | tet | PD+LE | 7.69×10$^{-7}$ | 7.69×10$^{-7}$ | – | 0.00 | – | PD+LE |
| xft | I4$_1$md | tet | PD-only | 1.75×10$^{-7}$ | – | 2.34×10$^{-7}$ | 1.00 | 33.4% | PD-only |
| xfu | P4$_2$/n | tet | $\mu$ | 3.21×10$^{-7}$ | 3.21×10$^{-7}$ | 2.36×10$^{-7}$ | 0.00 | – | $\mu$ |
| xfw | P$\bar{4}$c2 | tet | $\mu$ | 2.33×10$^{-7}$ | 2.33×10$^{-7}$ | 1.18×10$^{-7}$ | 0.00 | – | $\mu$ |
| xfx | I$\bar{4}$m2 | tet | $\mu$ | – | – | 2.28×10$^{-7}$ | – | – | $\mu$ |
| xfy | P$\bar{4}$n2 | tet | $\mu$ | 2.81×10$^{-7}$ | 2.81×10$^{-7}$ | 2.44×10$^{-7}$ | 0.00 | – | $\mu$ |
| xux | I$\bar{4}$m2 | tet | PD-only | 3.44×10$^{-8}$ | – | 3.59×10$^{-8}$ | 1.00 | 4.4% | PD-only |
| yan | I4$_1$22 | tet | $\mu$ | – | – | 1.45×10$^{-7}$ | – | – | $\mu$ |
| yav | P6/mmm | hex | PD+LE | 3.79×10$^{-7}$ | 3.79×10$^{-7}$ | – | 0.00 | – | PD+LE |
| ybh | P6$_3$/mmc | hex | $\mu$ | 1.97×10$^{-7}$ | 1.97×10$^{-7}$ | 1.90×10$^{-7}$ | 0.00 | – | $\mu$ |
| yfg | P4$_1$22 | tet | $\mu$ | 1.79×10$^{-7}$ | 1.79×10$^{-7}$ | 1.76×10$^{-7}$ | 0.00 | – | $\mu$ |
| yfh | P4$_1$2$_1$2 | tet | PD+LE | 2.07×10$^{-7}$ | 2.07×10$^{-7}$ | 2.08×10$^{-7}$ | 1.00 | – | PD+LE |
| yfx | I4/m | tet | $\mu$ | 6.35×10$^{-7}$ | 6.35×10$^{-7}$ | 5.38×10$^{-7}$ | 0.00 | – | $\mu$ |
| yfz | P4$_2$/nnm | tet | $\mu$ | – | – | 1.10×10$^{-7}$ | – | – | $\mu$ |
| yyz | P6$_1$22 | hex | PD-only | 3.15×10$^{-7}$ | 3.78×10$^{-7}$ | – | 1.00 | 20.1% | PD-only |
| zbb | C2/m | mono | PD-only | 1.75×10$^{-7}$ | 1.77×10$^{-7}$ | – | 1.00 | 1.3% | PD-only |
| zbc | C2/m | mono | PD+LE | 1.85×10$^{-7}$ | 1.85×10$^{-7}$ | – | 1.00 | – | PD+LE |
| zbj | P$\bar{1}$ | tric | PD+LE | 1.74×10$^{-7}$ | 1.75×10$^{-7}$ | – | 1.00 | – | PD+LE |
| zbm | P2$_1$/n | mono | $\mu$ | – | – | 2.74×10$^{-7}$ | – | – | $\mu$ |
| zbo | C2 | mono | PD-only | 1.98×10$^{-7}$ | 1.99×10$^{-7}$ | – | 1.00 | 0.7% | PD-only |
| zbt | Cmmm | orth | $\mu$ | 1.72×10$^{-7}$ | 1.72×10$^{-7}$ | 1.03×10$^{-7}$ | 0.00 | – | $\mu$ |
| zby | P2$_1$/n | mono | $\mu$ | – | – | 3.40×10$^{-7}$ | – | – | $\mu$ |

| Net | Space group | Sys. | Class | $\lambda_{PD}$ | $\lambda_{LE}$ | $\lambda_\mu^{num}$ | $f_{inv}$ | Window | Class, *r*/*l* = 0.05 |
|---|---|---|---|---|---|---|---|---|---|
| zca | P2$_1$/n | mono | $\mu$ | – | – | 2.27×10$^{-7}$ | – | – | $\mu$ |
| zce | P2$_1$/n | mono | $\mu$ | – | – | 3.90×10$^{-7}$ | – | – | $\mu$ |
| zfa | P4$_2$2$_1$2 | tet | $\mu$ | 4.24×10$^{-7}$ | 4.24×10$^{-7}$ | 3.81×10$^{-7}$ | 0.00 | – | $\mu$ |
| zfb | P4$_2$/mbc | tet | $\mu$ | 4.69×10$^{-7}$ | 4.69×10$^{-7}$ | 4.67×10$^{-7}$ | 0.00 | – | $\mu$ |
| zfr | P4$_1$22 | tet | PD+LE | 1.14×10$^{-7}$ | 1.14×10$^{-7}$ | 1.15×10$^{-7}$ | 0.00 | – | PD+LE |
| zfs | P4$_1$22 | tet | PD+LE | 1.83×10$^{-7}$ | 1.83×10$^{-7}$ | – | 0.00 | – | PD+LE |
| zft | P4$_1$2$_1$2 | tet | PD+LE | 4.70×10$^{-7}$ | 4.70×10$^{-7}$ | – | 0.00 | – | PD+LE |
| zfu | P4$_1$2$_1$2 | tet | PD+LE | 3.67×10$^{-7}$ | 3.67×10$^{-7}$ | – | 0.00 | – | PD+LE |
| zfz | P3$_2$21 | trig | $\mu$ | 3.29×10$^{-7}$ | 3.30×10$^{-7}$ | 1.64×10$^{-7}$ | 1.00 | – | $\mu$ |
| zlg | P4$_2$/mmc | tet | PD+LE | 3.56×10$^{-7}$ | 3.56×10$^{-7}$ | – | 0.00 | – | PD+LE |
| zll | I4/mmm | tet | PD+LE | 9.01×10$^{-7}$ | 9.01×10$^{-7}$ | – | 0.00 | – | PD+LE |
| zmc | P6m2 | hex | PD+LE | 1.49×10$^{-7}$ | 1.49×10$^{-7}$ | – | 0.00 | – | PD+LE |
| znz | P6$_3$/mmc | hex | $\mu$ | 4.47×10$^{-7}$ | 4.47×10$^{-7}$ | 3.29×10$^{-7}$ | 0.00 | – | $\mu$ |
| zra | P6/mmm | hex | $\mu$ | – | – | 6.63×10$^{-6}$ | – | – | $\mu$ |
| zsd | I2/a | mono | PD+LE | 5.02×10$^{-7}$ | 5.02×10$^{-7}$ | – | 0.00 | – | PD+LE |
| zst | Pnna | orth | $\mu$ | – | – | 1.86×10$^{-7}$ | – | – | $\mu$ |
| zyb | P2$_1$/n | mono | PD+LE | 1.83×10$^{-7}$ | 1.83×10$^{-7}$ | – | 0.00 | – | PD+LE |
| zyl | Cmc2$_1$ | orth | $\mu$ | – | – | 8.64×10$^{-8}$ | – | – | $\mu$ |
| zzz | P4/n | tet | PD+LE | 1.16×10$^{-7}$ | 1.16×10$^{-7}$ | – | 0.00 | – | PD+LE |

Sys., crystal system (cub, hex, tet, trig, orth, mono, tric). Uniform rigid-jointed circular struts, *r*/*l* = 0.02 unless stated; pressures in units of the strut Young's modulus $E_s$. $\lambda_\mu^{num}$: the lowest Bloch threshold found on the wavevector grid outside the zone-centre exclusion and of the Γ-optical modes, a numerical estimate of the finite-wavevector and optical channel (the mathematical $\lambda_\mu$ also includes the acoustic limit, so $\lambda_\mu \leq \lambda_{LE}$). Classes: PD+LE, positive definiteness and strong ellipticity lost together; PD-only, positive definiteness lost with ellipticity intact; $\mu$, finite-wavevector or unabsorbed Γ-optical mode first; PD+$\mu$, positive definiteness and a finite-wavevector mode lost at the same load. "–" under $\lambda_{PD}$ or $\lambda_{LE}$: the homogeneous root lies beyond an earlier Γ-optical instability of the cell and is not evaluated; under $\lambda_\mu^{num}$: the lowest Bloch root sits at the acoustic origin, so there is no separate finite-wavevector threshold. $f_{inv}$: fraction of the PD failing strain in the totally symmetric sector of the point group. Classes use the unrounded thresholds. Window: $\lambda_{loc}/\lambda_{PD} - 1$ for PD-only nets (78 of 78 have a defined $\lambda_{loc}$). The class is the same at both radii for 867 nets. Every net that is PD-only at either radius and 40 PD+LE controls were rechecked on an $8^3$ Bloch grid with a four times smaller zone-centre exclusion. Preload equilibrium residual relative to the member end forces: at most 1.4×10$^{-12}$. $\lambda_{LE}$ from 3,000 quasi-uniform directions refined from the ten lowest.

## Supplementary Table 11 | Sensitivity of the atlas classes to the tie tolerance

| Tie tolerance | PD-only | of which totally symmetric | PD+LE | PD+$\mu$ | $\mu$ | Window median | PD-only, $r/l$ = 0.05 |
|---|---|---|---|---|---|---|---|
| $10^{-8}$ | 104 | 104 | 258 | 0 | 514 | 3.6% | 99 |
| $10^{-6}$ | 98 | 98 | 264 | 0 | 514 | 4.2% | 95 |
| $10^{-4}$ | 87 | 87 | 275 | 0 | 514 | 5.6% | 86 |
| 0.1% | 80 | 80 | 284 | 0 | 512 | 7.1% | 77 |
| 0.3% | 78 | 78 | 290 | 0 | 508 | 7.6% | 74 |
| 1% | 70 | 70 | 307 | 4 | 495 | 8.9% | 67 |
| 3% | 54 | 54 | 337 | 9 | 476 | 18.0% | 52 |

A net is PD-only when $\lambda_{PD} < (1 - \text{tolerance}) \min(\lambda_{LE}, \lambda_{\mu})$; the main text uses 0.3%. Within the tolerance, a tie is classed PD+LE before PD+$\mu$, and $\mu$ only when $\lambda_{PD}$ exceeds $\lambda_{\mu}$ by more than the tolerance. Counts at $r/l = 0.02$ unless stated; PD-only, PD+LE, PD+$\mu$ and $\mu$ add up to 876 in every row, and the totally symmetric nets are a subset of PD-only; thresholds after the $8^3$ Bloch recheck where it was performed (Methods). Every PD-only net at every tolerance fails with a totally symmetric strain. At the tightest tolerance ($10^{-8}$) the cubic net pyu, whose $\lambda_{LE}$ and $\lambda_{\mu}$ lie within a relative $3\times10^{-6}$ of $\lambda_{PD}$, counts as PD-only by a pure dilation; at 0.3% it is a tie.

## Supplementary Table 12 | Mesh convergence of the atlas thresholds

| Threshold | Nets compared, 2→8 | Median, 2→8 | Max, 2→8 | Max, 4→8 |
|---|---|---|---|---|
| $\lambda_{PD}$ | 122 | 0.14% | 1.52% | 0.12% |
| $\lambda_{LE}$ | 111 | 0.15% | 1.52% | 0.12% |
| $\lambda_{\mu}$ | 71 | 0.16% | 1.68% | 0.14% |

$|\lambda(8)/\lambda(m) - 1|$ for $m = 2$ or 4 Hermite elements per parent strut, over 138 nets at $r/l = 0.02$: every net that is PD-only on the $8^3$ grid and 60 others drawn without replacement. Thresholds undefined at one mesh (beyond an earlier Γ-optical instability) are not compared. Class changes between 2, 4 and 8 elements: 0.

## Supplementary Table 13 | Mesh convergence of the beam initial-stress tangent

| Elements per strut | Euler column, rel. error | Portal column, rel. error | Loaded portal, critical factor |
|---|---|---|---|
| 1 | $2.16\times10^{-1}$ | $1.32\times10^{-2}$ | 7.4569 |
| 2 | $7.52\times10^{-3}$ | $7.52\times10^{-3}$ | 7.4136 |
| 4 | $5.12\times10^{-4}$ | $5.12\times10^{-4}$ | 7.3920 |
| 8 | $3.28\times10^{-5}$ | $3.28\times10^{-5}$ | 7.3905 |
| 16 | $2.06\times10^{-6}$ | $2.06\times10^{-6}$ | 7.3904 |

Pin-ended Euler column and portal frame with a rigid top beam against the exact load $\pi^2 EI/H^2$ per column; flexible portal (two unit columns and a top beam, $r = 0.03$) with vertical loads $-EI$ at both top nodes and a horizontal load $0.2EI$ at one, as a mesh benchmark. At one element the analytic Hessian of the loaded portal agrees with finite differences to $3.4\times10^{-8}$. The atlas uses two elements per strut; for 138 nets refining to eight changes the thresholds by at most 1.68% (Supplementary Table 12).

**Supplementary Table 14 | Element-level and self-stress identities of the consistent tangent**

| Check | Cases | Maximum error |
|---|---|---|
| Axial stiffness | 8 | $4.0\times10^{-16}$ |
| Tangent symmetry | 8 | $7.2\times10^{-17}$ |
| End reversal | 8 | $5.6\times10^{-16}$ |
| Linear stiffness | 8 | $4.8\times10^{-16}$ |
| Finite rigid rotation (objectivity) | 8 | $6.7\times10^{-16}$ |
| Self-stress force balance | 10 | $7.3\times10^{-17}$ |
| Self-stress rotation Ward identity | 10 | $9.7\times10^{-17}$ |
| Self-stress rigid-rotation energy | 10 | $1.8\times10^{-17}$ |
| Finite-rotation jet, $\varepsilon = 10^{-2}$ | 1 | $6.4\times10^{-6}$ |
| Finite-rotation jet, $\varepsilon = 10^{-3}$ | 1 | $6.3\times10^{-9}$ |
| Finite-rotation jet, $\varepsilon = 10^{-4}$ | 1 | $6.3\times10^{-12}$ |

Eight test elements carrying axial force, shears, torsion and end moments, and ten equilibrium self-stress states.

The finite-rotation jet error falls as $\varepsilon^3$, as expected for a tangent exact to second order.

**Supplementary Table 15 | Rotation invariance and long-wavelength consistency on loaded nets**

| Net | Load | Ward | Initial-stress Ward | Schur rel. error | Rotation energy | Bloch $k^2$ coefficient error | Bloch rel. error, $k$ = 0.02 / 0.0025 |
|---|---|---|---|---|---|---|---|
| pcu | $2.18\times10^{-4}$ | $5.5\times10^{-20}$ | $1.2\times10^{-16}$ | $5.0\times10^{-18}$ | $3.7\times10^{-21}$ | $6.4\times10^{-17}$ | $1.4\times10^{-3}$ / $2.3\times10^{-5}$ |
| utp | $3.83\times10^{-5}$ | $8.1\times10^{-18}$ | $1.3\times10^{-16}$ | $1.5\times10^{-15}$ | $2.5\times10^{-19}$ | $1.2\times10^{-15}$ | $5.3\times10^{-2}$ / $8.7\times10^{-4}$ |
| cco | $2.57\times10^{-4}$ | $8.4\times10^{-18}$ | $4.2\times10^{-17}$ | $3.0\times10^{-16}$ | $1.1\times10^{-18}$ | $4.3\times10^{-16}$ | $1.8\times10^{-3}$ / $2.8\times10^{-5}$ |
| cyg | $4.59\times10^{-3}$ | $5.7\times10^{-18}$ | $3.4\times10^{-15}$ | $1.4\times10^{-15}$ | $7.0\times10^{-19}$ | $2.4\times10^{-14}$ | $1.1\times10^{-3}$ / $1.7\times10^{-5}$ |
| tfk | $3.18\times10^{-3}$ | $9.3\times10^{-18}$ | $3.7\times10^{-15}$ | $7.8\times10^{-16}$ | $6.8\times10^{-19}$ | $7.9\times10^{-16}$ | $2.2\times10^{-2}$ / $3.5\times10^{-4}$ |
| srs | $6.06\times10^{-5}$ | $1.1\times10^{-17}$ | $1.1\times10^{-16}$ | $6.9\times10^{-16}$ | $6.1\times10^{-19}$ | $7.2\times10^{-16}$ | $1.1\times10^{-2}$ / $1.7\times10^{-4}$ |

Each net carries a hydrostatic preload at 0.4 times the lower of its $\lambda_{PD}$ and $\lambda_{\Gamma}$. Load: that load factor in the net-specific normalization of the solver, not in units of $E_s$. Ward: rigid-rotation residual of the full tangent; initial-stress Ward: the same residual of the initial-stress tangent alone; rotation energy: energy of a rigid rotation of the preloaded frame, relative to its elastic energy scale. Schur: macroscopic tangent from static condensation against the direct relaxed tangent; Bloch $k^2$ coefficient: analytic second-order Bloch term against the acoustic tensor; the last column shows the quadratic convergence of the lowest Bloch eigenvalue to the acoustic-tensor prediction.

**Supplementary Table 16 | Zero-velocity acoustic waves on the stability boundaries of repeated non-trivial representations**

| Class: irreps | Stability condition = 0 | Nullity | \|tr *e*\| | \|det *e*\| | Direction *n* | \|min eig *A*(*n*)\| | Other eig *A*(*n*) |
|---|---|---|---|---|---|---|---|
| tetragonal 4, $\bar{4}$, 4/m: 2B | $(c_{11} - c_{12})c_{66} - 2c_{16}^2$ | 1 | $2.8\times10^{-16}$ | $3.1\times10^{-16}$ | (0.924, 0.383, 0.000) | $4.4\times10^{-16}$ | 4.00, 13.00 |
| trigonal 32: 2E | $(c_{11} - c_{12})c_{44} - 2c_{14}^2 - 2c_{15}^2$ | 2 | $2.0\times10^{-16}$ | $1.7\times10^{-16}$ | (0.612, 0.354, 0.707) | $1.5\times10^{-15}$ | 4.38, 12.62 |
| trigonal 3: 2E | $(c_{11} - c_{12})c_{44} - 2c_{14}^2 - 2c_{15}^2$ | 2 | $7.9\times10^{-16}$ | $2.6\times10^{-17}$ | (0.953, −0.304, 0.000) | $2.6\times10^{-15}$ | 10.00, 12.00 |
| monoclinic 2/m, 2, m: 2B | $c_{44}c_{66} - c_{46}^2$ | 1 | 0 | 0 | (0.000, 1.000, 0.000) | $1.8\times10^{-15}$ | 11.00, 13.00 |

Nullity: dimension of the null space of the matrix on the boundary. Each row is an elastic matrix (Voigt order xx, yy, zz, yz, xz, xy; components in Supplementary Table 17) placed exactly on the stability boundary of the listed repeated representation, which Cowley[11] listed as type 0 and Terhune et al.[12] reassigned to type I. *e* is the zero-determinant strain in the null space, *n* the critical direction and *A*(*n*) the acoustic tensor. A further 96 random boundary matrices (24 per case) give $|\text{min eig } A| \leq 4.5\times10^{-15}$. Controls: on the totally symmetric boundaries the acoustic tensor stays positive (minimum eigenvalue 0.500 for cubic $A_1$, 0.091 for a tetragonal $A_1$ boundary mixing $e_{xx} + e_{yy}$ with $e_{zz}$).

**Supplementary Table 17 | Elastic matrices on the stability boundaries of the repeated representations**

| Class: irreps | Non-zero elastic constants $c_{ij}$ (Voigt, $i \le j$) |
|---|---|
| tetragonal 4, $\bar{4}$, 4/m: 2B | $c_{11}$ = 10, $c_{12}$ = 4, $c_{13}$ = 2, $c_{16}$ = 3, $c_{22}$ = 10, $c_{23}$ = 2, $c_{26}$ = −3, $c_{33}$ = 12, $c_{44}$ = 4, $c_{55}$ = 4, $c_{66}$ = 3 |
| trigonal 32: 2E | $c_{11}$ = 10, $c_{12}$ = 4, $c_{13}$ = 2, $c_{14}$ = 3, $c_{22}$ = 10, $c_{23}$ = 2, $c_{24}$ = −3, $c_{33}$ = 12, $c_{44}$ = 3, $c_{55}$ = 3, $c_{56}$ = 3, $c_{66}$ = 3 |
| trigonal 3: 2E | $c_{11}$ = 12, $c_{12}$ = 2, $c_{13}$ = 2, $c_{14}$ = 3, $c_{15}$ = 4, $c_{22}$ = 12, $c_{23}$ = 2, $c_{24}$ = −3, $c_{25}$ = −4, $c_{33}$ = 12, $c_{44}$ = 5, $c_{46}$ = −4, $c_{55}$ = 5, $c_{56}$ = 3, $c_{66}$ = 5 |
| monoclinic 2/m, 2, m: 2B | $c_{11}$ = 10, $c_{12}$ = 1, $c_{13}$ = 2, $c_{15}$ = 1, $c_{22}$ = 11, $c_{23}$ = 1, $c_{25}$ = 2, $c_{33}$ = 12, $c_{35}$ = 1, $c_{44}$ = 4, $c_{46}$ = 6, $c_{55}$ = 8, $c_{66}$ = 9 |

The matrices of Supplementary Table 16, in Voigt order xx, yy, zz, yz, xz, xy with engineering shear strains; each lies exactly on the stability boundary of the listed representation.

## Supplementary Table 18 | Prestress transfer over independent backgrounds: fixed central window versus domain size

| Sample $L$ | Domain | n | Fixed-window wrench error, mean ± s.d. (max) | Fixed-support form error | Growing-core wrench error | $s_R$ | Deviatoric fraction | Signed mode correction |
|---|---|---|---|---|---|---|---|---|
| 64 | 16×16 | 8 | 1.624 ± 0.577% (2.434%) | 0.497 ± 0.206% | 1.907% | 1.00037 | 0.0306% | −0.0123 ± 0.0282% |
| 64 | 32×32 | 8 | 0.663 ± 0.232% (0.970%) | 0.207 ± 0.113% | 1.254% | 1.00071 | 0.0116% | −0.0031 ± 0.0120% |
| 64 | 48×48 | 8 | 0.271 ± 0.080% (0.396%) | 0.073 ± 0.030% | 0.818% | 1.00051 | 0.0046% | −0.0060 ± 0.0046% |
| 96 | 16×16 | 8 | 1.507 ± 0.372% (2.097%) | 0.453 ± 0.098% | 1.879% | 1.00092 | 0.0346% | 0.0348 ± 0.0384% |
| 96 | 32×16 | 8 | 1.206 ± 0.282% (1.585%) | 0.373 ± 0.109% | 2.037% | 1.00076 | 0.0551% | −0.0107 ± 0.0359% |
| 96 | 32×32 | 8 | 0.879 ± 0.190% (1.210%) | 0.259 ± 0.079% | 1.463% | 1.00061 | 0.0142% | −0.0148 ± 0.0120% |
| 96 | 48×48 | 8 | 0.549 ± 0.164% (0.829%) | 0.160 ± 0.032% | 1.139% | 1.00065 | 0.0097% | −0.0107 ± 0.0064% |
| 128 | 48×48 | 8 | 0.520 ± 0.065% (0.597%) | 0.132 ± 0.036% | 1.145% | 1.00069 | 0.0050% | 0.0020 ± 0.0135% |

Eight independent random backgrounds per $L$; square domains of the checkerboard are nested in the same background, and the 32×16 domains reuse the $L = 96$ backgrounds. Fixed window: the central 4×4 tiles. Wrench error: largest deviation of the full member end wrench (axial and shear forces, end moments) from $s_R$ times that of the checkerboard, normalized by the largest reference wrench. Fixed-support form error: sup $|v^T(Q^\omega - Q^P \text{ at } s_R p)v|/(v^T K v)$ over fields supported inside the fixed window, with every touching interaction. Growing core: the domain without ⌈side/4⌉ edge layers. Deviatoric fraction: $\|\text{dev}\,\Sigma\|/(\sqrt{2}\,|p|)$ of the mean domain stress. Signed mode correction as in Supplementary Table 19. Fixed-window errors fall from $R = 16$ to 48 in all eight backgrounds at $L = 64$ and 96; the difference between $L = 64$ and $L \geq 96$ at $R = 48$ is a finite-box effect.

**Supplementary Table 19 | Prestress inside planted checkerboard domains: maximum, central, rms and mode-weighted deviations**

| Frame | Sample $L$ | Domain $R$ | Core max, $w = 1$ | $w = 2$ | $w = 4$ | Fixed centre max | Core rms, $w = 2$ | Signed mode correction |
|---|---|---|---|---|---|---|---|---|
| 2D | 64 | 8 | 1.90% | 1.87% | – | 1.87% | 0.66% | −0.1454% |
| 2D | 64 | 16 | 1.42% | 1.37% | 1.20% | 1.07% | 0.36% | −0.0068% |
| 2D | 64 | 24 | 2.10% | 1.93% | 1.46% | 0.67% | 0.36% | +0.0165% |
| 2D | 64 | 32 | 1.85% | 1.70% | 1.22% | 0.46% | 0.30% | −0.0234% |
| 2D | 64 | 40 | 1.99% | 1.83% | 1.60% | 0.23% | 0.28% | −0.0016% |
| 2D | 64 | 48 | 2.00% | 1.92% | 1.67% | 0.35% | 0.24% | −0.0026% |
| 2D | 96 | 8 | 1.98% | 1.74% | – | 1.74% | 0.71% | −0.0868% |
| 2D | 96 | 16 | 1.83% | 1.78% | 1.66% | 1.20% | 0.44% | −0.0816% |
| 2D | 96 | 24 | 2.20% | 1.96% | 1.73% | 1.17% | 0.46% | +0.0146% |
| 2D | 96 | 32 | 2.53% | 2.43% | 2.10% | 0.80% | 0.37% | −0.0396% |
| 2D | 96 | 40 | 2.52% | 2.26% | 1.92% | 0.44% | 0.31% | −0.0183% |
| 2D | 96 | 48 | 2.74% | 2.48% | 2.16% | 0.41% | 0.29% | −0.0133% |
| 3D | 10 | 4 | 0.84% | – | – | 0.84% | – | −0.0441% |
| 3D | 10 | 5 | 0.87% | 0.46% | – | 0.60% | 0.19% | −0.0010% |
| 3D | 10 | 6 | 0.91% | 0.83% | – | 0.83% | 0.22% | −0.0260% |
| 3D | 10 | 7 | 0.84% | 0.60% | – | 0.60% | 0.09% | −0.0320% |
| 3D | 10 | 8 | 0.73% | 0.50% | – | 0.28% | 0.08% | −0.0314% |

Hydrostatic preload on one random background per $L$ with a central domain of the weakest periodic motif. Member forces are divided by the applied pressure; deviations $|n^{\omega} - s_R n^P|$ use the domain mean-pressure factor $s_R$ and are normalized by the largest reference member force. The core excludes $w$ tiles (cells) from the domain edge; the fixed centre is the central 4×4 tiles (2×2×2 cells). Signed mode correction: $[Q^{\omega} - Q^P$ at $s_R p_c]$ evaluated on the compact critical mode $v$ of the domain, divided by $v^T K v$, with the full beam tangent. Lemma 3 remarks: for the finite-range form $\Phi = \rho_y + (1 - \lambda)\rho_x + \rho_x^2$, $\rho_i = 4 \sin^2(k_i/2)$, whose acoustic expansion along $x$ has the form $-\delta k_x^2 + k_x^4$, the exact square-patch excess times $R$ tends to a constant (6.41 at $R = 256$), an $R^{-1}$ law; for $\Phi = 1 - \lambda + \rho_x^2$, whose minimum has zero Hessian, the excess is positive and falls as $R^{-4}$. Theorem 4 remark: the Lemma 1 windows of the no-checker rule have negative Rayleigh values already at zero load (window radius 2–8), so they give no certificate.

**Supplementary Table 20 | Weakest 4×4 patterns that contain a 2×2 checkerboard block**

| Rank | Tile states (rows) | Threshold / checkerboard | First instability |
|---|---|---|---|
| 1 | 0101 / 1010 / 0101 / 1010 | 1.0000 | Γ-optical |
| 2 | 0001 / 1010 / 0101 / 1010 | 1.0537 | Γ-optical |
| 3 | 0001 / 1010 / 0101 / 1110 | 1.0824 | Γ-optical |
| 4 | 0001 / 0010 / 0101 / 1010 | 1.1053 | Γ-optical |
| 5 | 0000 / 0101 / 1010 / 0101 | 1.1072 | Γ-optical |
| 6 | 0001 / 1000 / 0101 / 1010 | 1.1088 | Γ-optical |
| 7 | 0001 / 1010 / 0101 / 0110 | 1.1107 | Γ-optical |
| 8 | 0001 / 1010 / 0101 / 1011 | 1.1112 | Γ-optical |

All 52,258 4×4 patterns with at least one checkerboard block, reduced to 511 orbits under translations, the point group and the tile-state exchange it induces, scanned at fixed hydrostatic pressure on a 4×4 wavevector grid; the 20 weakest were rescanned with local wavevector refinement, which changed no threshold. The 21st weakest unrefined orbit lies at 1.142. Rank 1 is the checkerboard itself; its $\lambda_{PD}$, $\lambda_{LE}$ and $\lambda_{\Gamma}$ agree to $10^{-13}$ relative, so its fixed-strain and fixed-pressure thresholds coincide. Together with the 158 orbits without a checkerboard block (Supplementary Table 9) this covers all 65,536 4×4 patterns.

## Supplementary Table 21 | Periodic configurations of the three-dimensional frame

| Supercell | Pattern | First class | Critical pressure / weakest |
|---|---|---|---|
| 2×2×1 | 0/1\|1/0 | Γ-optical | 1.0000 |
| 2×2×2 | 00/01\|11/00 | PD+LE | 1.1955 |
| 2×2×1 | 0/0\|0/1 | Γ-optical | 1.2243 |
| 2×1×1 | 0\|1 | finite *k* | 1.2306 |
| 1×1×1 | 0 | PD+LE | 1.2830 |
| 2×2×2 | 00/01\|11/10 | PD+LE | 1.3118 |
| 2×2×2 | 00/00\|01/11 | PD+LE | 1.3299 |
| 2×2×2 | 00/01\|01/00 | PD+LE | 1.3896 |
| 2×2×2 | 00/01\|11/01 | PD+LE | 1.3939 |
| 2×2×2 | 00/01\|10/00 | PD+LE | 1.4135 |
| 2×2×2 | 00/00\|00/01 | PD+LE | 1.4172 |
| 2×2×2 | 00/01\|01/10 | PD+LE | 1.4269 |
| 2×2×2 | 00/01\|01/11 | PD+LE | 1.4321 |
| 2×2×2 | 00/01\|10/11 | PD+LE | 1.4340 |
| 2×2×2 | 00/00\|01/10 | Γ-optical | 1.4353 |
| 2×2×2 | 01/10\|10/01 | Γ-optical | 1.4486 |
| 2×1×2 | 00\|01 | finite *k* | 1.5070 |
| 2×2×2 | 00/01\|10/01 | PD+LE | 1.5203 |
| 2×2×2 | 00/01\|01/01 | PD+LE | 1.5923 |
| 2×2×2 | 01/01\|01/10 | Γ-optical | 1.6167 |
| 2×1×2 | 01\|10 | finite *k* | 1.6228 |
| 1×1×2 | 01 | finite *k* | 1.7941 |

The 22 symmetry orbits of periodic binary states on supercells up to 2×2×2 (256 raw assignments), sorted by threshold. Patterns list states by $x$ index (|), $y$ index (/) and $z$ index. The weakest member, the layer-uniform checkerboard, fails by an in-plane Γ-optical mode.

## Supplementary Table 22 | Domain-size scan in the two-dimensional frame (Fig. 2b)

| Sample | Domain *R* | Full sample | Compact bound | Energy in domain | $s_P$ | Corrected compact excess |
|---|---|---|---|---|---|---|
| 32×32 | 2 | 1.2121 | 2.4127 | 0.6% | 0.9997 | 1.4120 |
| 32×32 | 3 | 1.2042 | 2.0329 | 10.3% | 1.0010 | 1.0349 |
| 32×32 | 4 | 1.1851 | 1.6674 | 31.6% | 1.0014 | 0.6697 |
| 32×32 | 6 | 1.1463 | 1.3275 | 63.1% | 1.0000 | 0.3275 |
| 32×32 | 8 | 1.0997 | 1.1910 | 80.9% | 1.0001 | 0.1911 |
| 32×32 | 12 | 1.0524 | 1.0862 | 91.6% | 0.9999 | 0.0860 |
| 32×32 | 16 | 1.0342 | 1.0484 | 96.7% | 1.0009 | 0.0493 |
| 32×32 | 24 | 1.0169 | 1.0213 | 98.8% | 1.0005 | 0.0218 |
| 64×64 | 16 | 1.0338 | 1.0483 | 96.6% | 1.0006 | 0.0489 |
| 64×64 | 24 | 1.0169 | 1.0212 | 98.9% | 1.0009 | 0.0221 |
| 64×64 | 32 | 1.0097 | 1.0115 | 99.5% | 1.0006 | 0.0121 |
| 64×64 | 40 | 1.0061 | 1.0072 | 99.7% | 1.0007 | 0.0079 |
| 64×64 | 48 | 1.0043 | 1.0049 | 99.8% | 1.0006 | 0.0055 |
| 96×96 | 64 | 1.0022 | 1.0024 | 99.9% | 1.0006 | 0.0031 |

Checkerboard domains of size *R*×*R* planted in random backgrounds (one background per sample size), at fixed hydrostatic pressure. Thresholds in units of the checkerboard threshold. Compact bound: trial fields confined to the domain. Without a domain, the 32×32 background of Fig. 2a gives 1.2128.

**Supplementary Table 23 | Periodic configurations of the two-dimensional decorated Truchet frame**

| Pattern | Critical load / checkerboard (prescribed strain) | Critical wavevector *k* |
|---|---|---|
| 01/10 | 1.0000 | (0, 0) |
| 001/110 | 1.0817 | (0, 0.5) |
| 001/010/100 | 1.1550 | (0.5, 0.5) |
| 001/010 | 1.1570 | (0, 0.5) |
| 001/001/110 | 1.1612 | (0.5, 0.5) |
| 001/010/101 | 1.1613 | (0.5, 0.5) |
| 001/011/110 | 1.1651 | (0.5, 0.5) |
| 001/100/010 | 1.1663 | (0.5, 0.5) |
| 000/001/110 | 1.2114 | (0.5, 0.5) |
| 000/011/101 | 1.2122 | (0.5, 0.5) |
| 000/011/110 | 1.2182 | (0.5, 0.5) |
| 00/01 | 1.2316 | (0, 0) |
| 00/11 | 1.2391 | (0, 0) |
| 000/111 | 1.2391 | (0, 0.5) |
| 01 | 1.2391 | (0.5, 0) |
| 001/011 | 1.2392 | (0, 0.5) |
| 000/011 | 1.2397 | (0, 0.5) |
| 000/001/010 | 1.2545 | (0.5, 0.5) |
| 000/001/100 | 1.2617 | (0.5, 0.5) |
| 000/001/101 | 1.2644 | (0.5, 0.5) |
| 000/001/011 | 1.2651 | (0.5, 0.5) |
| 000/011/011 | 1.2654 | (0.5, 0.5) |
| 000/001/111 | 1.2655 | (0.5, 0.5) |
| 000/001 | 1.3032 | (0, 0.5) |
| 000/000/011 | 1.3097 | (0.5, 0.5) |
| 000/000/111 | 1.3105 | (0.5, 0.5) |
| 001/001 | 1.3105 | (0, 0.5) |
| 001 | 1.3105 | (0.5, 0.5) |
| 000/000/001 | 1.3515 | (0.5, 0.5) |
| 000 | 1.4348 | (0.5, 0.5) |
| 000/000 | 1.4348 | (0, 0.5) |
| 000/000/000 | 1.4348 | (0.5, 0.5) |
| 00/00 | 1.4348 | (0, 0) |
| 00 | 1.4348 | (0.5, 0) |
| 0 | 1.4348 | (0.5, 0.5) |

35 periodic-cell representatives with periods up to 3×3 (rows separated by "/"; repeated supercells included), at prescribed macroscopic strain, sorted by threshold. The checkerboard 01/10 is the weakest and sets the reference load. *k* in units of the reciprocal supercell vectors.

**Supplementary Table 24 | Random samples and a planted domain in the three-dimensional frame**

| Sample ($10^3$ cubes) | Domain | Pressure / weakest periodic | Participating cells | Energy in domain |
|---|---|---|---|---|
| random 1 | – | 1.4222 | 95.0% | – |
| random 2 | – | 1.4265 | 93.0% | – |
| random 3 | – | 1.4247 | 93.5% | – |
| random 4 | – | 1.4360 | 87.8% | – |
| random 5 | – | 1.4212 | 93.0% | – |
| random 6 | – | 1.4179 | 92.2% | – |
| random 7 | – | 1.4180 | 90.1% | – |
| random 8 | – | 1.4262 | 91.7% | – |
| random 9 | – | 1.4305 | 94.7% | – |
| random 10 | – | 1.4144 | 94.4% | – |
| background, no domain | – | 1.4108 | 92.1% | – |
| background + $8^3$ domain | $8^3$ | 1.1377 | 49.7% | 81.6% |

Mean of the ten random samples: 1.4238. The planted samples share one random background with a central layer-uniform checkerboard domain. Participating cells: participation ratio of the cell-resolved critical-mode energy, as a fraction of all cells.

**Supplementary Table 25 | TRUMOF-1 thresholds under the objective (−p) and published (−2.5p) pressure conventions**

| Structure | Soft, $-p$ (GPa) | Soft, $-2.5p$ (GPa) | Hard, $-p$ (GPa) | Hard, $-2.5p$ (GPa) |
|---|---|---|---|---|
| $Pc$ (1×1×1) | 0.746 (PD+LE) | 0.372 (PD-only) | 2.611 (PD+LE) | 1.286 (PD-only) |
| $P\bar{4}$ (1×1×1) | 0.924 (PD+LE) | 0.445 (PD+LE) | 3.487 (PD+LE) | 1.578 (PD+LE) |
| $P2_1$ (1×1×1) | 1.746 (PD+LE) | 0.630 (PD-only) | 4.860 (PD+LE) | 2.254 (PD-only) |
| random, $2^3$ (mean) | 1.673 | 0.966 | 5.892 | 2.958 |
| random, $3^3$ (mean) | 1.801 | 1.055 | 6.665 | 3.206 |
| random, $4^3$ (mean) | 1.846 | 1.124 | 6.660 | 3.352 |

The published convention is applied as the isotropic correction $\Delta C = p\delta_{ij}\delta_{kl} - (3p/2)(\delta_{ik}\delta_{jl} + \delta_{il}\delta_{jk})$ to the macroscopic relaxed tangent only; it defines no finite-wavevector operator. Under it the order $Pc < P\bar{4} < P2_1$ and the position of every random sample above the weakest approximant are kept, but $Pc$ and $P2_1$ first lose positive definiteness alone, so the convention does not fix a localizable cap.

**Supplementary Table 26 | First instabilities of the TRUMOF-1 approximants in the coarse-grained model**

| Approximant | First class (soft) | Pressure, soft (GPa) | Pressure, hard (GPa) | Neutral zero modes |
|---|---|---|---|---|
| $Pc$ (1×1×1) | PD+LE | 0.746 | 2.611 | 0 |
| $P\bar{4}$ (1×1×1) | PD+LE | 0.924 | 3.487 | 3 |
| $P2_1$ (1×1×1) | PD+LE | 1.746 | 4.860 | 0 |
| 20 × $P1$ (2×2×2), soft | PD-only 18, finite $k$ 2 | 1.396–2.107 | – | 0 |
| 20 × $P1$ (2×2×2), hard | PD-only 20 | – | 4.869–7.943 | 0 |

Approximant connectivities from the 30 published DFT structures, ten 1×1×1 and twenty 2×2×2 (Data S6 of ref.[13]); all satisfy the reconstructed rule (13,824 origin-fixed assignments give 4 zero-energy states in 3 symmetry classes). Soft and hard: the two parameter sets of the coarse-grained potential[14]. Pressures from the objective Gibbs tangent with the $-p$ correction; the published Source Data correspond to $-2.5p$ (Supplementary Section S7.2). The three neutral modes of $P\bar{4}$ are relative translations of its two non-interacting interpenetrated frameworks and stay at zero at all pressures.

## Supplementary Table 27 | Random and planted TRUMOF-1 samples (soft parameters)

| Sample | n | Mean (GPa) | S.d. (GPa) | Range (GPa) | $s_P$ | Energy in domain |
|---|---|---|---|---|---|---|
| random, $2^3$ | 10 | 1.673 | 0.144 | 1.412–1.832 | – | – |
| random, $3^3$ | 10 | 1.801 | 0.137 | 1.584–2.026 | – | – |
| random, $4^3$ | 10 | 1.846 | 0.074 | 1.730–1.952 | – | – |
| $4^3$ + $P\bar{4}$ domain, $R$ = 1 | 1 | 1.755 | – | – | 0.773 | 2.0% |
| $4^3$ + $P\bar{4}$ domain, $R$ = 2 | 1 | 1.741 | – | – | 1.149 | 37.8% |
| $4^3$ + $P2_1$ domain, $R$ = 1 | 1 | 1.818 | – | – | 1.050 | 0.6% |
| $4^3$ + $P2_1$ domain, $R$ = 2 | 1 | 1.754 | – | – | 0.941 | 16.3% |
| $4^3$ + $Pc$ domain, $R$ = 1 | 1 | 1.504 | – | – | 1.236 | 0.5% |
| $4^3$ + $Pc$ domain, $R$ = 2 | 1 | 1.536 | – | – | 1.031 | 29.3% |
| $6^3$ + $P\bar{4}$ domain, $R$ = 2 | 1 | 1.740 | – | – | 1.166 | 0.5% |
| $6^3$ + $P\bar{4}$ domain, $R$ = 4 | 1 | 1.459 | – | – | 1.117 | 62.2% |

Random samples satisfy the rule exactly (zero rule energy, Metropolis annealing); planted samples carry a cubic approximant domain of side length $R$ cells in a random background. Thresholds from Γ of the supercell and four supercell wavevectors; they are not certified Brillouin-zone minima. The $L$ = 6, $R$ = 4 sample fails first at a sampled finite wavevector. Compact rule-preserving switches of 14 to 26 clusters embed each approximant in a random background. Random samples yield 29 roots of the augmented equilibrium and zero-eigenvalue equations of the relaxed cell tangent (largest equilibrium residual $2.5\times10^{-11}$, largest critical eigenvalue $1.9\times10^{-9}$ GPa) and one sign change at the wavevector (1/2, 1/2, 1/2) of the supercell (one $L$ = 3 sample, at 1.6545 GPa, before its root at 1.6549 GPa), searched at Γ and four supercell wavevectors. Each root lies within 0.4 MPa of the last load reached by continuation.

**Supplementary Table 28 | TRUMOF-1 rule: correlated configurations biased toward a weak or a strong approximant**

| Biased toward | Bias *b* | n | Fraction of clusters | Largest cluster (of 864) | Located onsets (GPa) | Continuation endpoints, mean (range) (GPa) | Iteration limit (GPa) |
|---|---|---|---|---|---|---|---|
| *Pc* | 0.0 | 5 | 0.04 | 2 | – | 5: 1.82 (1.75–1.85) | – |
| *Pc* | 0.8 | 5 | 0.10 | 12 | – | 5: 1.83 (1.73–1.90) | – |
| *Pc* | 1.0 | 5 | 0.13 | 29 | – | 5: 1.81 (1.73–1.90) | – |
| *Pc* | 1.2 | 5 | 0.16 | 58 | – | 5: 1.71 (1.55–1.87) | – |
| *Pc* | 1.3 | 5 | 0.19 | 96 | – | 5: 1.62 (1.53–1.69) | – |
| *Pc* | 1.4 | 5 | 0.28 | 218 | – | 5: 1.34 (1.22–1.50) | – |
| *Pc* | 1.5 | 5 | 0.64 | 553 | – | 5: 0.88 (0.73–1.07) | – |
| *Pc* | 1.6 | 5 | 0.89 | 768 | – | 5: 0.71 (0.68–0.76) | – |
| *Pc* | 1.8 | 5 | 0.99 | 859 | 2: 0.746, 0.737 | 3: 0.74 (0.73–0.74) | – |
| $P2_1$ | 0.0 | 5 | 0.04 | 3 | – | 5: 1.82 (1.75–1.85) | – |
| $P2_1$ | 0.8 | 5 | 0.09 | 9 | – | 5: 1.84 (1.69–1.93) | – |
| $P2_1$ | 1.0 | 5 | 0.12 | 25 | – | 5: 1.80 (1.74–1.84) | – |
| $P2_1$ | 1.2 | 5 | 0.15 | 42 | – | 5: 1.82 (1.62–1.92) | – |
| $P2_1$ | 1.3 | 5 | 0.18 | 75 | – | 5: 1.83 (1.72–1.91) | – |
| $P2_1$ | 1.4 | 5 | 0.24 | 172 | – | 5: 1.74 (1.69–1.78) | – |
| $P2_1$ | 1.5 | 5 | 0.52 | 446 | – | 5: 1.42 (1.31–1.65) | – |
| $P2_1$ | 1.6 | 5 | 0.89 | 767 | – | 2: 1.23 (1.13–1.33) | 3: 1.26, 1.29, 1.30 |
| $P2_1$ | 1.8 | 5 | 1.00 | 861 | 3: 1.746, 1.746, 1.746 | 1: 1.46 | 1: 1.37 |
| *Pc*, tiled | – | 1 | 1.00 | 864 | 1: 0.746 | – | – |
| $P2_1$, tiled | – | 1 | 1.00 | 864 | 1: 1.746 | – | – |

$6^3$ rule-satisfying samples (864 clusters) drawn by the Metropolis annealing of Supplementary Table 27 with a fixed bias *b* on clusters that differ from the biased approximant (*b* = 0 is the unbiased ensemble, listed under both; its order fraction and largest cluster are evaluated for each target). Soft parameters, objective tangent, straight-angle kernel of S7.2. Located onsets: a recorded sign change of the relaxed cell tangent or a Γ internal mode. Continuation endpoints: the last converged load where the continuation could not be extended; iteration limit: the continuation stopped after 120 load steps. Endpoints are not onsets: they are not averaged with onsets, and positivity at the four supercell wavevectors tested does not certify the full Bloch spectrum.

## Supplementary References